\documentclass[11pt,a4paper]{article}
\usepackage{graphicx}
\usepackage{etoolbox}
\usepackage{float}
\usepackage{amsmath, amssymb, amsthm, accents}
\usepackage{centernot}
\usepackage{bm}
\usepackage{booktabs}
\usepackage{hyperref}
\ExplSyntaxOn
\cs_if_free:NT \pdfannot_link_off:
  {
    \sys_if_engine_luatex:TF
      {
        \cs_new_protected:Npn \pdfannot_link_off: { \tex_pdfextension:D linkstate 1 ~ }
        \cs_new_protected:Npn \pdfannot_link_on:  { \tex_pdfextension:D linkstate 0 ~ }
      }
      {
        \cs_if_exist:cT { pdfrunninglinkoff }
          {
            \cs_new_protected:Npn \pdfannot_link_off: { \use:c { pdfrunninglinkoff } }
            \cs_new_protected:Npn \pdfannot_link_on:  { \use:c { pdfrunninglinkon } }
          }
      }
  }
\ExplSyntaxOff
\makeatletter
\@ifundefined{@outputbox@appendfootnotes}{}{%
  \patchcmd\@outputbox@appendfootnotes
    {\unvbox \footins}{\vbox{\unvbox\footins}}{}{%
    \PackageWarningNoLine{contiguity}{Footnote link patch not applied}}}
\makeatother
\usepackage{enumitem}

\usepackage{natbib}
\defcitealias{berliant1992InternationalEconomicReview}{BF}

\makeatletter
\def\widebar{\accentset{{\cc@style\underline{\mskip10mu}}}}
\makeatother

\newcommand{\vertiii}[1]{{\left\vert\kern-0.25ex\left\vert\kern-0.25ex\left\vert #1 \right\vert\kern-0.25ex\right\vert\kern-0.25ex\right\vert}}

\usepackage{setspace}
\newtheorem{rem}{Remark}

\newtheorem{assumption}{Assumption}

\newtheorem{theorem}{Theorem}
\newtheorem{proposition}{Proposition}
\newtheorem{lemma}{Lemma}
\newtheorem{definition}{Definition}	
\newtheorem{corollary}{Corollary}
\newcommand{\lasttheoremnumber}{0}

\newcommand{\setcorollaryparent}[1]{%
    \xdef\lasttheoremnumber{#1}%
    \setcounter{corollary}{0}%
}
\AtEndEnvironment{theorem}{\setcorollaryparent{\thetheorem}}
\AtEndEnvironment{proposition}{\setcorollaryparent{\theproposition}}
\AtEndEnvironment{lemma}{\setcorollaryparent{\thelemma}}
\theoremstyle{definition}
\newtheorem{example}{Example}
\theoremstyle{plain}

\makeatletter
\let\example@orig\example
\let\endexample@orig\endexample
\renewenvironment{example}[1][]{%
    \pushQED{\qed}%
    \if\relax\detokenize{#1}\relax
        \example@orig
    \else
        \example@orig[#1]%
    \fi
}{%
    \popQED
    \endexample@orig
}
\makeatother

\title{Endogenous Parcel Contiguity and Rent Indeterminacy in a Finite-Household Monocentric City\footnote{%
This work was supported by JSPS KAKENHI Grant Numbers 26K00334 and 24K03007. All remaining errors are my own.}}
\date{November 13, 2026}
\author{Shota Fujishima\footnote{Graduate School of Economics, Hitotsubashi University, 2-1 Naka, Kunitachi, Tokyo 186-8601, Japan. Email: {\tt s.fujishima@r.hit-u.ac.jp}; Website: \texttt{https://sfujishima.github.io/}}}

\begin{document}

\maketitle

\begin{abstract}
We study a monocentric city with finitely many households, each of which may hold any measurable set of land and incurs commuting costs that depend on the spatial distribution of its holdings. In equilibrium, every parcel is an interval and parcels are ordered by lot size from the center outward. The freedom to adjust land holdings at any location restricts supporting rents to a band between the competing bid rents and the assigned household's bid rent. This band excludes rents that support the same allocation when lots are restricted to intervals, including the one constructed by Berliant and Fujita (1992), and makes the existence of a supporting rent a substantive question even with identical households. We give sufficient conditions for the existence of a locally supporting rent that meets every household's budget and deters all small deviations. We also show that any equilibrium minimizes the real resource cost of attaining equilibrium utilities and extend the sorting result to heterogeneous households under a single-crossing condition. Finally, we discuss the relation to hedonic models, distinguishing the prices of whole parcels from location-specific rents that govern the cost of changes in parcel boundaries and fragmentation.


\end{abstract}

\section{Introduction}

The monocentric city model, which originates in the work of Alonso, Muth,
and Mills, is the workhorse of urban land-use theory
\citep[see, e.g.,][]{fujita1989}.  In its standard form, a continuum of
households commutes to a central business district (CBD), and land use is
described locally by schedules of rent, lot size, and population density,
with the market rent at each location equal to the highest bid rent.  This
local description is the source of the model's tractability, but it does
not assign a positive-measure parcel to any individual household.  Urban
land, by contrast, is both spatially divisible and privately consumed in
positive-measure parcels, and when households are finite, parcel
boundaries, fragmentation, and deviations that rearrange an entire parcel
become part of the economic problem.  Indeed,
\citet{berliant1985RegionalScienceandUrbanEconomics} shows that a continuum
of consumers cannot each hold a parcel of positive area, because only
countably many disjoint parcels of positive area fit in a Euclidean space.
In addition,
\citet{berliant2008RegionalScienceandUrbanEconomics} demonstrates that
finite-agent and continuum-agent land models can yield qualitatively
different comparative statics.

Parcels are also the units on which much of land policy operates.
Eminent domain often takes only part of a parcel, as when a strip is taken
to widen a road \citep{bell2017ColumbiaLawReview}, redevelopment often
requires the assembly of several parcels
\citep{brooks2016AmericanEconomicJournal:EconomicPolicy}, and subdivision
regulations govern how, and how quickly, parcels may be divided
\citep{white1988JournalofUrbanEconomics,wrenn2015RegionalScienceandUrbanEconomics}.
These policies value or reshape land that need not coincide with any
traded parcel.  Valuing such
land requires a model in which both parcels and the rents within them are
determined in equilibrium: the continuum model has no parcels, and models in
which lots or dwellings of given size and location are traded as indivisible
units
\citep[e.g.,][]{kaneko1983JournalofUrbanEconomics,miyake2003Reg.Sci.UrbanEcon.}
price only whole units.

The natural benchmark for such a model is
\citet[hereafter BF]{berliant1992InternationalEconomicReview}, who formalize the
monocentric city model with finitely many households and establish
existence and both welfare theorems by constructing supporting rent
functions explicitly.  In their model each household's lot is restricted to
be an interval.  This restriction leaves two questions open.  Contiguity is
assumed, so the model cannot say whether a land market eliminates fragmented
holdings, which are common in agriculture, where they raise the cost of
travel between plots
\citep{blarel1992WorldBankEconRev,ali2019WorldBankEcon.Rev.}.  Moreover, a household can adjust its land only by moving the
endpoints of its lot and cannot buy a detached piece, so the rent inside a
lot is disciplined only by deviations to other intervals and need not
reflect what any household would pay for a small piece of land there.  We therefore
let each household choose an
arbitrary measurable land assignment, represented by an indicator function
$f_i(r)\in\{0,1\}$ over locations $r\geq0$.  The choice set therefore
includes disconnected and mosaic-like holdings as well as intervals.  We
evaluate commuting cost at the centroid of the assigned land, which is the
expected commuting cost when the trip starts from a point drawn uniformly
from the household's land.\footnote{Section \ref{subsec:relation-bf}
explains why the front of the lot, at which
\citetalias{berliant1992InternationalEconomicReview} evaluate commuting cost, is
no longer a harmless normalization once lots need not be intervals.}
Each unit of land thus contributes to commuting cost in proportion to its
distance from the CBD, wherever the rest of the household's land lies.

We first show that this property leads equilibrium to eliminate
fragmentation.  A household's bid rent declines with distance from
the CBD, and more steeply the smaller its lot, so the bid rents of two
households cross at most once and each household outbids the others on a
single interval.  Households that share a bid-rent function would be
indifferent across their common region, so the pointwise
first-order conditions allow them to hold fragmented or interlaced land there,
but global utility maximization rules this out.  Every equilibrium parcel is
consequently a nonempty interval, and parcels are strictly ordered by lot size
from the CBD outward.  The ordering itself parallels
\citetalias{berliant1992InternationalEconomicReview}; what is new is that the
interval structure is an equilibrium outcome rather than a restriction on the
feasible set.  Fragmented holdings may persist because of factors absent from our model, such as risk diversification or transaction costs that impede consolidation.

The enlarged choice set matters chiefly for prices.  Because a household can
add or drop a small piece of land at any location, its first-order conditions
bind pointwise, and not only at the endpoints of its lot.  A first-order
supporting rent must therefore lie between the highest competing bid rent and
the assigned household's bid rent, a band that closes at parcel boundaries.
The band excludes some of the rents that support the same allocation when
lots are restricted to intervals, including the one constructed by
\citetalias{berliant1992InternationalEconomicReview}.  Rent indeterminacy thus
survives, but part of it in the interval model reflects the restriction of
lots to intervals rather than the discreteness of the locational ordering.

The band also makes the existence of a supporting price a substantive
question, even with identical households and a common income.  Equilibrium
requires that every household attain the common utility level with the
common income, which we call budget consistency.  With finitely many
households, the upper envelope of bid rents, the market rent of the continuum
model, violates it whenever the marginal rate of substitution is strictly
convex in land, so rents must be discounted below the assigned household's
bid rent.  Because the band keeps the rent above the
competing bid rents, budget consistency restricts the common income.
Moreover, first-order support and
budget consistency do not imply household optimality: we give an example in
which both hold but a second-order condition for interval deviations fails.
We provide sufficient conditions under which a strictly convex rent satisfies
first-order support and budget consistency and rules out all small
deviations, including measurable ones.  Whether it also rules out large
deviations is left open, although a grid search finds none in our numerical
examples.

We also show that the first welfare theorem survives the enlargement of the
choice set.  Following the Herbert--Stevens formulation in
\citet{fujita1989}, an equilibrium allocation minimizes the real resource
cost of providing households with their equilibrium utilities.  The argument
requires neither identical households nor convex choice sets.  Whenever an
equilibrium exists, it is therefore efficient, and the conditions above are
needed for existence rather than for efficiency.

We then extend the analysis to households that differ in preferences,
income, or commuting costs.  Once commuting costs differ, as when the value
of commuting time rises with income, lot size alone no longer determines
spatial order, and the relevant sorting index is bid-rent steepness, the
ratio of commuting cost to lot size.  Under a pairwise
single-crossing condition on parcels, the finite-parcel counterpart of the
income-elasticity condition in continuum models, each type occupies a
contiguous block, so sorting occurs at two levels, across types and by lot
size within each type.  We do not establish existence in this case, but we
outline a strategy for constructing locally supporting prices and give a
numerical example in which it yields an explicit price.  The welfare result above covers
this case as well.

Finally, these results clarify what rent indeterminacy means for empirical
work and for the policies discussed above.  Supporting rents can differ in
two ways.  One is the level: the common utility level determines how the
surplus is divided between landlords and households.  The other is the
profile: given the utility level, budget consistency pins down the total
payment for each parcel but not how it is spread over the locations in the
parcel.  The level layer is familiar from hedonic models,
whereas the profile layer arises because land prices are additive over
locations and parcels are endogenous.  Although parcel transactions do not
reveal the profile, the price of any union of existing parcels does not
depend on it.  The profile matters whenever parts of
parcels are priced, as in compensation for partial takings and in
land-value surfaces
\citep{mcmillen1996JournalofUrbanEconomics,ahlfeldt2018TheReviewofEconomicsandStatistics}.
The model nonetheless restricts the profile: rents are pinned down at parcel
boundaries and bounded between them, so some of these prices are determined
or bounded without selecting a particular profile (Remark
\ref{rem:parcel-payments-hedonic} and Appendix
\ref{app:hedonic-estimation}).

The remainder of the paper is organized as follows.  Section
\ref{subsec:relation-bf} relates the model to
\citetalias{berliant1992InternationalEconomicReview} in detail, and Section
\ref{subsec:related-literature} reviews the related literature.  Section
\ref{sec:homogeneous} develops the homogeneous-household benchmark, from
household optimization and the structure of equilibrium allocations,
including the exclusion of degenerate equilibria, to supporting prices.  Section
\ref{sec:heterogeneous} extends the analysis to heterogeneous households,
Section \ref{sec:welfare} establishes the first welfare theorem, and Section
\ref{sec:conclusion} concludes.  Appendix \ref{app:endpoint-stability}
collects the endpoint stability conditions, Appendix
\ref{app:auxiliary-lemmas} the auxiliary lemmas for the supporting-price
construction, and Appendix \ref{app:hedonic-estimation} the implications for
hedonic estimation.

\subsection{Relation to \texorpdfstring{\citet{berliant1992InternationalEconomicReview}}{Berliant and Fujita (1992)}}
\label{subsec:relation-bf}

The present model is closest to that of
\citetalias{berliant1992InternationalEconomicReview}, so this subsection
discusses the relation in detail.  The model departs from theirs in two
respects: households may hold any measurable set of land rather than an
interval, and commuting cost is evaluated at the centroid of the land held
rather than at the front of the lot.  We first explain why the larger choice
set requires the change in commuting cost and then summarize how the two
departures affect the results.

In \citetalias{berliant1992InternationalEconomicReview}, commuting cost is
evaluated at the front of the lot, the point closest to the CBD, and they
note that their main results extend to a fixed weighted average of its front
and back, including its centroid.  On intervals, the choice is indeed a
harmless normalization: moving the commuting origin from the front to the
centroid acts as a uniform surcharge on rent (Lemma
\ref{lem:bf-correspondence}), so the equilibrium lot-size profile derived
below is the one that obtains in the front-based interval model with the
agricultural rent raised by this surcharge.  The front-based rule, however,
does not extend
to measurable assignments.  If commuting cost were evaluated at the innermost
point of the assigned set, a household could move its commuting origin to the
CBD by acquiring an arbitrarily small piece of land there, and it could place
the rest of its land wherever rent is lowest without affecting its commuting
cost, so commuting costs would cease to discipline location.  The difficulty
is that this rule makes commuting cost discontinuous in the land assignment
and makes the location of most of a household's land irrelevant to its
commuting cost.  Extending the model beyond intervals therefore requires a
commuting cost that is continuous in the land assignment and increases with
the distance of every unit of land from the CBD.\footnote{%
\label{fn:equivalent-front}The problem lies in defining the front by a single
point of the assigned set.  A front defined from the whole distribution of
land, such as $r_c(A)-|A|/2$ for a measurable set $A$ with centroid $r_c(A)$
and measure $|A|$, coincides with the usual front on intervals and has both
properties.  It is hard to interpret, however, because it is the left end of a
hypothetical interval rather than the origin of any trip.}  The centroid is an interpretable
benchmark with both properties.

The enlarged choice set leaves the lot-size profile unchanged but changes
the set of supporting rents.
Convex rents support an allocation here exactly when, with the uniform
surcharge added, they support it in the interval model (Lemmas
\ref{lem:convex-interval-reduction} and \ref{lem:bf-correspondence}).  The
bid-rent band, however, excludes nonconvex rents such as the one constructed
by \citetalias{berliant1992InternationalEconomicReview}, which is flat just inside
each interior parcel boundary: the household located just outside the
boundary would gain by acquiring a small detached piece of land just inside
it (Remark \ref{rem:bf-prices-excluded}).

The band also makes the existence of a supporting price harder to
establish.  In the interval model, the rent inside a lot can be shaped freely, so it can be discounted enough to
meet each household's budget and, with the flat segments that
\citetalias{berliant1992InternationalEconomicReview} use, to make each lot
globally optimal among intervals.  A supporting price therefore exists
without further conditions.  Here, the band limits the discount and rules
out the flat segments.  The conditions on the common income and on
preferences in Theorem \ref{thm:supporting-price-existence} are thus the
price of endogenizing contiguity.  Global optimality still reduces to the
same interval problem (Lemma \ref{lem:bf-correspondence}), but it must be
verified for a rent that satisfies the bid-rent bounds, and we leave this
step open (Section \ref{subsubsec:interval-local-stability}).

The two papers also differ in how far they treat heterogeneous households.
\citetalias{berliant1992InternationalEconomicReview} allow incomes to differ and
establish existence for any profile of positive incomes, but assume a common
utility function and a common commuting cost, so that lot size still orders
households.  They conjecture that their techniques extend to different
utility functions when incomes and bid-rent steepness rank households in the
same order, and leave this extension for future research.
Neither paper treats the general case completely.  We establish two-level
sorting under a pairwise single-crossing condition and outline a strategy for
constructing locally supporting prices, but we do not establish existence.

Finally, the connection to hedonic and housing-market models discussed in
Section \ref{subsec:related-literature} and the distinction between the
level and the profile of rents do not rely on either departure.  Both
rest on arguments that use only interval deviations, so they apply to the
interval model as well.
\citetalias{berliant1992InternationalEconomicReview} identify a continuum of
supporting rents but do not relate it to these models.  What the enlarged
choice set adds is the band, which bounds rents pointwise within parcels.
The profile is therefore restricted more tightly than in the interval model,
which narrows the range of prices consistent with the same parcel data
(Appendix \ref{app:hedonic-estimation}).

\subsection{Related Literature}
\label{subsec:related-literature}

The closest literature can be organized around two alternative descriptions of
the spatial commodity.  The first is the Alonso--Berliant--Fujita tradition of
finite-population land-use models, of which
\citetalias{berliant1992InternationalEconomicReview} is the benchmark discussed
above.  Subsequent work studies the equilibrium--core relationship and
alternative ownership institutions
\citep{berliant2007IntEconomicReview,lafountain2008IntJofEconomicTheory}, while
\citet{berliant2013RegionalScienceandUrbanEconomics} analyzes bid-rent
equilibrium and information revelation.  Throughout this tradition, households
consume positive-measure parcels and pay additive spatial rents, and each lot
is restricted to be an interval.

A complementary foundation is provided by the measurable-subset approach.
\citet{berliant1985RegionalScienceandUrbanEconomics} represents land consumption
as subsets of a land space, and \citet{berliant2004JournalofMathematicalEconomics}
develop a general location-theoretic framework based on a $\sigma$-algebra of
measurable subsets.  These contributions establish general-equilibrium and
welfare foundations but do not impose a monocentric urban structure with
commuting costs and spatial bid-rent competition.  The present
paper combines their commodity-space formulation with the finite-population
monocentric environment of the Alonso--Berliant--Fujita tradition.

The second closely related literature treats dwellings as differentiated
products whose characteristics are given, either as a finite collection of
indivisible units or as points in a continuum of characteristics.  In the
first strand, \citet{kaneko1983JournalofUrbanEconomics} develops a
competitive rental-housing model in which households choose one apartment
from a finite collection of indivisible units.\footnote{%
\citet{kaneko2006JournalofUrbanEconomics} describe their model as an
extension of the assignment market of \citet{shapley1971IntJGameTheory},
in which the core coincides with the set of competitive outcomes and
contains a seller-optimal and a buyer-optimal extreme point.  More
generally, price indeterminacy and the possible nonexistence of equilibrium
are familiar features of markets for indivisible goods: with general utility
functions, the set of equilibrium price vectors is a lattice with smallest
and largest elements \citep{demange1985Econometrica}, and existence is
guaranteed only for certain classes of preferences \citep{baldwin2019ECTA}.
Both features arise because supporting prices are defined by inequalities
rather than tangency conditions, so their set can be large or empty.  When
nonconvexities cause Walrasian equilibrium to fail,
\citet{milgrom2026Rev.Econ.Stud.} propose markup equilibria as an
alternative.}  Building on this framework,
\citet{kaneko2006JournalofUrbanEconomics} classify apartments into categories
defined by attributes such as size, commuting time, and region.  Because
equilibrium rents are generally not unique, they study the comparative
statics of the maximal competitive rent vector.
\citet{kaneko2017SpatialInteractionModels} then
connect this equilibrium framework to econometric analysis using data on
rents, apartment sizes, and commuting time-distance.  In the second strand,
the hedonic model of \citet{rosen1974JournalofPoliticalEconomy}, the market
price of a differentiated product is a function of its characteristics.
The two strands are closely related:
\citet{chiappori2010EconTheory} show that, with quasilinear preferences,
hedonic equilibria correspond to stable matchings with transferable utility,
of which the assignment market is a finite version.

The present model has a counterpart of both.  The price of a parcel, as a
function of its size and centroid, coincides with the households' common bid
function at the equilibrium parcels (Remark
\ref{rem:parcel-payments-hedonic}), as a hedonic price function coincides
with each consumer's bid function at the product that the consumer buys.
The prices of given products are not unique in either strand; in hedonic
models, marginal payoffs are unique but their levels are not
\citep{chiappori2010EconTheory}.  This indeterminacy corresponds to the
level layer here.

What differs is the commodity, and this difference gives rise to the profile
layer.  In both strands, a dwelling is a primitive product whose size and
accessibility are given, and its price can be set unit by unit or as an
arbitrary function of its characteristics.  Here, households assemble
location-specific units of land into parcels, so parcel size, centroid,
boundaries, and contiguity are equilibrium outcomes, and the price of a
parcel is the integral of location-specific rents over it.  The
indeterminacy of the rent profile within parcels therefore has no
counterpart when products are fixed.

\section{The Homogeneous-Household Benchmark}
\label{sec:homogeneous}

This section develops the homogeneous-household benchmark. We first specify
the environment and derive,
under centroid-based commuting costs, the pointwise first-order condition and
the resulting affine bid rents. We then show that, in a non-degenerate
equilibrium, parcels are intervals ordered by lot size from the CBD, and we
determine their lot sizes at a given utility level. We also show that
degenerate equilibria do not exist, so that every equilibrium has this
ordered, contiguous structure. Next, we study the rents
that support such an allocation. We characterize first-order support by
bid-rent bounds, examine budget consistency, derive second-order conditions
under which each parcel is a strict local expenditure minimizer, and relate
the household problem to the front-based interval model of
\citetalias{berliant1992InternationalEconomicReview}. We then construct a rent that
satisfies all of these conditions; whether it also rules out large
deviations is left open. Welfare properties, which also cover the heterogeneous
extension, are discussed in Section \ref{sec:welfare}.

\subsection{Environment and Preferences}

Consider a linear region $[0,\hat r]$, with the CBD at $r=0$, and a
discrete set of identical households $\mathcal{N} = \{1, 2, \dots, N\}$,
where $N\geq2$ and each household has a positive mass $m>0$. Land that is not
occupied by households is used for agriculture and earns the agricultural rent
$p_A>0$. The urban boundary $\bar r\in(0,\hat r)$, the outer edge of
residential land, is determined in equilibrium. The outer edge $\hat r$ of the
region only keeps the choice set bounded and is taken large enough never to
bind in equilibrium. The finite
model is written with a general mass $m$ to make its connection to the
standard continuum formulation transparent. A purely discrete normalization
sets $m=1$. Alternatively, a unit-mass continuum can be approximated by
setting $m=1/N$ and taking $N\to\infty$, so that each finite household
represents a vanishingly small group of otherwise identical consumers.

Let $f_i(r) \in \{0,1\}$ be a measurable indicator function that equals 1 if household $i$ occupies land at location $r$, and 0 otherwise.\footnote{%
Throughout, measurability and ``almost everywhere'' (a.e.)\ refer to Lebesgue
measure on $[0,\hat r]$, and $|A|$ denotes the Lebesgue measure of a
measurable set $A\subseteq[0,\hat r]$.  We do not distinguish between land
assignments that differ only on a set of measure zero.  Rent functions are taken to
be integrable; footnote \ref{fn:pointwise-foc} explains how we treat rent
functions that agree almost everywhere.}
The total land occupied by household $i$ is
\begin{equation}\label{eq:land_consumption}
    S_i = \int_0^{\hat r} f_i(r)\,dr.
\end{equation}
The per-capita land consumption for household $i$ is therefore
\(
    s_i = S_i/m.
\)

The per-capita utility of household $i$ is
\begin{equation}
    u_i = U(z_i, s_i),
\end{equation}
where $z_i$ is per-capita consumption of the numeraire good.  We maintain
two assumptions on preferences.  The first collects standard regularity
conditions.

\begin{assumption}\label{ass:primitive}
\begin{enumerate}[label=(A\arabic*)]
    \item $U$ is twice continuously differentiable and strictly quasi-concave,
    with $U_z>0$ and $U_s>0$.
    \item For every $s>0$, $\lim_{z\downarrow0} U(z,s)=-\infty$ and
    $\lim_{z\to\infty} U(z,s)=\infty$.
\end{enumerate}
\end{assumption}

Under Assumption \ref{ass:primitive}, for each $s>0$ and $u\in\mathbb R$,
there exists a unique $Z(s,u)$ such that $U(Z(s,u),s)=u$. By the Implicit Function Theorem,
\[
    Z_s(s,u)
    =
    -\frac{U_s(Z(s,u),s)}{U_z(Z(s,u),s)}.
\]
We define the marginal rate of substitution by
\[
    MRS(s,u)
    :=
    -Z_s(s,u)
    =
    \frac{U_s(Z(s,u),s)}{U_z(Z(s,u),s)}.
\]
The second assumption makes land essential, which ensures that every
household holds a positive amount of land in equilibrium.

\begin{assumption}[Land essentiality]\label{ass:land-essentiality}
For every $z>0$, $\lim_{s\downarrow0}U(z,s)=-\infty$.
\end{assumption}

The following additional assumptions will be invoked only where explicitly
stated.
\begin{assumption}\label{ass:mrs-boundary}
    For each utility level $u$, $MRS(s,u)$ satisfies $\lim_{s\to 0} MRS(s,u)=\infty$ and $\lim_{s\to\infty} MRS(s,u)=0$. That is, the utility function represents a smooth preference \citep{hayashi2021}.
\end{assumption}

\begin{assumption}[Convex marginal willingness to pay]
\label{ass:mrs-convexity}
For every utility level $u$, the function $s\longmapsto MRS(s,u)$ is convex on $(0,\infty)$.
\end{assumption}

\begin{assumption}[Normality of land]
\label{ass:land-normality}
For every $s>0$ and $u\in\mathbb R$, $MRS_u(s,u)>0$.
Under the maintained regularity conditions, this is equivalent to requiring
that the income effect on the Marshallian demand for land be positive.
\end{assumption}

The per-capita budget constraint is
\begin{equation}\label{eq:bc}
z_i + \frac{1}{m}\int_0^{\hat r} p(r) f_i(r)\,dr + t\,r_{c,i} = w,
\end{equation}
where $p(r)$ is the land rent at location $r$, $w$ is per-capita income, and
$t\,r_{c,i}$ is the commuting cost specified below.

All land is owned by absentee landlords. Rental income is not rebated to
resident households and therefore does not appear in the household budget
constraint. The landlord sector is otherwise suppressed.

Assuming that the starting point of the commute is drawn uniformly at random
from the allocated land, the expected commuting cost is $t\,r_{c,i}$, where
$t>0$ is the marginal commuting cost and, for $S_i>0$,
\begin{equation}\label{eq:commuting_cost}
r_{c,i} := \frac{1}{S_i}\int_0^{\hat r} r f_i(r)\,dr
\end{equation}
is the center of mass (centroid) of household $i$'s land assignment.  As a
formal convention, a household occupying no land incurs no commuting cost.
More generally, for a measurable set $A\subseteq[0,\hat r]$ with $|A|>0$, we
write $r_c(A):=|A|^{-1}\int_A r\,dr$ for its centroid, so that
$r_{c,i}=r_c(\{r:f_i(r)=1\})$.

\subsection{Household Optimization and Bid Rents}

Because the household's choice variable $f_i(r)$ is a function defined over the continuum of locations $r\in[0,\hat r]$, rather than a finite-dimensional vector, household $i$'s problem is naturally formulated as an optimal control problem, with $r$ playing the role of the control problem's independent variable and $f_i(r)$ the control. To apply standard optimal control methods, we relax the control constraint from $f_i(r)\in\{0,1\}$ to the convex set $f_i(r)\in[0,1]$. The resulting Hamiltonian will turn out to be linear in $f_i(r)$, so the Pontryagin Maximum Principle yields a pointwise condition under which the optimal control equals $0$ or $1$ wherever the household is not indifferent. As explained below, the same condition is necessary for optimality in the original $\{0,1\}$-valued problem. By the standard duality between utility maximization and expenditure minimization, this problem can equivalently be characterized as an expenditure-minimization problem: given a target utility level $u$, household $i$ chooses the land assignment $f_i(r)\in[0,1]$ to minimize total expenditures.

The one complication is that the commuting-cost term depends on the centroid $r_{c,i}$, a ratio of integrals over the entire distribution $f_i(r)$, so total expenditure is initially a non-local functional of $f_i(r)$ rather than a simple integral of a per-location cost. We resolve this by introducing three state variables: the cumulative land area, the cumulative rent payment, and the cumulative first moment of land about the CBD, which is the numerator of the centroid in \eqref{eq:commuting_cost}. The centroid, and hence total expenditure, is then recovered from their values at the outer edge $r=\hat r$. With this construction, the objective becomes a function of the state only at the terminal point $\hat r$, so the household's problem is naturally posed as a Mayer problem and can be solved in a single step using optimal control theory.\footnote{Commuting from the innermost point $\inf\{r:f_i(r)>0\}$, as in
\citet{berliant2026} and \citetalias{berliant1992InternationalEconomicReview},
makes commuting cost discontinuous in the land assignment once lots need not
be intervals, as explained in Section \ref{subsec:relation-bf}.  It also cannot be recovered
from the terminal values of the state variables below, so representing it as
an optimal control problem requires a different formulation, for example one
in which the innermost point is chosen separately as a free initial point.
The distribution-based front in footnote \ref{fn:equivalent-front}, by
contrast, equals $M_i(\hat r)/S_i(\hat r)-S_i(\hat r)/2$ and fits the
formulation below.}%

Using the expenditure function $Z(s,u)$ defined above, the state variables for household $i$ are defined as:
\begin{align}
    S_i'(r) &= f_i(r), \quad S_i(0) = 0 & \text{(cumulative land area)} \label{eq:state_S}\\
    M_i'(r) &= r f_i(r), \quad M_i(0) = 0 & \text{(cumulative first moment)} \label{eq:state_M}\\
    Q_i'(r) &= \frac{1}{m} p(r) f_i(r), \quad Q_i(0) = 0 & \text{(cumulative rent payment)} \label{eq:state_E}
\end{align}
Taking the rent function $p$ as given, household $i$ minimizes total expenditure $\Phi_i$, evaluated at $r=\hat r$:
\begin{equation}
    \min_{f_i(r) \in [0,1]} \Phi_i \equiv Z\left(\frac{S_i(\hat r)}{m}, u\right) + Q_i(\hat r) + t \frac{M_i(\hat r)}{S_i(\hat r)}
\end{equation}

To solve this as a Mayer problem, we define the Hamiltonian $\mathcal{H}_i = \lambda_{S_i} S_i' + \lambda_{M_i} M_i' + \lambda_{Q_i} Q_i'$. Since the objective is minimization, the adjoint variables $\lambda$ (shadow prices) are given by the negative gradient of $\Phi_i$ at the terminal state:
\begin{align}
    \lambda_{Q_i} &= -\frac{\partial \Phi_i}{\partial Q_i} = -1 \label{eq:costate_E}\\
    \lambda_{M_i} &= -\frac{\partial \Phi_i}{\partial M_i} = -\frac{t}{S_i} \label{eq:costate_M}\\
    \lambda_{S_i} &= -\frac{\partial \Phi_i}{\partial S_i} = -\frac{1}{m} \frac{\partial Z(s_i, u)}{\partial s_i} + \frac{t M_i}{S_i^2} = -\frac{1}{m} Z_s + \frac{t r_{c,i}}{S_i} \label{eq:costate_S}
\end{align}
since $r_{c,i} = M_i/S_i$ by equation \eqref{eq:commuting_cost}.

Substituting the state dynamics \eqref{eq:state_S}--\eqref{eq:state_E} and the costate values \eqref{eq:costate_E}--\eqref{eq:costate_S} into the Hamiltonian, we obtain:
\begin{equation}
    \mathcal{H}_i = \left[ -\frac{1}{m} p(r) - \frac{t}{S_i} r - \frac{1}{m} Z_s + \frac{t r_{c,i}}{S_i} \right] f_i(r)
\end{equation}
Factoring out $\frac{1}{m}$, the optimal control implies the following first-order condition:
\begin{equation}\label{eq:foc}
f_i^*(r)=
\begin{cases}
1 & \text{if} \quad -Z_s(s_i, u) > p(r) + \frac{t}{s_i}(r - r_{c,i}), \\
\in [0, 1] & \text{if} \quad -Z_s(s_i, u) = p(r) + \frac{t}{s_i}(r - r_{c,i}), \\
0 & \text{if} \quad -Z_s(s_i, u) < p(r) + \frac{t}{s_i}(r - r_{c,i}).
\end{cases}
\end{equation}
Thus $f_i^*(r)$ takes the extreme values $\{0,1\}$ wherever the two sides differ. Where they coincide, the household is indifferent and the condition does not determine $f_i^*(r)$. This indifference set need not have measure zero: when several households share a bid-rent function, the rent can equal their common bid rent on a region of positive measure, the case examined in Theorem \ref{thm:non-degenerate-equilibria}. The relaxation is only a device: \eqref{eq:foc}, with $f_i^*(r)\in\{0,1\}$ in the middle case, is also necessary for optimality when $f_i$ must take values in $\{0,1\}$.\footnote{%
Without the relaxation, let $A$ be the household's land, with
$|A|=ms_i>0$, let $\Phi_i(A)$ be its expenditure, and write
$g_i(r):=p(r)+\frac{t}{s_i}(r-r_{c,i})+Z_s(s_i,u)$.
For measurable sets $B\subseteq[0,\hat r]\setminus A$ and $D\subseteq A$,
\[
    \Phi_i(A\cup B)-\Phi_i(A)=\frac1m\int_B g_i(r)\,dr+o(|B|),
    \qquad
    \Phi_i(A\setminus D)-\Phi_i(A)=-\frac1m\int_D g_i(r)\,dr+o(|D|).
\]
Optimality of $A$ therefore requires $g_i\ge0$ a.e.\ outside $A$ and
$g_i\le0$ a.e.\ on $A$, which is \eqref{eq:foc} with
$f_i^*=\mathbf{1}_A$.}

The first-order condition implies that the maximum rent household $i$ is willing to pay at location $r$ is determined by the right-hand side of equation \eqref{eq:foc}. Because $MRS(s_i,u)=-Z_s(s_i,u)$, we formally define this as household $i$'s \textit{bid rent}, denoted by $\Psi_i(r, u)$:
\begin{equation}\label{eq:bid_rent}
    \Psi_i(r, u) = MRS(s_i,u) - \frac{t}{s_i}(r - r_{c,i}).
\end{equation}
The bid-rent function is affine in location. We define household $i$'s
\emph{bid-rent steepness} by
\begin{equation}\label{eq:bid-rent-steepness}
    b_i:=\frac{t}{s_i}>0,
\end{equation}
so that $(\Psi_i)'=-b_i$. Thus a household with a smaller lot has a steeper
bid-rent function.

\subsection{Ordered Candidate Allocations}

Having established the household's optimization behavior, we next introduce
the Walrasian equilibrium concept. Consistently with Assumptions
\ref{ass:primitive} and \ref{ass:land-essentiality}, we extend utility to the
boundary by setting \(U(0,s)=U(z,0)=-\infty\). We call an allocation
$a=\{z_i,f_i\}_{i=1}^N$
\textit{feasible} if numeraire consumption is nonnegative, $z_i\geq0$ for
every $i$, and land assignments do not overlap, $\sum_{i=1}^N f_i(r)\leq1$
for a.e. $r\in[0,\hat r]$.

\begin{definition}[Walrasian Equilibrium]\label{def:walrasian-equilibrium}
An urban boundary $\bar r\in(0,\hat r)$, a land rent function
$p:[0,\hat r]\to\mathbb R$, and a feasible
allocation $a^\ast=\{z_i^\ast, f_i^\ast\}_{i=1}^N$ constitute a Walrasian
equilibrium if the following conditions are satisfied:
\begin{enumerate}
    \item[(i)] Utility Maximization: Given $p$, for each
    household $i \in \{1, \dots, N\}$, the allocation
    $\{z_i^\ast, f_i^\ast\}$ maximizes utility $u_i$ subject to
    \eqref{eq:land_consumption} and budget constraint \eqref{eq:bc}.

    \item[(ii)] Market Clearing: The land market clears at every location: $\sum_{i=1}^N f_i^\ast(r) = 1$ for a.e. $r \in [0,\bar r]$, and land beyond the urban boundary remains in agriculture, $\sum_{i=1}^N f_i^\ast(r) = 0$ for a.e. $r\in(\bar r,\hat r]$.

    \item[(iii)] Land Supply and Boundary Condition: The urban land rent weakly exceeds the agricultural rent, $p(r)\geq p_A$ for a.e. $r\in[0,\bar r]$, and land beyond the urban boundary rents at the agricultural rent, $p(r)=p_A$ for a.e. $r\in[\bar r,\hat r]$.
\end{enumerate}
\end{definition}

A direct and fundamental consequence of this standard definition is the strict equalization of utility among all identical households. If there were any spatial arrangement yielding a discrepancy in utility, say $u_i > u_j$, household $j$—acting as a price taker—could simply mimic the location plan of household $i$ (i.e., choose $f_j(r) = f_i^*(r)$) to attain the exact same higher utility. This would contradict the assumption that household $j$ had already maximized its utility. Consequently, in any Walrasian equilibrium, we must have $u_1 = u_2 = \dots = u_N$.

With utility equalized across households, the bid-rent functions $\Psi_i(r,u)$ defined in equation \eqref{eq:bid_rent} may or may not coincide across households, since two distinct households can in principle share the same lot size and centroid. We refer to an equilibrium as \textit{non-degenerate} if the bid-rent functions are pairwise distinct, $\Psi_i \neq \Psi_j$ for all $i\neq j$, and as \textit{degenerate} otherwise, i.e., when at least two households share the same lot size, centroid, and hence the same bid-rent function. We first characterize the equilibrium land allocation in the non-degenerate case; we then show in Theorem \ref{thm:non-degenerate-equilibria} that degenerate equilibria cannot in fact arise, so that every Walrasian equilibrium is non-degenerate and exhibits the ordered, contiguous structure established below.

By equation \eqref{eq:bid_rent}, the bid rent function of household $i$ is $\Psi_i(r, u)=\text{MRS}(s_i, u)- \frac{t}{s_i}(r-r_{c,i})$ for each $i=1, ..., N$. For convenience, we define $\Psi_{N+1}(r, u) \equiv p_A$. Let $$\Psi_{-i}(r, u)=\max_{j\in \{1, \ldots, N+1\} \setminus \{i\}} \Psi_j(r, u)$$ be the next-highest bid rent after household $i$ at location
$r$, including the agricultural rent as an outside option. The {\it dominance region} of household $i$ is defined as $$\Omega_i= \{r \in [0, \bar r] : \Psi_i(r, u) \geq \Psi_{-i}(r, u)\},$$ which is the set of locations where household $i$ has the highest bid rent.

We now show that, in a non-degenerate equilibrium, these dominance regions are ordered and contiguous intervals, sorted by lot size from the CBD outward.

\begin{lemma}[Non-degenerate ordering of dominance regions]\label{lem:nondegenerate-ordering}
In a non-degenerate equilibrium, no two
households have the same lot size. Hence households can be relabeled so that
\[
    s_1<s_2<\cdots<s_N.
\]
Under this labeling, the dominance regions are ordered
sequentially from the CBD to the urban boundary. In particular, each
dominance region is a nonempty interval.
\end{lemma}

\begin{proof}
For each household, the bid-rent function \eqref{eq:bid_rent} is affine in location. Suppose, toward a contradiction, that two distinct households $i\neq j$ have
identical lot sizes, $s_i=s_j$. Then $b_i=b_j$, so $\Psi_i$ and $\Psi_j$ have
the same slope.
Since the equilibrium is non-degenerate, the two bid-rent functions cannot
coincide identically. Hence one of them lies strictly above the other on the
entire city. Without loss of generality, suppose $\Psi_i(r,u)>\Psi_j(r,u)$ for all $r\in[0,\bar r]$.

By Assumption \ref{ass:land-essentiality} and household optimality, $s_j>0$, and hence
\(
    \int_0^{\bar r} f_j(r)\,dr>0.
\)
Since $f_j(r)\in\{0,1\}$, the set
\[
    A_j:=\{r\in[0,\bar r]: f_j(r)=1\}
\]
has positive measure. On $A_j$, the first-order optimality condition
\eqref{eq:foc} requires $p(r)\le \Psi_j(r,u)$ for a.e. $r\in A_j$. But
$\Psi_i(r,u)>\Psi_j(r,u)$ for all $r\in[0,\bar r]$, so
$p(r)<\Psi_i(r,u)$ for a.e. $r\in A_j$. Hence household $i$ strictly demands
all locations in $A_j$, and \eqref{eq:foc} requires $f_i(r)=1$ for a.e.
$r\in A_j$. Since $f_j(r)=1$ on $A_j$, this implies
\(
    f_i(r)+f_j(r)= 2
\)
on a set of positive measure, contradicting the land-market clearing condition
$\sum_k f_k(r)=1$ for a.e. $r\in[0,\bar r]$. Thus the supposition
$s_i=s_j$ is inconsistent with land-market clearing in equilibrium. Therefore
$s_i\neq s_j$ for all $i\neq j$.

We may consequently relabel households so that $s_1<s_2<\cdots<s_N$, or
equivalently $b_1>b_2>\cdots>b_N$. Since the bid-rent functions are affine
with slopes $-b_i$, they satisfy the single-crossing property, and the upper
envelope can switch from one household to another only sequentially. Hence the
dominance regions appear in the same order,
from the CBD to the urban boundary: $\Omega_1,\Omega_2,\ldots,\Omega_N$.
Equivalently, households with smaller lot sizes occupy locations closer to the
CBD, while households with larger lot sizes occupy locations farther away.

This sequential ordering implies that each nonempty dominance region is
connected. Assumption \ref{ass:land-essentiality} and household optimality
imply that every household consumes a strictly positive amount of land.
Therefore, each dominance region has positive measure.
\end{proof}

Lemma \ref{lem:nondegenerate-ordering} establishes only the qualitative ordering of dominance regions; it does not pin down their boundaries or the lot sizes they support. The next lemma shows that, for a given target utility level $u$, adjacent bid-rent functions must match at their shared boundary, and this adjacent-boundary condition uniquely determines the dominance-region lot sizes via a backward difference equation anchored at the urban fringe.

\begin{lemma}[Dominance-region lot sizes at a fixed utility level]
\label{lem:fixed-utility-lot-sizes-dominance-regions}
Under Assumption \ref{ass:mrs-boundary}, for a given utility level $u$, let
\begin{equation}\label{eq:dominance-region-lot-size}
    s_i^\ast
    =
    MRS(\cdot,u)^{-1}
    \left(
        p_A+\left(N-i+\frac12\right)mt
    \right),
\end{equation}
and set $x_1^\ast=0$ and
$x_{i+1}^\ast=x_i^\ast+ms_i^\ast$ for $i=1,\ldots,N$. Then, the dominance region is given by
$\Omega_i^\ast=[x_i^\ast, x_{i+1}^\ast]$ for $i=1, ..., N$.
\end{lemma}

\begin{proof}
By Lemma \ref{lem:nondegenerate-ordering}, we may set $\Omega_i=[x_i,x_{i+1}]$ for some $0=x_1<x_2<\cdots<x_N<x_{N+1}=\bar r$. For household $i$, contiguity implies
$r_{c,i}=x_i+ms_i/2=x_{i+1}-ms_i/2$. Similarly, $r_{c,i+1}=x_{i+1}+ms_{i+1}/2$. Evaluating the two bid-rent functions at their common boundary $x_{i+1}$ gives $\Psi_i(x_{i+1})=MRS(s_i,u)-mt/2$ and $\Psi_{i+1}(x_{i+1})=MRS(s_{i+1},u)+mt/2$. The adjacent boundary condition $\Psi_i(x_{i+1})=\Psi_{i+1}(x_{i+1})$ therefore yields
\begin{equation}\label{eq:pinned-down-mrs-profile}
    MRS(s_i,u)-\frac{mt}{2}
    =
    MRS(s_{i+1},u)+\frac{mt}{2}.
\end{equation}

At the urban fringe $x_{N+1}$, the center of mass of the outermost parcel satisfies $x_{N+1}-r_{c,N}=ms_N/2$. Thus the boundary condition $\Psi_N(x_{N+1})=p_A$ gives $MRS(s_N,u)-mt/2=p_A$. Iterating the difference equation \eqref{eq:pinned-down-mrs-profile} backward from $i=N$ yields
\begin{equation}\label{eq:pinned-down-lot-sizes}
    MRS(s_i,u)
    =
    p_A+\left(N-i+\frac12\right)mt.
\end{equation}
Finally, strict monotonicity of $MRS(\cdot,u)$ implies that equation \eqref{eq:pinned-down-lot-sizes} has at most one solution in $s_i$, and Assumption \ref{ass:mrs-boundary} gives existence for each positive right-hand side. Hence the dominance-region lot-size profile is uniquely pinned down by \eqref{eq:pinned-down-lot-sizes}.
\end{proof}

\subsection{Excluding Degenerate Equilibria}

We now consider a situation in which multiple households share the same bid-rent function. Since a household's bid-rent function $\Psi_i(r)$ is uniquely parameterized by its lot size $s_i$ and center of mass $r_{c,i}$, identical bid rents imply identical values of $(s_i,r_{c,i})$. We therefore define a \emph{group} $G_k\subseteq\mathcal N$ as a set of households that share the same lot size $s_k$, center of mass $r_{c,k}$, and bid-rent function $\Psi_k(r)$. A group is degenerate if $\lvert G_k\rvert\ge2$.

Let $\Omega_k$ denote the aggregate winning region associated with group $k$.
In contrast to the non-degenerate case, no household in the group can be the
unique highest bidder within $\Omega_k$. In the non-degenerate case, the rent
may satisfy $p(r)<\Psi_i(r)$ because the unique winner already fully
occupies the location and cannot further increase land consumption. In a
degenerate group, however, such a strict inequality cannot be sustained. If
$p(r)<\Psi_k(r)$ were to hold at some location $r\in\Omega_k$, then every
household belonging to group $k$ would strictly prefer to occupy that location.
Since the households share the same bid-rent function, no member of the group
can be excluded through price differences. This contradicts the land market
clearing condition $\sum_{i\in G_k} f_i(r)\le 1$. Consequently, the market rent within the group region must satisfy
\[
p(r)=\Psi_k(r)
\qquad
\text{for all } r\in\Omega_k.
\]

Under this equality, each household in the group becomes locally indifferent at every location within $\Omega_k$. The first-order condition \eqref{eq:foc} therefore permits either value $f_i(r)\in\{0,1\}$. However, the first-order condition only provides a local necessary condition for optimality. As we show below, any allocation within a degenerate group fails to satisfy the household's global utility maximization, implying that degenerate equilibria do not exist.

\begin{theorem}[Non-Existence of Degenerate Equilibria]\label{thm:non-degenerate-equilibria}
    In any Walrasian equilibrium, there are no degenerate groups. That is, $\lvert G_k\rvert = 1$ for every group $G_k$, and the equilibrium allocation is strictly non-degenerate with a unique, contiguous spatial ordering.
\end{theorem}

\begin{proof}
Suppose, toward a contradiction, that a degenerate group $G$ exists with $\lvert G\rvert \ge 2$. Let $s_G$ be its common per-capita lot size and $r_{c,G}$ its common center of mass. The group occupies an aggregate region $\Omega_G$. By the same single-crossing logic as in Lemma \ref{lem:nondegenerate-ordering}, this aggregate region must be an interval $\Omega_G = [x, x + m\lvert G\rvert s_G]$ for some inner boundary $x$. The market rent within this region must equal the group's bid rent, $p(r) = \Psi_G(r) = MRS(s_G, u) - \frac{t}{s_G}(r - r_{c,G})$ for $r \in [x, x + m\lvert G\rvert s_G]$. Moreover, the aggregate center of mass must match the individual ones, which implies $r_{c,G} = x + \frac{m\lvert G\rvert s_G}{2}$.

Consider a household in group $G$ that acts as a price taker facing $p(r)$. Suppose the household deviates from its assigned (potentially fragmented) allocation and instead chooses a per-capita lot size $s$ in a contiguous parcel starting exactly at the inner boundary $x$. Its physical parcel is $[x,x+ms]$: the household chooses $f(r)=1$ on this interval and $f(r)=0$ otherwise. We evaluate whether $s = s_G$ is a global optimum.

For this contiguous deviation parcel, the center of mass is $r_c([x,x+ms]) = x + \frac{ms}{2}$, and the per-capita land and commuting expenditure is $E(s) = \frac{1}{m}\int_x^{x+ms} p(r)\,dr + t\left(x + \frac{ms}{2}\right)$. The household's utility maximization problem is dual to minimizing the total expenditure $Z(s, u) + E(s)$ required to achieve utility $u$. By Leibniz's rule, $E'(s)=p(x+ms)+mt/2$. Since $\lvert G\rvert\ge2$, the back end $x+ms$ lies inside $\Omega_G$ for $s$ near $s_G$, so the market rent there equals the group's bid rent, $p(x+ms)=\Psi_G(x+ms)$. We evaluate the derivative at the candidate optimum $s=s_G$, using $r_{c,G}=x+m\lvert G\rvert s_G/2$. The marginal expenditure then simplifies as
\begin{align*}
    \frac{\partial}{\partial s}\Bigl[Z(s,u)+E(s)\Bigr]\Biggr|_{s=s_G}
    &= -MRS(s_G,u) + \left[MRS(s_G,u)-\frac{t}{s_G}(x+ms_G-r_{c,G})\right] + \frac{mt}{2}\\
    &= -\frac{t}{s_G}\left(ms_G-\frac{m\lvert G\rvert s_G}{2}\right) + \frac{mt}{2}\\
    &= -mt\left(1-\frac{\lvert G\rvert}{2}\right) + \frac{mt}{2}
    = \frac{mt}{2}(\lvert G\rvert - 1).
\end{align*}
Since the objective is expenditure minimization, the household desires to decrease $s$ if this derivative is positive. Because group $G$ is degenerate, $\lvert G\rvert \ge 2$, so $\frac{mt}{2}(\lvert G\rvert-1)>0$.
This strict inequality implies that the household can strictly decrease its total expenditure (and thus strictly increase its utility) by choosing a smaller per-capita lot size $s < s_G$ in a contiguous parcel originating at $x$. Therefore, the candidate allocation fails global utility maximization. Every household in the group will attempt to deviate toward the inner boundary, violating the land market clearing condition. Hence, no such degenerate equilibrium can exist.
\end{proof}

This theorem provides a powerful spatial implication: the market mechanism itself endogenously guarantees contiguity. Pooling several tied households into one shared block prices land according to the group's average position, rather than each household's own marginal advantage from concentrating its holdings near the inner edge of the block. A fragmented or mosaic-like allocation within the group satisfies the pointwise first-order condition \eqref{eq:foc}, because each household is indifferent at every location in the group region. It nevertheless fails household optimality: a household can move its land to the inner edge of the region at no cost and then gain by shrinking it, because the rent at its new back end is set by the group's centroid rather than its own. This deviation changes the household's holding at many locations at once, so the first-order condition cannot detect it.

Ruling out these arbitrage opportunities is precisely what rules out degeneracy: any equilibrium must therefore be non-degenerate, with all households holding distinct lot sizes. Lemma \ref{lem:nondegenerate-ordering} then implies that the dominance regions are nonempty intervals strictly ordered from the CBD outward, and since each household holds land only where its bid rent is highest, its equilibrium assignment coincides with its dominance region up to a null set. Contiguity is thus a necessary property of every equilibrium rather than a restriction imposed on the choice set.

\subsection{Supporting Prices}

The results above determine the candidate parcels at a given utility level,
but not a rent that supports them.  Such a rent must satisfy the pointwise
first-order condition, meet every household's budget at the common income,
and make each parcel optimal against both small and large deviations.  We
examine these requirements separately.

\subsubsection{First-Order Support and Budget Consistency}

We distinguish two notions of support for a
candidate allocation $\Omega^\ast=(\Omega_1^\ast,\ldots,\Omega_N^\ast)$ at a
fixed utility level $u$, with induced land assignment
$f_i:=\mathbf{1}_{\Omega_i^\ast}$ and centroid
$r_{c,i}^\ast:=r_c(\Omega_i^\ast)$ for each $i=1,\ldots,N$. We call $p$ a {\it
first-order supporting price} for $\Omega^\ast$ at $u$ if $p$ satisfies
market clearing and the boundary condition---conditions (ii) and (iii) of
Definition \ref{def:walrasian-equilibrium}---and each household's assignment
satisfies the pointwise first-order condition \eqref{eq:foc} at price $p$ and
utility level $u$ in place of full utility maximization. This notion
therefore captures only local, pointwise optimality conditional on $u$, and
does not impose the household budget constraints; equivalently, it does not
guarantee that households' utilities equalize under $p$. Budget consistency at a
common utility level $u$ requires
\[
    Z(s_i^\ast,u)
    +\frac{1}{m}\int_{\Omega_i^\ast}p(r)\,dr
    +t r_{c,i}^\ast
    =
    w
    \quad\text{for all } i=1,\ldots,N.
\]
We call $p$ a {\it supporting price} for $\Omega^\ast$ if it is a first-order
supporting price, satisfies budget consistency, and each household's assigned
land consumption is globally utility maximizing at $p$ among all feasible
measurable land assignments. Thus, a supporting price rules out nonlocal
deviations, rather than merely satisfying the local second-order conditions
used below as tractable checks for interval deviations.

Because global optimality is difficult to verify directly, we also use a local
notion.  For a measurable set $A\subseteq[0,\hat r]$ with $|A|>0$, let
\[
    \Phi(A;p,u)
    :=
    Z\left(\frac{|A|}{m},u\right)
    +\frac1m\int_A p(r)\,dr
    +t\,r_c(A)
\]
denote the expenditure required to attain utility $u$ with the land assignment
$\mathbf 1_A$.  We say that the parcel $\Omega_i^\ast$ is a {\it strict local
expenditure minimizer among measurable assignments} at $(p,u)$ if there
exists $\delta_i>0$ such that
$\Phi(A;p,u)>\Phi(\Omega_i^\ast;p,u)$ for every measurable
$A\subseteq[0,\hat r]$ with $0<|A\triangle\Omega_i^\ast|<\delta_i$, where
$A\triangle B:=(A\setminus B)\cup(B\setminus A)$ denotes the symmetric
difference, so that
$|A\triangle\Omega_i^\ast|=\int_0^{\hat r}|\mathbf 1_A(r)-\mathbf 1_{\Omega_i^\ast}(r)|\,dr$
is the total area of land added to or removed from the assigned parcel.  We
call $p$ a {\it locally supporting price} for $\Omega^\ast$ at $u$ if it is a
first-order supporting price, satisfies budget consistency, and makes each
assigned parcel a strict local expenditure minimizer among measurable
assignments at $(p,u)$.  Since
budget consistency gives $\Phi(\Omega_i^\ast;p,u)=w$ and $Z$ is strictly
increasing in $u$, no nearby land assignment then yields a
utility above $u$ at income $w$.  A locally supporting price therefore rules
out all small deviations, including measurable ones, but not necessarily large
ones.  Table \ref{tab:price-support-notions} summarizes the three notions.

\begin{table}[H]
\centering
\caption{Notions of price support for a candidate allocation $\Omega^\ast$ at $u$}
\label{tab:price-support-notions}
\begin{tabular}{lcccc}
\toprule
 & Pointwise FOC & Budget & Small & Large \\
 & (with (ii), (iii)) & consistency & deviations & deviations \\
\midrule
First-order supporting price & \checkmark & & & \\
Locally supporting price & \checkmark & \checkmark & \checkmark\textsuperscript{a} & \\
Supporting price & \checkmark & \checkmark & \checkmark\textsuperscript{b} & \checkmark \\
\bottomrule
\end{tabular}

\smallskip
\begin{minipage}{0.92\textwidth}\footnotesize
\textsuperscript{a}\,Every measurable assignment $A$ with
$0<|A\triangle\Omega_i^\ast|<\delta_i$ requires strictly higher expenditure.
\textsuperscript{b}\,Implied by global optimality, in the weak sense that no
nearby assignment requires lower expenditure.
\end{minipage}
\end{table}

Let
\[
\Psi^\ast_i(r,u)
=
\mathrm{MRS}(s_i^\ast,u)
-\frac{t}{s_i^\ast}(r-r_{c,i}^\ast).
\]
By Lemma \ref{lem:nondegenerate-ordering} and the single-crossing property of
the bid-rent functions, the maximum in the definition of $\Psi_{-i}$ given
above reduces, on $\Omega_i^\ast$, to
\begin{equation}
    \Psi_{-i}^\ast(r,u)
    =
    \begin{cases}
    \Psi_{i+1}^\ast(r,u), & i=1,\\
    \max\{\Psi_{i-1}^\ast(r,u),\Psi_{i+1}^\ast(r,u)\},
        & 1<i<N,\\
    \max\{\Psi_{i-1}^\ast(r,u),p_A\}, & i=N.
    \end{cases}
    \qquad r\in\Omega_i^\ast.
\end{equation}
The next lemma establishes that any first-order supporting price must
lie between the assigned household's bid rent and the highest competing bid
rent.\footnote{%
\label{fn:pointwise-foc}Changing $p$ on a set of measure zero affects no
household's expenditure, so household optimality restricts $p$ only almost
everywhere.  We nevertheless state conditions on $p$, including
\eqref{eq:foc} and these bid-rent bounds, at every location.  This is
without loss of generality: a rent that satisfies these conditions almost
everywhere can be changed on a set of measure zero so that it satisfies them
everywhere.  Its value at each interior parcel boundary, where the upper and
lower bounds coincide, then equals their common value.}
If these inequalities hold strictly in the interior, the optimal land assignment
uniquely induces the contiguous interval $\Omega_i$ up to measure-zero boundary
points.

\begin{lemma}[Bid-Rent Bounds for First-Order Support]\label{lem:foc-support-bid-rent-bounds}
$p$ is a first-order supporting price for $\Omega^\ast$ at
$u$ if and only if $p(r)=p_A$ for $r\in[\bar r,\hat r]$ and
$\Psi_{-i}^\ast(r,u)\le p(r)\le
\Psi_i^\ast(r,u)$ for $r\in\Omega_i^\ast$ for each $i=1,\ldots,N$.
\end{lemma}
\begin{proof}
For notational simplicity, suppress the utility arguments of the bid-rent
functions. Suppose that $p(r)>\Psi_i^\ast(r)$ or
$p(r)<\Psi_{-i}^\ast(r)$ for some $r\in \Omega_i^\ast$ and some
$i \in \{1,\dots,N\}$. If $p(r)>\Psi_i^\ast(r)$, then $f_i(r)=0$ by
\eqref{eq:foc}, which contradicts the assigned land allocation. If
$p(r)<\Psi_{-i}^\ast(r)$, then either there exists a household
$j\ne i$
with $p(r)<\Psi_j^\ast(r)$, or $p(r)<p_A$. In the former case,
$f_j(r)=1$ by \eqref{eq:foc}, which contradicts market clearing because $r$
is already assigned to household $i$. In the latter case, the agricultural
rent is the outside option, landowners would not supply this land to the urban
sector, violating market clearing. Therefore, $p$ is not a first-order supporting price for $\Omega^\ast$.

Conversely, suppose $p(r)=p_A$ for $r\in[\bar r,\hat r]$ and $\Psi_{-i}^\ast(r) \le p(r) \le \Psi_i^\ast(r)$ for $r \in \Omega_i^\ast$ for each $i=1,\ldots,N$. Then, by \eqref{eq:foc}, we may set $f_i(r)=1$ for all $r\in \Omega_i^\ast$. Beyond the urban boundary, $\Psi_N^\ast(\bar r)=p_A$ by \eqref{eq:dominance-region-lot-size}, and $\Psi_j^\ast(\bar r)\le\Psi_{-N}^\ast(\bar r)\le p_A$ for $j<N$. Since bid rents decrease in $r$, $\Psi_i^\ast(r)<p_A=p(r)$ for $r>\bar r$, so \eqref{eq:foc} requires $f_i(r)=0$ there. For $r\in[0,\bar r]$ outside $\Omega_i^\ast$, $\Psi_{-j}^\ast(r)\le p(r)\le\Psi_j^\ast(r)$ for some $j$. Since $i\neq j$, we have $\Psi_i^\ast(r)\le\Psi_{-j}^\ast(r)\le p(r)$, so \eqref{eq:foc} requires $f_i(r)=0$ whenever the first inequality is strict, and otherwise permits $f_i(r)=0$. Hence, $f_i(r)=\mathbf{1}_{\Omega_i^\ast}(r)$ while it satisfies the first-order condition. Since the intervals $\Omega_i^\ast$ are disjoint up to boundary points and cover the urban land, this assignment also satisfies $\sum_i f_i(r)=1$ for a.e. $r\in[0,\bar r]$. Therefore, $p$ is a first-order supporting price for $\Omega^\ast$.
\end{proof}

\begin{rem}[Front-based supporting rents are excluded]
\label{rem:bf-prices-excluded}\normalfont
Let $p$ be a continuous first-order supporting price for $\Omega^\ast$ at $u$
and let $1\le i<N$.  At the shared boundary $x_{i+1}^\ast$,
$\Psi_i^\ast(x_{i+1}^\ast,u)=\Psi_{i+1}^\ast(x_{i+1}^\ast,u)$, so Lemma
\ref{lem:foc-support-bid-rent-bounds} forces
$p(x_{i+1}^\ast)=\Psi_{i+1}^\ast(x_{i+1}^\ast,u)$.  Since $\Psi_{i+1}^\ast$ is
strictly decreasing in $r$, $\Psi_{i+1}^\ast(r,u)>p(x_{i+1}^\ast)$ for every
$r<x_{i+1}^\ast$.  Hence $p$ cannot be constant on any interval
$(x_{i+1}^\ast-\varepsilon,\,x_{i+1}^\ast]$: otherwise $p<\Psi_{i+1}^\ast$
there, violating the lower bound $p\ge\Psi_{-i}^\ast$.  The supporting rent
function constructed by \citetalias{berliant1992InternationalEconomicReview} in their equation~(4.17)
is constant at the level $p_i$ on a left neighborhood of the back end of each
lot $i<n$, and so is any translation of it.  It therefore satisfies the
endpoint conditions of the interval model but fails first-order support in the
present model: the household located just outside would gain by acquiring a
small detached piece of land just inside the boundary, which lowers its
centroid by more than the piece costs net of its land value.\qed
\end{rem}

We next ask when a first-order supporting price can also be
budget-consistent.  The bid-rent bounds limit the rent payment for each
parcel from above and below, and hence the expenditure with which its
occupant attains utility $u$.  Let
\begin{equation}\label{eq:upper-envelope-expenditure}
    E_i(u) = Z(s_i^\ast, u) + s_i^\ast MRS(s_i^\ast, u) + t \cdot r_{c,i}^\ast.
\end{equation}
This is the expenditure required to attain the utility level $u$ when the market rent function is the upper envelope of bid rent functions: i.e., $p(r)=\max_j \Psi_j^\ast(r, u)$ for all $r$.

Define the maximal rent discount compatible with the bid-rent bounds by
\begin{equation}\label{eq:rent-gap}
    T_i(u)
    =
    \frac{1}{m}\int_{\Omega_i^\ast}
    \left\{\Psi_i^\ast(r,u)-\Psi_{-i}^\ast(r,u)\right\}\,dr.
\end{equation}
Any first-order supporting and budget-consistent price must therefore satisfy
\begin{equation}\label{eq:first-order-budget-bounds}
    E_i(u)-T_i(u)\leq w\leq E_i(u),
    \qquad i=1,\ldots,N.
\end{equation}
These bounds are also sufficient, because the proportional-discount
construction in Example \ref{ex:numerical-soc-failure} below yields a first-order supporting
and budget-consistent price whenever they hold at a common utility level.
The next lemma orders the upper bounds $E_i(u)$ across households when the
marginal rate of substitution is convex in land.

\begin{lemma}[Monotonicity of upper-envelope expenditure]\label{lem:upper-envelope-expenditure-monotonicity}
Under Assumption \ref{ass:mrs-convexity}, $E_i(u)$ is weakly increasing in $i$, i.e., $E_{i+1}(u) \ge E_i(u)$ for all $i=1, \dots, N-1$.
If, in addition, $MRS(s,u)$ is strictly convex in $s$, then the inequality is strict:
$E_{i+1}(u)>E_i(u)$ for all $i=1,\dots,N-1$.
\end{lemma}

\begin{proof}
Fix adjacent households \(i\) and \(i+1\), and write \(a=s_i^\ast\),
\(b=s_{i+1}^\ast\), and \(g(s)=MRS(s,u)\).
By Lemma \ref{lem:fixed-utility-lot-sizes-dominance-regions}, their centroids
satisfy \(r_{c,i+1}^\ast-r_{c,i}^\ast=m(a+b)/2\), and the adjacent boundary
condition gives \(g(a)-g(b)=mt\). Since \(Z_s(s,u)=-g(s)\), we have
\(Z(b,u)-Z(a,u)=-\int_a^b g(s)\,ds\). Therefore
\[
\begin{aligned}
E_{i+1}(u)-E_i(u)
={}&
Z(b,u)-Z(a,u)+bg(b)-ag(a)+t(r_{c,i+1}^\ast-r_{c,i}^\ast) \\
={}&
-\int_a^b g(s)\,ds
    +bg(b)-ag(a)
    +\frac{a+b}{2}\{g(a)-g(b)\} \\
={}&
\frac{b-a}{2}\{g(a)+g(b)\}
    -\int_a^b g(s)\,ds .
\end{aligned}
\]
The last expression is the difference between the trapezoidal approximation to
the area under \(g\) on \([a,b]\) and the exact integral. If \(g\) is convex,
the trapezoidal approximation is weakly larger than the integral, so
\(E_{i+1}(u)\ge E_i(u)\). If \(g\) is strictly convex and \(a<b\), the
trapezoidal approximation is strictly larger, so
\(E_{i+1}(u)>E_i(u)\).
\end{proof}

\begin{corollary}[Exactness condition for upper-envelope support]
\label{cor:upper-envelope-trapezoid-exact}
The upper-envelope price $p(r)=\max_j \Psi_j^\ast(r,u)$ can be both
first-order supporting and budget-consistent for $\Omega^\ast$ only if
\[
    \int_{s_i^\ast}^{s_{i+1}^\ast} MRS(s,u)\,ds
    =
    \frac{s_{i+1}^\ast-s_i^\ast}{2}
    \left\{MRS(s_i^\ast,u)+MRS(s_{i+1}^\ast,u)\right\},
    \qquad i=1,\ldots,N-1.
\]
\end{corollary}

Lemma \ref{lem:upper-envelope-expenditure-monotonicity} has a direct implication
for standard utility functions such as Cobb--Douglas preferences. For these
preferences, the marginal rate of substitution is strictly convex in land
consumption. Hence the upper-envelope expenditure is strictly increasing across
households. The upper envelope of bid-rent functions therefore cannot satisfy
budget consistency for identical households with a common income, and a
supporting rent must be discounted below the assigned household's bid rent
on its own parcel. By contrast,
under quasilinear preferences with a quadratic land-utility term, the marginal
rate of substitution is affine in \(s\). In that case the trapezoidal formula in
Corollary \ref{cor:upper-envelope-trapezoid-exact} is exact, so the
upper-envelope expenditures are equalized and, at the corresponding common
income level, the upper envelope can itself be both first-order supporting and
budget-consistent.\footnote{The exactness result should not be interpreted as
saying that quadratic land utility is intrinsically special. Rather, it is
special relative to the linear commuting-cost specification used here. With a
different commuting-cost technology, the corresponding spatial averaging
formula would generally differ, and a different class of utility functions
would be exact.}

When a discount is required, whether it fits within the bid-rent bounds is
the question of whether \eqref{eq:first-order-budget-bounds} holds at some
utility level.
When the marginal rate of substitution is convex, Lemma
\ref{lem:upper-envelope-expenditure-monotonicity} reduces the upper bounds
to the single bound $w\le E_1(u)$.  Building on this, a supplementary
appendix, available from the author upon request, shows that, under
Assumptions \ref{ass:mrs-convexity} and \ref{ass:land-normality}, such a
utility level always exists, so first-order support and budget consistency
can always be met.  It also gives an example in which they cannot be met
when the marginal rate of substitution is not convex.

The pointwise lower envelope $\Psi_{-i}^\ast$ is the other extreme permitted by
the bid-rent bounds on parcel $\Omega_i^\ast$. It provides the largest possible
discount from the upper envelope, but neither a lower price nor budget
consistency alone guarantees household optimality. To determine whether this
lower-envelope candidate, or any price between the two envelopes, can support
the allocation, we next derive the relevant second-order conditions.

\subsubsection{Interval Deviations and Local Stability}
\label{subsubsec:interval-local-stability}

This section derives conditions under which each assigned parcel is a
strict local expenditure minimizer: no small change in the household's land
holding, including the addition or removal of detached pieces, lowers its
expenditure. Because land assignments are measurable sets, this is a priori an
infinite-dimensional problem. Under a convex rent, however, every measurable
assignment is weakly more expensive than the interval with the same measure and
centroid (Lemma \ref{lem:convex-interval-reduction}), so the problem reduces to
the choice of an interval, that is, of two endpoints. Lemma
\ref{lem:local-measurable-optimality} makes this reduction precise for local
optimality, and Proposition \ref{prop:local-interval-stability} then gives
second-order conditions at the endpoints under which an interior parcel is a
strict local expenditure minimizer among all nearby measurable assignments.

Every rent function considered below equals \(p_A\) on
\([\bar r,\hat r]\), so it cannot be strictly convex on the whole region.  We
therefore call such a rent function strictly convex if it is strictly convex
on the city \([0,\bar r]\).  It is then convex on \([0,\hat r]\) by the
boundary condition \(p\ge p_A=p(\bar r)\) on the city.

\begin{lemma}[Interval domination under convex rents]
\label{lem:convex-interval-reduction}
Let $p:[0,\hat r]\to\mathbb R$ be convex.  Let $A\subseteq[0,\hat r]$ be
measurable with measure $S>0$ and centroid
$r_c(A)$, and let $I=[r_c(A)-S/2,\,r_c(A)+S/2]$.  Then
$I\subseteq[0,\hat r]$ and
\[
    \int_A p(r)\,dr\;\ge\;\int_I p(r)\,dr .
\]
If $p$ is strictly convex and $I\subseteq[0,\bar r]$, the inequality is
strict unless $A=I$ up to a null set.  Consequently, for any household and any target utility level, the
expenditure of the assignment $\mathbf 1_A$ is weakly larger than that of
$\mathbf 1_I$, which has the same land consumption and the same centroid.
\end{lemma}

\begin{proof}
Since $A\subseteq[0,\hat r]$ has measure $S$, $\int_A r\,dr$ lies between
$\int_0^S r\,dr$ and $\int_{\hat r-S}^{\hat r}r\,dr$ (that is, the uniform
distribution on $A$ first-order stochastically dominates that on $[0,S]$ and
is dominated by that on $[\hat r-S,\hat r]$), so
$S/2\le r_c(A)\le\hat r-S/2$ and $I\subseteq[0,\hat r]$.  Let $F_A$ and $F_I$
be the distribution functions of the uniform distributions on $A$ and $I$.
Both have mean $r_c(A)$.  Moreover, $F_A-F_I$ is nonnegative to the left of
$I$, nonincreasing on $I$, where $F_I$ has the maximal slope $1/S$, and
nonpositive to the right of $I$, so the two distribution functions cross once.
By \citet[Proposition 6.D.2]{mas-colell1995} and the discussion preceding it,
$F_A$ is therefore a mean-preserving spread of $F_I$.  Since $p$ is convex,
and bounded on $[0,\hat r]$, Jensen's inequality gives
$\int p\,dF_A\ge\int p\,dF_I$, that is, $\int_A p\,dr\ge\int_I p\,dr$.

Now let $p$ be strictly convex, let $I\subseteq[0,\bar r]$, and suppose that
$A\neq I$ on a set of positive measure.  If almost all of $A$ lay to the right of
$\min I$, the uniform distribution on $A$ would first-order stochastically
dominate that on $I$.  Since the two distributions have the same mean, they
would then coincide, a contradiction.  Hence $F_A>F_I=0$ on a set of
positive measure to the left of $I$.  Let
$\Delta(x):=\int_0^x\{F_A(y)-F_I(y)\}\,dy$.  Since $F_A$ is a
mean-preserving spread of $F_I$, $\Delta\ge0$, and $\Delta(\hat r)=0$
because the means coincide.  Moreover, since $F_I=0$ to the left of $I$,
$\Delta(\min I)=\int_0^{\min I}F_A(y)\,dy>0$, so $\Delta>0$ on a right
neighborhood $J\subseteq I$ of $\min I$.  Integrating by parts twice
gives $\int p\,dF_A-\int p\,dF_I=\int\Delta\,dp'$, where $p'$ is the
nondecreasing right derivative of $p$.  The interval $J$ lies in the city,
because $J\subseteq I\subseteq[0,\bar r]$, and $\Delta>0$ on $J$.  Since $p$ is
strictly convex on the city, $p'$ strictly increases on $J$, so the part of the
integral over $J$ is strictly positive.  Elsewhere, $\Delta\ge0$ and $p'$ is
nondecreasing, so the remaining part is nonnegative.  Hence the integral is
strictly positive.  The
final statement follows because $A$ and $I$ have the same land consumption and
the same centroid.
\end{proof}

The next lemma shows that, under a strictly convex rent, local optimality among
intervals implies local optimality among all nearby measurable assignments. It
therefore suffices to check local optimality among intervals.

\begin{lemma}[From interval to measurable local optimality]
\label{lem:local-measurable-optimality}
Let $p:[0,\hat r]\to\mathbb R$ be strictly convex, fix $u$, and let
$\Omega_i^\ast=[x_i^\ast,x_{i+1}^\ast]$.  Suppose that $\Omega_i^\ast$ is a
strict local minimizer of the interval expenditure: there exists
$\varepsilon>0$ such that $\Phi([a,b];p,u)>\Phi(\Omega_i^\ast;p,u)$ for every
interval $[a,b]\subseteq[0,\hat r]$ with
$0<\|(a,b)-(x_i^\ast,x_{i+1}^\ast)\|<\varepsilon$.  Then $\Omega_i^\ast$
is a strict local expenditure minimizer among measurable assignments at
$(p,u)$.
\end{lemma}

\begin{proof}
Write $S_i^\ast=|\Omega_i^\ast|$.  Let $\delta>0$, fix a measurable
$A\subseteq[0,\hat r]$ with $0<|A\triangle\Omega_i^\ast|<\delta$, and set
$D_+:=A\setminus\Omega_i^\ast$ and $D_-:=\Omega_i^\ast\setminus A$.  These
sets are disjoint, $A\triangle\Omega_i^\ast=D_+\cup D_-$, and
$A=(A\cap\Omega_i^\ast)\cup D_+$ and
$\Omega_i^\ast=(A\cap\Omega_i^\ast)\cup D_-$ are disjoint unions.  The
contributions of $A\cap\Omega_i^\ast$ therefore cancel, and
$|A|-S_i^\ast=|D_+|-|D_-|$ and
$\int_A r\,dr-\int_{\Omega_i^\ast}r\,dr=\int_{D_+}r\,dr-\int_{D_-}r\,dr$.
Since $0\le r\le\hat r$ on $[0,\hat r]$, we have
$0\le\int_{D_\pm}r\,dr\le\hat r|D_\pm|$.  Hence
$\bigl||A|-S_i^\ast\bigr|\le|D_+|+|D_-|=|A\triangle\Omega_i^\ast|<\delta$ and
$\bigl|\int_A r\,dr-\int_{\Omega_i^\ast}r\,dr\bigr|
\le\hat r(|D_+|+|D_-|)<\hat r\delta$.
Now let $\delta<S_i^\ast$, so that $|A|>S_i^\ast-\delta>0$.  Because
$\int_{\Omega_i^\ast}r\,dr=S_i^\ast r_{c,i}^\ast$,
$r_c(A)-r_{c,i}^\ast
=|A|^{-1}\bigl(\int_A r\,dr-\int_{\Omega_i^\ast}r\,dr\bigr)
+|A|^{-1}r_{c,i}^\ast(S_i^\ast-|A|)$.
Using the two bounds above and $0\le r_{c,i}^\ast\le\hat r$, we obtain
$|r_c(A)-r_{c,i}^\ast|<2\hat r\delta/(S_i^\ast-\delta)$.
Hence $r_c(A)\to r_{c,i}^\ast$ as $\delta\to0$, uniformly in such $A$.  Let $I(A)=[r_c(A)-|A|/2,\,r_c(A)+|A|/2]$ be the interval with the same
measure and centroid as $A$.  Its endpoints converge to
$(r_{c,i}^\ast-S_i^\ast/2,\,r_{c,i}^\ast+S_i^\ast/2)=(x_i^\ast,x_{i+1}^\ast)$, so we can choose
$\delta$ small enough that they lie within distance $\varepsilon$ of
$(x_i^\ast,x_{i+1}^\ast)$.  By Lemma \ref{lem:convex-interval-reduction},
$I(A)\subseteq[0,\hat r]$ and $\Phi(A;p,u)\ge\Phi(I(A);p,u)$, because $A$
and $I(A)$ have the same measure and centroid.  If $I(A)\neq\Omega_i^\ast$, the hypothesis gives
$\Phi(I(A);p,u)>\Phi(\Omega_i^\ast;p,u)$, and hence
$\Phi(A;p,u)\ge\Phi(I(A);p,u)>\Phi(\Omega_i^\ast;p,u)$.  If
$I(A)=\Omega_i^\ast$, then $|A\triangle\Omega_i^\ast|>0$ implies that
$A\neq I(A)$ on a set of positive measure.  Since
$I(A)=\Omega_i^\ast\subseteq[0,\bar r]$, the inequality from Lemma
\ref{lem:convex-interval-reduction} is then strict:
$\Phi(A;p,u)>\Phi(I(A);p,u)=\Phi(\Omega_i^\ast;p,u)$.  In either case,
$\Phi(A;p,u)>\Phi(\Omega_i^\ast;p,u)$.
\end{proof}

\begin{proposition}[Local stability of an interior parcel]
\label{prop:local-interval-stability}
Let $p$ be a strictly convex first-order supporting price for $\Omega^\ast$ at
utility level $u^\ast$, and fix an interior parcel
\(\Omega_i^\ast=[x_i^\ast,x_{i+1}^\ast]\), where \(1<i<N\). Suppose that, for
every \((\sigma,\tau)\in\{-,+\}^2\),
\begin{gather*}
Z_{ss}(s_i^\ast,u^\ast)-m p'_\sigma(x_i^\ast)>0, \\
Z_{ss}(s_i^\ast,u^\ast)+m p'_\tau(x_{i+1}^\ast)>0, \\
\{Z_{ss}(s_i^\ast,u^\ast)-m p'_\sigma(x_i^\ast)\}
\{Z_{ss}(s_i^\ast,u^\ast)+m p'_\tau(x_{i+1}^\ast)\}
-Z_{ss}(s_i^\ast,u^\ast)^2>0.
\end{gather*}
Then \(\Omega_i^\ast\) is a strict local expenditure minimizer among measurable
assignments at \((p,u^\ast)\).
\end{proposition}

\begin{proof}
Since $p$ is finite and convex on $[0,\bar r]$, both one-sided derivatives of $p$ exist and are finite
at every point in $(0,\bar r)$, in particular at $x_i^\ast$ and
$x_{i+1}^\ast$.

For a contiguous interval $[a,b]$, define
\[
    V(a,b;u^*)
    =
    Z\left(\frac{b-a}{m},u^*\right)
    +\frac{1}{m}\int_a^b p(r)\,dr
    +t\frac{a+b}{2},
    \qquad
    s=\frac{b-a}{m}.
\]
Differentiating
$V(a,b;u^*)$ with respect to the two endpoints gives
\begin{align}
    \frac{\partial V}{\partial a}
    &=
    -\frac{1}{m}Z_s\left(\frac{b-a}{m},u^*\right)
    -\frac{1}{m}p(a)+\frac{t}{2},
    \label{eq:Va}\\
    \frac{\partial V}{\partial b}
    &=
    \frac{1}{m}Z_s\left(\frac{b-a}{m},u^*\right)
    +\frac{1}{m}p(b)+\frac{t}{2}.
    \label{eq:Vb}
\end{align}
Thus, when both endpoints are free, the first-order conditions are
\[
    p(a)=MRS(s,u^*)+\frac{mt}{2},
    \qquad
    p(b)=MRS(s,u^*)-\frac{mt}{2},
\]
since $MRS(s,u^*)=-Z_s(s,u^*)$. Hence
the first-order supporting property of \(p\) implies that
\((x_i^\ast,x_{i+1}^\ast)\) is a stationary point of the interval expenditure problem for
an interior parcel.

Let
\[
    A_{\sigma}
    :=
    Z_{ss}(s_i^\ast,u^*)-m p'_{\sigma}(x_i^\ast),
    \qquad
    B_{\tau}
    :=
    Z_{ss}(s_i^\ast,u^*)+m p'_{\tau}(x_{i+1}^\ast),
\]
for $(\sigma,\tau)\in\{-,+\}^2$. Consider a small endpoint deviation
\[
    a=x_i^\ast+\Delta a,
    \qquad
    b=x_{i+1}^\ast+\Delta b.
\]
Because the one-sided derivatives exist, the price function admits the
corresponding directional first-order expansions. Using the endpoint
first-order conditions, the linear terms cancel, and the second-order change in
expenditure is
\[
    V(a,b;u^*)-V(x_i^\ast,x_{i+1}^\ast;u^*)
    =
    \frac{1}{2m^2}
    \Bigl[
        A_{\sigma}(\Delta a)^2
        +B_{\tau}(\Delta b)^2
        -2Z_{ss}(s_i^\ast,u^*)\Delta a\Delta b
    \Bigr]
    +o(\|(\Delta a,\Delta b)\|^2),
\]
where $\sigma=\operatorname{sgn}(\Delta a)$ and
$\tau=\operatorname{sgn}(\Delta b)$. The strict SOC conditions make the
quadratic form positive definite in every directional regime. Therefore
every sufficiently small nonzero interval deviation strictly raises
expenditure. Since \(V(a,b;u^\ast)=\Phi([a,b];p,u^\ast)\), the parcel
\(\Omega_i^\ast\) is a strict local minimizer of the interval expenditure, and
Lemma \ref{lem:local-measurable-optimality} yields the conclusion.
\end{proof}

The boundary parcels require separate one-sided arguments.  The innermost
parcel cannot extend below the CBD, and its first-order inequality there may
be slack, while the rent becomes flat beyond the outer end of the outermost
parcel. Lemmas
\ref{lem:appendix-innermost-local-stability} and
\ref{lem:appendix-outermost-local-stability} in Appendix
\ref{app:endpoint-stability} give conditions under which the innermost and
outermost parcels are strict local minimizers of the interval expenditure. When the rent is strictly convex, Lemma
\ref{lem:local-measurable-optimality} then extends the conclusion of
Proposition \ref{prop:local-interval-stability} to these parcels.

The local second-order conditions in Proposition \ref{prop:local-interval-stability}
can be interpreted in terms of the curvature of the supporting price in the
smooth case where \(p\) is \(C^1\). The first condition,
\(Z_{ss}(s_i^\ast,u^*)-m p'(x_i^\ast)>0\),
is automatically satisfied under Assumptions \ref{ass:mrs-boundary} and
\ref{ass:mrs-convexity} whenever the supporting price is nonincreasing. Indeed,
these assumptions imply $MRS_s(s,u)<0$, and hence
$Z_{ss}(s,u)=-MRS_s(s,u)>0$. The second condition,
\(Z_{ss}(s_i^\ast,u^*)+m p'(x_{i+1}^\ast)>0\),
requires that the rent schedule not decline too steeply, namely
\(p'(x_{i+1}^\ast)> -Z_{ss}(s_i^\ast,u^*)/m\).
The determinant condition,
\[
    Z_{ss}(s_i^\ast,u^*)\{p'(x_{i+1}^\ast)-p'(x_i^\ast)\}
    -m p'(x_i^\ast)p'(x_{i+1}^\ast)>0,
\]
captures the interaction between the curvature of preferences and that of
the rent schedule. If \(p\) is strictly convex, then
\(p'(x_{i+1}^\ast)>p'(x_i^\ast)\), implying that the first term is
strictly positive. Thus, strict convexity of the supporting price tends to
stabilize interval deviations by increasing the curvature of the
finite-dimensional expenditure problem. In particular, when the decline of
the rent schedule is sufficiently gradual and its slope increases with
distance from the CBD, every sufficiently small interval deviation has a
strictly positive second-order expenditure effect, so the assigned interval is
a strict local expenditure minimum.

We close this subsection by relating the household problem to the
front-based interval model of
\citetalias{berliant1992InternationalEconomicReview}.  The following identity
holds for every interval.

\begin{lemma}[Correspondence with the front-based interval model]
\label{lem:bf-correspondence}
For any rent function $p$, any interval $[a,b]\subseteq[0,\hat r]$ with
$s=(b-a)/m$, and any utility level $u$,
\[
    Z(s,u)+\frac1m\int_a^b p(r)\,dr+t\,\frac{a+b}{2}
    \;=\;
    Z(s,u)+\frac1m\int_a^b P(r)\,dr+t\,a,
    \qquad P:=p+\frac{mt}{2}.
\]
The left-hand side is the household's expenditure in the present model; the
right-hand side is its expenditure in the front-based interval model of
\citetalias{berliant1992InternationalEconomicReview} under the rent function $P$.
Moreover, $p(\bar r)=p_A$ if and only if $P(\bar r)=p_A+mt/2$.  Consequently,
if $p$ is convex, an interval parcel minimizes a household's expenditure among
all measurable assignments at $p$ if and only if it minimizes the front-based
expenditure among all intervals at $P$.
\end{lemma}

\begin{proof}
The identity follows from $\frac1m\int_a^b\frac{mt}{2}\,dr=\frac{t(b-a)}{2}$ and
$a+\frac{b-a}{2}=\frac{a+b}{2}$.  The boundary statement is immediate.  For the
last statement, Lemma \ref{lem:convex-interval-reduction} implies that, at a
convex $p$, the infimum of expenditure over measurable assignments equals its
infimum over intervals, and the identity equates the latter with the
front-based interval problem at $P$.
\end{proof}

Lemma \ref{lem:bf-correspondence} shows how the enlarged choice set bears on
household optimality.  On intervals, the present model and the front-based
interval model differ only by a uniform rent shift, whatever the rent
function.  Convexity of the rent is sufficient to reduce the enlarged choice
problem to the interval problem: if the rent is convex, Lemma
\ref{lem:convex-interval-reduction} rules out any gain from detached land, and
a household's problem is exactly the interval problem of
\citetalias{berliant1992InternationalEconomicReview}.

The correspondence also shows what is missing for global optimality.  At a
convex rent, a household's assigned parcel is globally optimal if and only if
no interval yields a lower expenditure.  We establish only the local version:
the second-order conditions above make each parcel a strict local minimizer.
\citetalias{berliant1992InternationalEconomicReview} obtain the global version by
shaping the rent inside each lot, but the bid-rent bounds here rule out the
flat rent segments that their construction uses (Remark
\ref{rem:bf-prices-excluded}).  We leave a general argument for future
work.\footnote{%
\label{fn:bf-two-stage}\citetalias{berliant1992InternationalEconomicReview}
proceed in two stages.
In our notation, they first minimize the expenditure \(V(a,b;u)\) of an
interval \([a,b]\) over \(a\) at each fixed \(b\), with minimizer
\(a^\ast(b)\), then study
\(
    E^\ast(b):=V(a^\ast(b),b;u).
\)
Their price construction makes \(E^\ast\) quasi-convex: it is
nonincreasing before a minimizing interval and nondecreasing after it.
Households with the same utility share the same \(E^\ast\), which is
constant between their assigned back ends. The construction uses flat
rent segments excluded by our bid-rent bounds. Our construction instead
yields multiple strict local minima of the common \(E^\ast\), all with
value \(w\), so it is not quasi-convex. Global support requires
\(E^\ast(b)\ge w\) for every feasible \(b\), including those beyond
the urban boundary; local second-order conditions alone do not
establish this bound.}

\begin{example}[Why the SOCs matter]
\label{ex:numerical-soc-failure}
Consider the model with \(N=2\), \(m=1\), \(t=1\), \(p_A=1/2\), \(w=10\),
and log-quasilinear utility \(U(z,s)=z+\log s\).\footnote{Python code that
reproduces the numerical examples and figures in this paper is available from
the author upon request.} The first-order conditions
give \(1/s_i=p_A+(N-i+1/2)t\), and hence \(s_1=1/2\) and \(s_2=1\).\footnote{%
\label{fn:quasilinear-examples}Log-quasilinear utility violates condition
(A2) of Assumption \ref{ass:primitive} and satisfies Assumption
\ref{ass:land-normality} only weakly, because \(MRS(s,u)=1/s\) does not
depend on \(u\).  We use it because every expression is then explicit.  The
results used in the examples still apply: each \(E_i(u)\) is strictly
increasing and unbounded in \(u\), which is all that the proofs of Lemma
\ref{lem:anchor-utility} and Theorem \ref{thm:supporting-price-existence}
use these assumptions for.}
The corresponding centers of mass are \(r_{c,1}=1/4\) and \(r_{c,2}=1\), the
urban boundary is \(\bar r=3/2\), and the assigned parcels are
\(\Omega_1^\ast=[0,1/2]\) and \(\Omega_2^\ast=[1/2,3/2]\).

The bid-rent functions are \(\Psi_1^\ast(r)=5/2-2r\) and
\(\Psi_2^\ast(r)=2-r\).
They intersect at \(x_2=1/2\), where both equal \(3/2\), and
\(\Psi_2^\ast(\bar r)=p_A=1/2\).

To build a price that satisfies first-order support and budget consistency,
we use a construction that is not specific to this example. Whenever a utility
level $u^\ast$ satisfies the bounds \eqref{eq:first-order-budget-bounds},
discount the upper envelope in proportion to the parcelwise bid-rent gap: for
$r\in\Omega_i^\ast$, let
\[
    p(r)=\Psi_i^\ast(r,u^\ast)
    -\lambda_i
    \left\{\Psi_i^\ast(r,u^\ast)-\Psi_{-i}^\ast(r,u^\ast)\right\},
    \qquad
    \lambda_i
    =
    \frac{E_i(u^\ast)-w}{T_i(u^\ast)}.
\]
The bounds imply $0\le\lambda_i\le1$, so this price satisfies the bid-rent
bounds. By \eqref{eq:rent-gap}, the discount lowers household $i$'s
expenditure by $\lambda_iT_i(u^\ast)=E_i(u^\ast)-w$, so the price is budget
consistent. This is a convenient proportional-discount construction rather
than the general form of every budget-consistent price.

To illustrate why the SOCs impose substantive restrictions beyond
first-order support and budget consistency, choose $u_L=8.07$. At this utility
level, \(E_1(u_L)=10.01315\), \(T_1(u_L)=0.125\), \(E_2(u_L)=10.07\), and
\(T_2(u_L)=0.25\), so \(\lambda_1=0.1052\) and \(\lambda_2=0.28\) lie strictly
between zero and one. Applying the construction above defines the price
$p^L$. On the first parcel, \(\Psi_{-1}^\ast=\Psi_2^\ast\); on the second,
\(\Psi_{-2}^\ast\) equals \(\Psi_1^\ast\) for \(r\le1\) and \(p_A\) for
\(r\ge1\). Hence
\[
p^L(r)=
\begin{cases}
2.4474-1.8948r,
&0\le r\le1/2,\\
2.14-1.28r,
&1/2\le r\le1,\\
1.58-0.72r,
&1\le r\le3/2.
\end{cases}
\]
It is continuous and convex, satisfies
\(\Psi_{-i}^\ast(r)\le p^L(r)\le\Psi_i^\ast(r)\), and gives each household an
expenditure of exactly \(w=10\).
Thus $p^L$ satisfies first-order support and budget consistency.
Figure \ref{fig:theta-one-lemma7-price} plots $p^L$ together with the
bid-rent functions.

\begin{figure}[H]
\centering
\includegraphics[width=.9\textwidth]{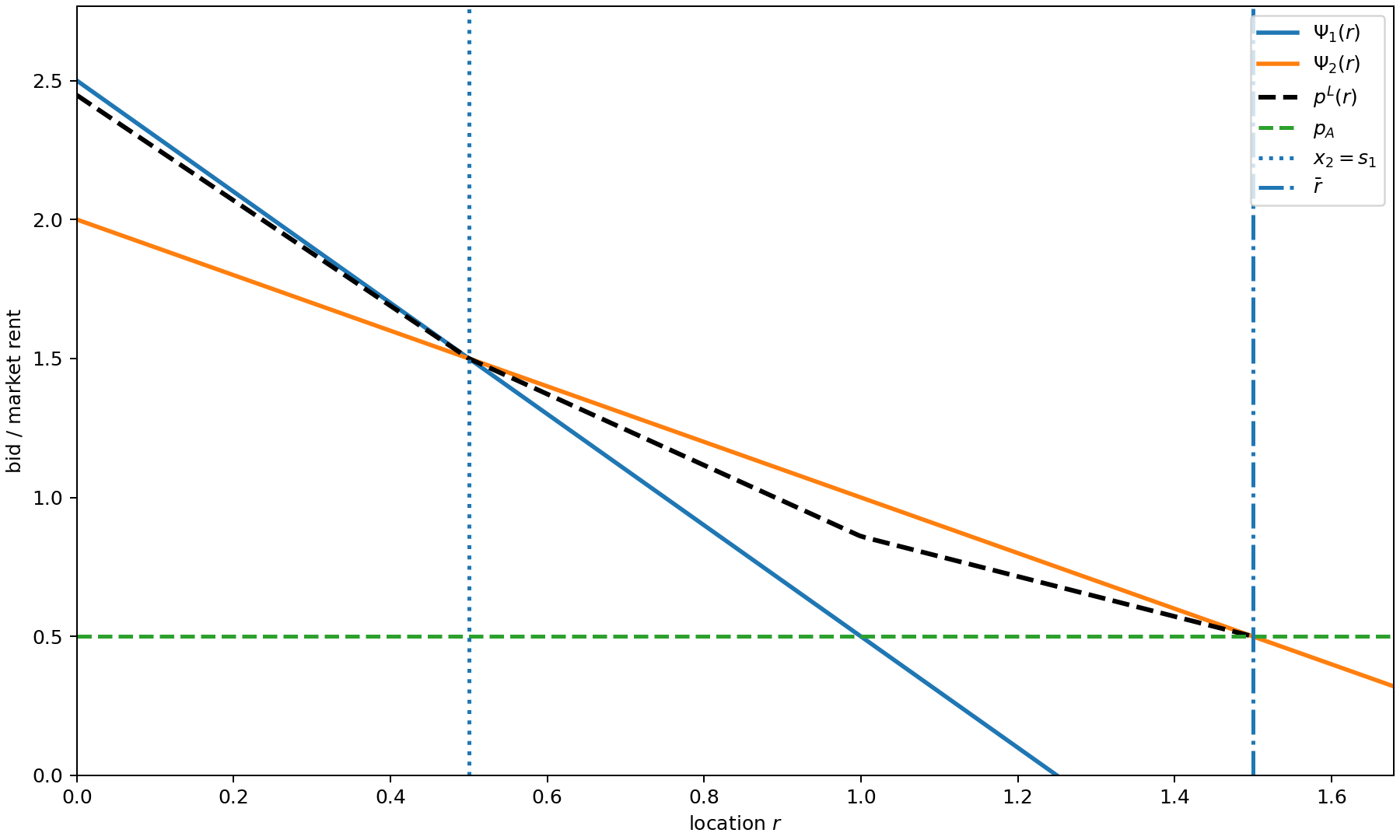}
\caption{
A first-order supporting and budget-consistent price that fails the
outer-parcel SOC.
The solid colored lines are the bid-rent schedules, and the dashed black line
is \(p^L\). The price satisfies first-order support, budget consistency, and
the boundary conditions, but its endpoint slopes violate the determinant SOC.
}
\label{fig:theta-one-lemma7-price}
\end{figure}

It does not, however, satisfy the outer-parcel SOC. At the common boundary and
the fringe, its relevant slopes are
\((p^L)'_-(x_2)=-1.8948\), \((p^L)'_+(x_2)=-1.28\), and
\((p^L)'_-(\bar r)=-0.72\).
Since \(Z_{ss}(s_2^\ast,u_L)=1\), the two feasible left-endpoint directions
give
\[
\begin{array}{c|c|c|c}
\sigma & A_\sigma & B_-
& A_\sigma B_- - Z_{ss}^2\\
\hline
- & 2.8948 & 0.28 & -0.1894\\
+ & 2.28 & 0.28 & -0.3616
\end{array}
\]
so the determinant condition fails in both directions. By contrast, on the
innermost parcel \(p^L(0)<\Psi_1^\ast(0)\), and the relevant right-endpoint
diagonal conditions are positive. Hence the failure is localized to the
outer-parcel SOC. The price is piecewise affine rather than strictly convex,
so it is not the supporting price constructed below.

\begin{figure}[H]
\centering
\includegraphics[width=.82\textwidth]{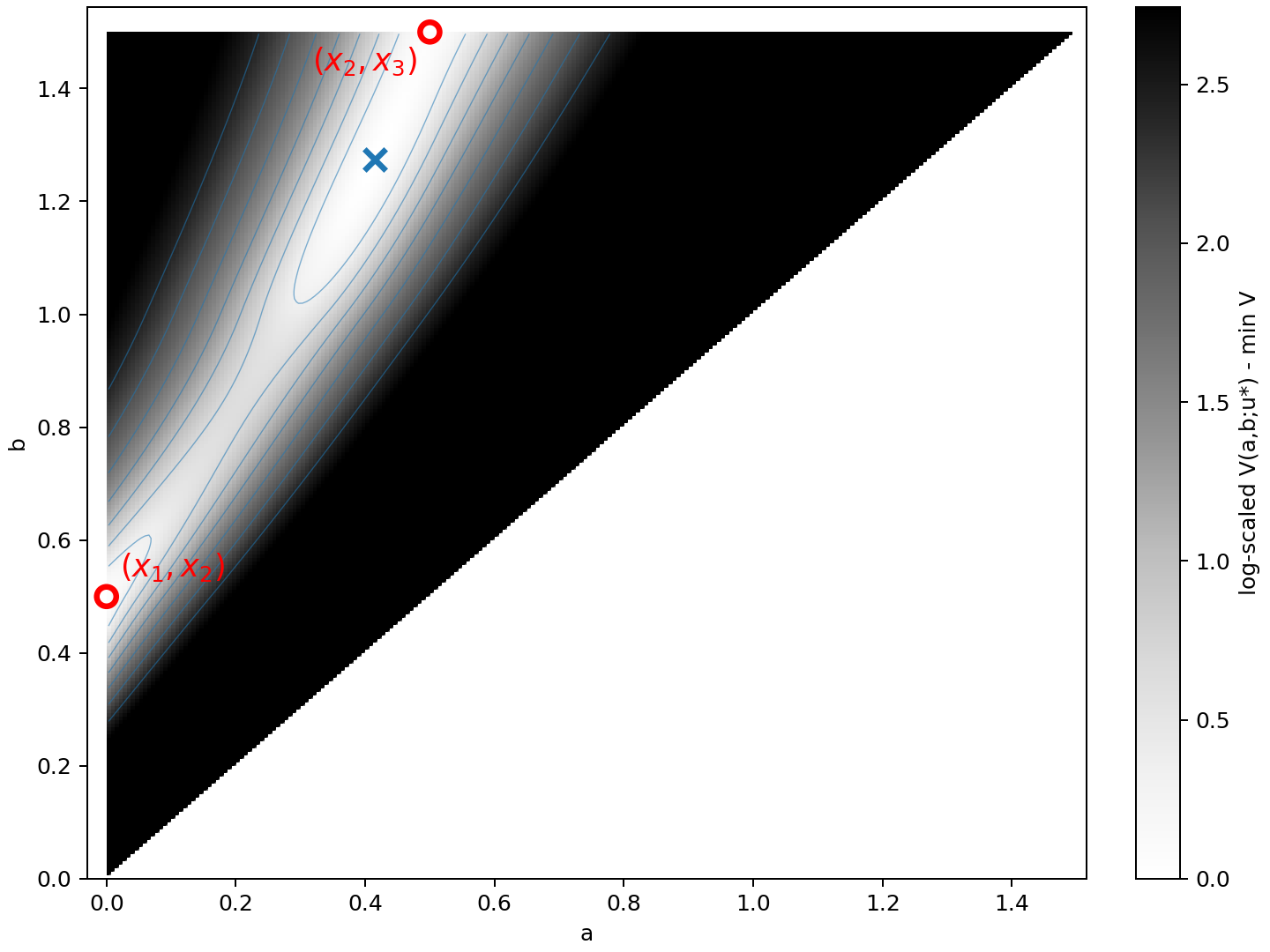}
\caption{
Interval expenditure under the first-order supporting and budget-consistent
price $p^L$.
The horizontal and vertical axes are the interval endpoints \(a\) and \(b\),
respectively. The red circles mark the assigned endpoint pairs, and the blue
cross marks the grid minimizer. Lighter shading indicates lower expenditure
relative to the grid minimum. The minimizer lies below the household budget,
revealing a profitable interval deviation.
}
\label{fig:theta-one-lemma7-heatmap}
\end{figure}

The expenditure surface in Figure \ref{fig:theta-one-lemma7-heatmap} confirms
that the SOC failure is economically consequential. On a \(240\times240\)
grid, the minimum is approximately \((a,b)=(0.414,1.274)\), where
\(V(a,b;u_L)=9.9994<10\), so a profitable interval deviation exists.
\end{example}

In the continuum model, a household's optimal location is where its bid-rent
curve touches the market rent curve from below \citep[Rule 2.1]{fujita1989},
and with identical households the market rent equals the common bid rent at
every occupied location.  Each household is then indifferent among all
occupied locations, and no second-order condition arises.  In the present
model, a household's bid rent depends on its own lot size and centroid, so the
bid-rent bounds are only first-order conditions, and budget consistency forces
the rent below the occupant's bid rent inside its parcel (Lemma
\ref{lem:upper-envelope-expenditure-monotonicity}).  Example
\ref{ex:numerical-soc-failure} shows that first-order support and budget
consistency then no longer imply household optimality.

This failure is unrelated to the extension to measurable assignments.  The
profitable deviation is itself an interval, and since $p^L$ is convex, Lemma
\ref{lem:convex-interval-reduction} implies that detached land offers no
further gain.  The same failure would therefore arise if lots were restricted
to intervals: by Lemma \ref{lem:bf-correspondence}, $p^L+mt/2$ satisfies the
endpoint conditions of the front-based interval model of
\citetalias{berliant1992InternationalEconomicReview} but does not make household
2's lot optimal.

\subsubsection{Existence and Construction}

Example \ref{ex:numerical-soc-failure} shows that first-order support and
budget consistency alone do not guarantee local stability. Motivated by this failure, we now ask whether there
exists a price function that also satisfies the directional SOCs in Proposition
\ref{prop:local-interval-stability}. We have already seen that, under strict
convexity of \(MRS(s,u)\) in \(s\), the upper-envelope price fails budget
consistency. Consider now the other extreme permitted by the bid-rent bounds.
The price obtained by pasting together the parcelwise lower envelopes cannot
satisfy all the coupled directional SOCs: at a shared boundary its one-sided
slopes are reversed, and the SOC associated with moving both endpoints outward
has determinant
\(
    -m^2t^2/(s_i^\ast)^2<0.
\)
Thus neither envelope by itself provides the desired locally supporting
price.

Our construction instead combines the upper envelope with the
pointwise midpoint between the two envelopes, which we call the
\emph{midpoint schedule}.  The midpoint schedule plays two roles.  First, it
fixes the slopes of the rent at the parcel boundaries: among averages of the
two envelopes with a common weight, it is the only one whose one-sided slopes
agree at shared boundaries (Remark \ref{rem:midpoint-role}), and under the
curvature condition below these slopes satisfy the SOCs.  Second, it provides
a lower benchmark for rent payments, while the upper envelope provides an
upper one.  On each parcel, the rent is a weighted combination of smooth
approximations to the two schedules that share the midpoint boundary slopes,
with the weight chosen to meet the budget.  Because the boundary slopes do not
depend on the weights, the SOCs continue to hold.  The midpoint schedule
itself is generally not budget consistent and is not the price we construct.

The following assumption gives a sufficient condition
for the midpoint boundary slopes to satisfy the relevant SOCs. This condition relates
the curvature of preferences directly to the equilibrium lot sizes of
neighboring households.
\begin{assumption}[Curvature condition for the midpoint construction]
\label{ass:midpoint-curvature}
At the utility level \(u\) under consideration, let
\[
    Z_i(u):=Z_{ss}(s_i^\ast(u),u),
\]
and suppose that the following three conditions hold. For the innermost
household, \(i=1\),
\[
    2Z_1(u)s_1^\ast(u)s_2^\ast(u)
    >
    mt\{s_1^\ast(u)+s_2^\ast(u)\};
\]
for every interior household, \(1<i<N\),
\[
    \begin{aligned}
    &2Z_i(u)\{s_i^\ast(u)\}^2
      \{s_{i+1}^\ast(u)-s_{i-1}^\ast(u)\}\\
    &\qquad >
      mt\{s_{i-1}^\ast(u)+s_i^\ast(u)\}
        \{s_i^\ast(u)+s_{i+1}^\ast(u)\};
    \end{aligned}
\]
and for the outermost household, \(i=N\),
\[
    2Z_N(u)\{s_N^\ast(u)\}^2
    >
    mt\{s_{N-1}^\ast(u)+s_N^\ast(u)\}.
\]
\end{assumption}

Under the midpoint boundary-slope choice used below, Assumption
\ref{ass:midpoint-curvature} ensures that the remaining initial slope can be
chosen so that every endpoint second-order matrix is positive definite, and
hence that all directional second-order conditions in Proposition
\ref{prop:local-interval-stability} hold strictly.

\begin{rem}\normalfont
Assumption \ref{ass:midpoint-curvature} can be assessed explicitly for
standard preferences, but it is not guaranteed merely by choosing a familiar
functional form. For example, consider Cobb--Douglas utility
\(U(z,s)=\alpha\log z+(1-\alpha)\log s\) with \(0<\alpha<1\). Writing
\(k=1/\alpha>1\), its expenditure function satisfies
\(MRS(s,u)=K(u)s^{-k}\) and \(Z_{ss}(s,u)=(k/s)MRS(s,u)\) for some
\(K(u)>0\). The adjacent-boundary equations imply
\(MRS(s_i^\ast,u)-MRS(s_{i+1}^\ast,u)=mt\). For an interior household, set
\(a=s_{i-1}^\ast/s_i^\ast\) and \(b=s_{i+1}^\ast/s_i^\ast\). Then
\(a^{-k}+b^{-k}=2\), and the interior inequality in Assumption
\ref{ass:midpoint-curvature} reduces to
\[
    2k(b-a)
    >
    (1-b^{-k})(1+a)(1+b),
\]
which holds for \(0<a<1<b\) satisfying the preceding relation. The
innermost inequality is likewise automatic. At the fringe, however, writing
\(a=s_{N-1}^\ast/s_N^\ast\), the remaining condition becomes
\(2k>(a^{-k}-1)(1+a)\). It holds when the last increase in lot size is
sufficiently moderate, but it need not hold for an arbitrarily small \(a\).
Thus, under Cobb--Douglas preferences, the only potentially binding part of
the assumption is a restriction on the outermost lot-size jump. The same
calculation applies to log-quasilinear preferences
\(U(z,s)=z+\eta\log s\), for which \(k=1\).\qed
\end{rem}

Having seen when Assumption \ref{ass:midpoint-curvature} can be expected to
hold, we now turn to constructing an actual locally supporting price from the midpoint
schedule. The construction below pins the schedule down at a single reference
utility level, tied to the innermost household's budget, which we call the
\emph{anchor utility}. The next lemma establishes that this utility level
exists and is unique.

\begin{lemma}[Existence of the anchor utility]\label{lem:anchor-utility}
Under Assumptions \ref{ass:mrs-boundary}--\ref{ass:land-normality},
$E_1(u)$ is continuous and strictly increasing, and there exists a unique
utility level $u_0$ satisfying
\[
    E_1(u_0)=w.
\]
\end{lemma}

\begin{proof}
For the innermost household, $r_{c,1}^\ast=m s_1^\ast(u)/2$ and
$MRS(s_1^\ast(u),u)=p_A+(N-1/2)mt\equiv C_1>0$. Hence
$E_1(u)=Z(s_1^\ast(u),u)+(C_1+mt/2)s_1^\ast(u)$.
Assumptions \ref{ass:mrs-convexity} and \ref{ass:mrs-boundary} imply that
$MRS_s(s,u)<0$ and that $MRS(s,u)=C_1$ has a unique solution. The Implicit
Function Theorem and Assumption \ref{ass:land-normality} give
\[
    \frac{d s_1^\ast}{du}
    =-\frac{MRS_u(s_1^\ast(u),u)}{MRS_s(s_1^\ast(u),u)}>0.
\]
Since $Z_s(s_1^\ast(u),u)=-C_1$ and $Z_u=1/U_z>0$, we obtain
\[
    \frac{dE_1}{du}
    =Z_u(s_1^\ast(u),u)
    +\frac12 mt\frac{d s_1^\ast}{du}>0.
\]
The boundary conditions in Assumption \ref{ass:primitive} imply that $E_1(u)$
ranges from below $w$ to above $w$. Continuity, the Intermediate Value Theorem,
and strict monotonicity yield existence and uniqueness of $u_0$.
\end{proof}

\begin{theorem}[Existence of a locally supporting price]
\label{thm:supporting-price-existence}
Under Assumptions \ref{ass:mrs-boundary}--\ref{ass:land-normality}, let
\(u_0\) be the utility level satisfying \(E_1(u_0)=w\). Suppose that Assumption
\ref{ass:midpoint-curvature} holds at \(u_0\), and that
\begin{equation}\label{eq:midpoint-expenditure-slack}
    \max_{i=1,\ldots,N}
    \left\{
    E_i(u_0)-\frac{T_i(u_0)}{2}
    \right\}
    <E_1(u_0).
\end{equation}
Then, $\Omega^\ast$ admits a strictly convex locally supporting price.
\end{theorem}
\begin{proof}
At \(u_0\), condition \eqref{eq:midpoint-expenditure-slack} and
\(E_1(u_0)=w\) give a strict margin between \(w\) and every midpoint-discounted
expenditure. The functions \(E_i\), \(T_i\), and the expressions in Assumption
\ref{ass:midpoint-curvature} are continuous in \(u\).
Moreover, Lemma \ref{lem:anchor-utility} shows that $E_1$ is strictly
increasing.
Hence, increasing the utility level slightly yields a nearby \(u^\ast\) such
that Assumption \ref{ass:midpoint-curvature} continues to hold and
\[
    E_i(u^\ast)-\frac{T_i(u^\ast)}{2}
    <w<E_1(u^\ast)\le E_i(u^\ast),
    \qquad i=1,\ldots,N .
\]
Here the last inequality follows from Lemma
\ref{lem:upper-envelope-expenditure-monotonicity}.

The required total rent payment over the parcel
\(\Omega_i^\ast=[x_i^\ast,x_{i+1}^\ast]\) must equal a target value
\(I_i^\ast:=m\{w-Z(s_i^\ast,u^\ast)-t r_{c,i}^\ast\}\).
Define the midpoint schedule on this parcel by
\(q_i^\ast(r):=\frac12\{\Psi_i^\ast(r,u^\ast)+\Psi_{-i}^\ast(r,u^\ast)\}\).
Recalling the definition of \(T_i(u^\ast)\) in \eqref{eq:rent-gap},
the preceding expenditure inequalities are equivalent to
\begin{equation}\label{eq:integral-bounds}
    \int_{\Omega_i^\ast} q_i^\ast(r)\,dr
    <
    I_i^\ast
    <
    \int_{\Omega_i^\ast} \Psi_i^\ast(r,u^\ast)\,dr,
    \qquad i=1,\ldots,N.
\end{equation}

Since Assumption \ref{ass:midpoint-curvature} holds at \(u^\ast\), Lemma
\ref{lem:appendix-midpoint-approximations} yields a boundary-slope vector
\((c_1,\ldots,c_{N+1})\), with the midpoint slopes
\(c_i=-(b_{i-1}^\ast+b_i^\ast)/2\) for \(i=2,\ldots,N\) and
\(c_{N+1}=-b_N^\ast/2\), where \(b_i^\ast:=t/s_i^\ast(u^\ast)\), that satisfies the slope ordering and all the joint
determinant conditions. Lemma
\ref{lem:appendix-upper-envelope-approximations} applies to any such slope
vector. Lemmas
\ref{lem:appendix-upper-envelope-approximations} and
\ref{lem:appendix-midpoint-approximations} therefore provide two globally
\(C^1\), parcelwise \(C^2\) and strictly convex sequences with this same
boundary-slope vector. Write their parcelwise restrictions as
\(\underline h_{i,n}\) and \(\overline h_{i,n}\). They satisfy the envelope
bounds and all coupled directional SOCs, and
\(\underline h_{i,n}\to q_i^\ast\) and \(\overline h_{i,n}\to\Psi_i^\ast\)
uniformly on every parcel. Since the inequalities in
\eqref{eq:integral-bounds} are strict and there are finitely many parcels, one
common sufficiently large \(n\) satisfies
\[
    \int_{\Omega_i^\ast}\underline h_{i,n}(r)\,dr
    <I_i^\ast<
    \int_{\Omega_i^\ast}\overline h_{i,n}(r)\,dr,
    \qquad i=1,\ldots,N.
\]

For each parcel choose the unique \(\lambda_i\in(0,1)\) such that
\[
    \int_{\Omega_i^\ast}
    \{\lambda_i\overline h_{i,n}(r)
      +(1-\lambda_i)\underline h_{i,n}(r)\}\,dr
    =
    I_i^\ast,
\]
and define
\[
    p^\ast(r)
    :=
    \lambda_i\overline h_{i,n}(r)
    +(1-\lambda_i)\underline h_{i,n}(r),
    \qquad r\in\Omega_i^\ast.
\]
Both component functions have the same value and the same prescribed slope at
every shared parcel boundary. Consequently, the weights \(\lambda_i\) may vary
across parcels without disturbing either continuity or differentiability:
\(p^\ast\) is \(C^1\) on \([0,\bar r]\); beyond the urban boundary the rent
is constant at \(p_A\), so its slope changes at \(\bar r\). On \([0,\bar r]\),
\(p^\ast\) is strictly convex, satisfies
\(\Psi_{-i}^\ast\le p^\ast\le\Psi_i^\ast\), and has the midpoint boundary
slopes chosen above. Therefore all coupled directional SOCs hold strictly. On
the first parcel,
\(\underline h_{1,n}(0)<\Psi_1^\ast(0)\) and \(\lambda_1<1\), so
\(p^\ast(0)<\Psi_1^\ast(0)\), so the first case of Lemma
\ref{lem:appendix-innermost-local-stability} applies.

The definition of \(I_i^\ast\) gives budget consistency, and the envelope
inequalities give first-order support. Since \(p^\ast\) is strictly convex and the endpoint SOCs hold strictly,
Proposition \ref{prop:local-interval-stability} makes each interior parcel a
strict local expenditure minimizer among measurable assignments. For the
innermost and outermost parcels, Lemmas
\ref{lem:appendix-innermost-local-stability} and
\ref{lem:appendix-outermost-local-stability} give strict local minimality
among intervals, which Lemma \ref{lem:local-measurable-optimality} extends to
nearby measurable assignments. Hence \(p^\ast\) is a locally supporting price.
\end{proof}

Theorem \ref{thm:supporting-price-existence} establishes local support only
and does not rule out large deviations (Section
\ref{subsubsec:interval-local-stability}).  We check global optimality
numerically in Example \ref{ex:numerical-supporting-price}.

\begin{rem}\normalfont
\label{rem:midpoint-role}
To clarify the role of the midpoint schedule in Theorem
\ref{thm:supporting-price-existence}, consider a common
parcel-independent weight \(\theta\in(0,1)\) and define
\(q_i^\theta:=\theta\Psi_i^\ast+(1-\theta)\Psi_{-i}^\ast\).
At a shared boundary \(x_{i+1}^\ast\), its one-sided derivatives from the two
adjacent parcels are
\((q_i^\theta)'_-(x_{i+1}^\ast)=-\{\theta b_i^\ast+(1-\theta)b_{i+1}^\ast\}\) and
\((q_{i+1}^\theta)'_+(x_{i+1}^\ast)=-\{\theta b_{i+1}^\ast+(1-\theta)b_i^\ast\}\).
Since \(b_i^\ast\ne b_{i+1}^\ast\), these derivatives agree if and only if
\(\theta=1/2\).  For this value, the midpoint schedule
\(q_i^\ast=\frac12(\Psi_i^\ast+\Psi_{-i}^\ast)\) has the common derivative
\((q_i^\ast)'_-(x_{i+1}^\ast)=(q_{i+1}^\ast)'_+(x_{i+1}^\ast)
=-(b_i^\ast+b_{i+1}^\ast)/2\).
Thus, among convex combinations using a common parcel-independent weight, the
midpoint is the unique choice for which the pasted schedule is differentiable
at every shared boundary.  This is a uniqueness statement within the common-weight construction, rather than
among all continuous convex supporting prices.
If \(\theta<1/2\), the derivative jumps downward and global convexity fails;
if \(\theta>1/2\), it jumps upward, so global convexity may be preserved but
differentiability fails.  If parcel-specific weights \(\theta_i\) are
allowed, differentiable pasting instead requires \(\theta_i+\theta_{i+1}=1\)
at every shared boundary. It is not immediate that weights chosen to satisfy
parcelwise budget balance can simultaneously satisfy these cross-parcel
pasting restrictions.\qed
\end{rem}

\begin{example}[Constructing a locally supporting price]
\label{ex:numerical-supporting-price}
We now return to Example \ref{ex:numerical-soc-failure} and implement the
construction in Theorem \ref{thm:supporting-price-existence}, which combines
smooth approximations to the midpoint schedule and to the upper envelope with
parcel-specific weights chosen to meet each budget. The utility level
satisfying \(E_1(u_0)=w\) is \(u_0=8.0569\). At this utility, the
midpoint-expenditure slack and the margins in Assumption
\ref{ass:midpoint-curvature} are
\[
\begin{array}{c|c|c}
i
&
w-\{E_i(u_0)-T_i(u_0)/2\}
&
\text{curvature margin}
\\
\hline
1 & 0.0625 & 2.5\\
2 & 0.0681 & 0.5
\end{array}
\]
and are strictly positive. Thus both the expenditure-slack condition and the
curvature condition in Theorem \ref{thm:supporting-price-existence} hold.

Choose \(u^\ast=8.0881\). The bid-rent steepnesses are \(b_1^\ast=2\)
and \(b_2^\ast=1\). Following Theorem
\ref{thm:supporting-price-existence}, set
\(c_2=-(b_1^\ast+b_2^\ast)/2=-1.5\) and
\(c_3=-b_2^\ast/2=-0.5\), and choose
\(c_1=-2.52<-b_1^\ast\). Thus
\(c_1<-b_1^\ast<c_2<-b_2^\ast<c_3<0\).

For the numerical implementation, take approximation index \(n=32\).
Let \(\underline h_{i,n}\) denote the smooth midpoint-schedule approximation
from Lemma \ref{lem:appendix-midpoint-approximations}, and let
\(\overline h_{i,n}\) denote the smooth upper-envelope approximation from
Lemma \ref{lem:appendix-upper-envelope-approximations}. The target rent
integrals and the two bracketing integrals are
\[
\begin{array}{c|c|c|c}
i
&
\displaystyle\int_{\Omega_i^\ast}\underline h_{i,n}(r)\,dr
&
I_i^\ast
&
\displaystyle\int_{\Omega_i^\ast}\overline h_{i,n}(r)\,dr
\\
\hline
1 & 0.94546 & 0.96875 & 0.99630\\
2 & 0.88280 & 0.91190 & 0.98574
\end{array}
\]
so each target lies strictly between the corresponding lower and upper
integrals. Define
\[
    p^\ast(r)
    =
    \lambda_i\overline h_{i,n}(r)
    +(1-\lambda_i)\underline h_{i,n}(r),
    \qquad r\in\Omega_i^\ast.
\]
The integral-matching weights are \(\lambda_1=0.458\) and
\(\lambda_2=0.283\).
Although these weights differ across parcels, both component functions have
the same value and slope at \(x_2\). Hence the resulting price is \(C^1\)
on \([0,\bar r]\), parcelwise strictly convex, and therefore strictly convex
on \([0,\bar r]\).
It also satisfies
\(\Psi_{-i}^\ast(r)\le p^\ast(r)\le\Psi_i^\ast(r)\) on every
\(\Omega_i^\ast\); see Figure \ref{fig:theta-one-theorem2-price}.

\begin{figure}[H]
\centering
\includegraphics[width=.9\textwidth]{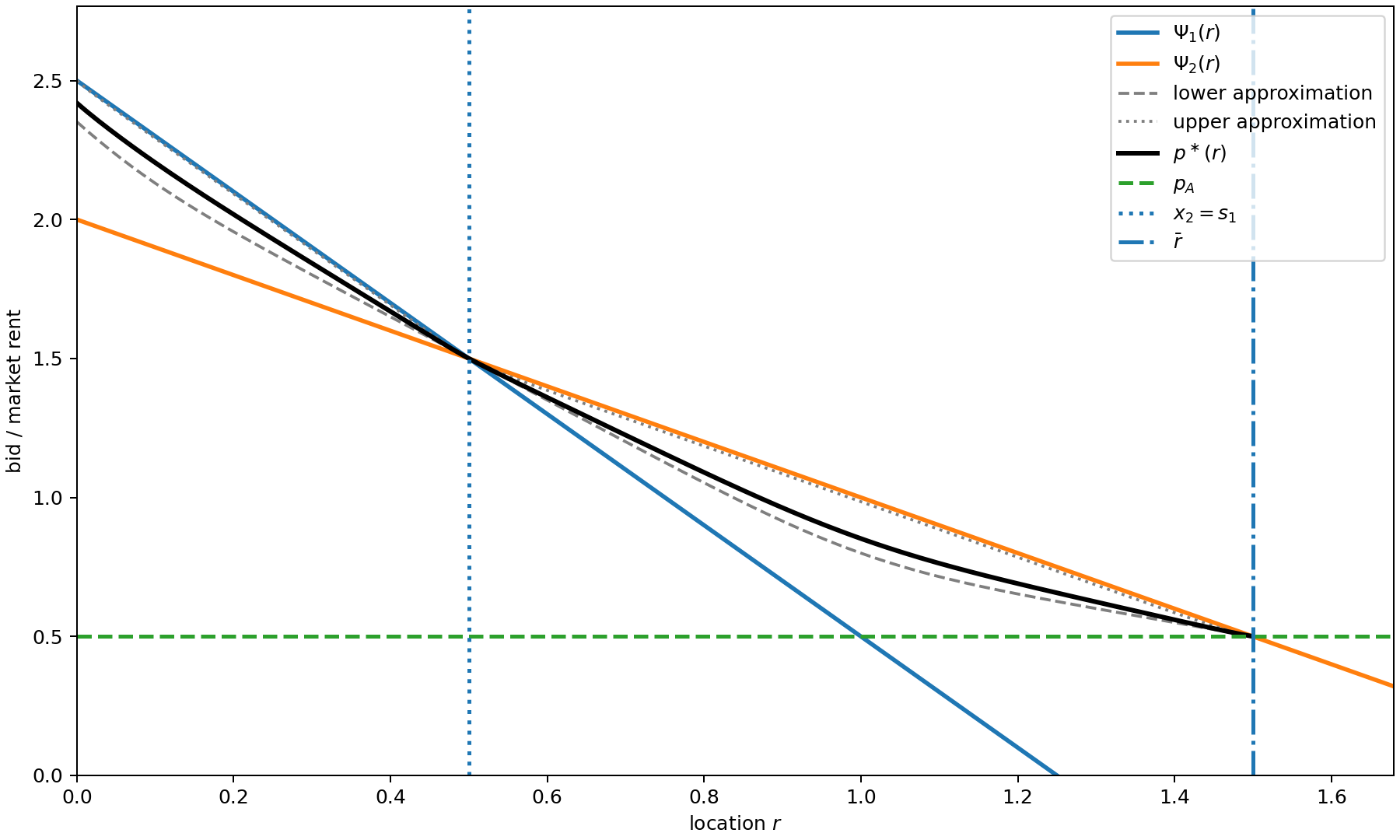}
\caption{
A locally supporting price constructed from smooth approximations to the
midpoint schedule and the upper envelope.
The gray dashed and dotted curves are the lower and upper bracketing
approximations, and the solid black curve is their parcel-specific convex
combination \(p^\ast\). Shared boundary slopes make \(p^\ast\) continuously
differentiable at \(x_2\).
}
\label{fig:theta-one-theorem2-price}
\end{figure}

The budget identities are satisfied numerically to the reported precision:
\[
    Z(s_i^\ast,u^\ast)
    +\int_{\Omega_i^\ast}p^\ast(r)\,dr
    +t r_{c,i}^\ast
    =10,
    \qquad i=1,2.
\]
Finally, because \(p^\ast\) is \(C^1\), all directional SOCs at a given
boundary use the same derivative. The resulting values are
\[
\begin{array}{c|c|c|c|c|c|c}
i & Z_{ss} & c_i & c_{i+1}
& A_i & B_i & A_iB_i-Z_{ss}^2\\
\hline
1 & 4 & -2.52 & -1.5
  & 6.52 & 2.5 & 0.3\\
2 & 1 & -1.5 & -0.5
  & 2.5 & 0.5 & 0.25
\end{array}
\]
so the coupled directional SOCs hold strictly for both parcels, and \(p^\ast\)
is a locally supporting price.

\begin{figure}[H]
\centering
\includegraphics[width=.82\textwidth]{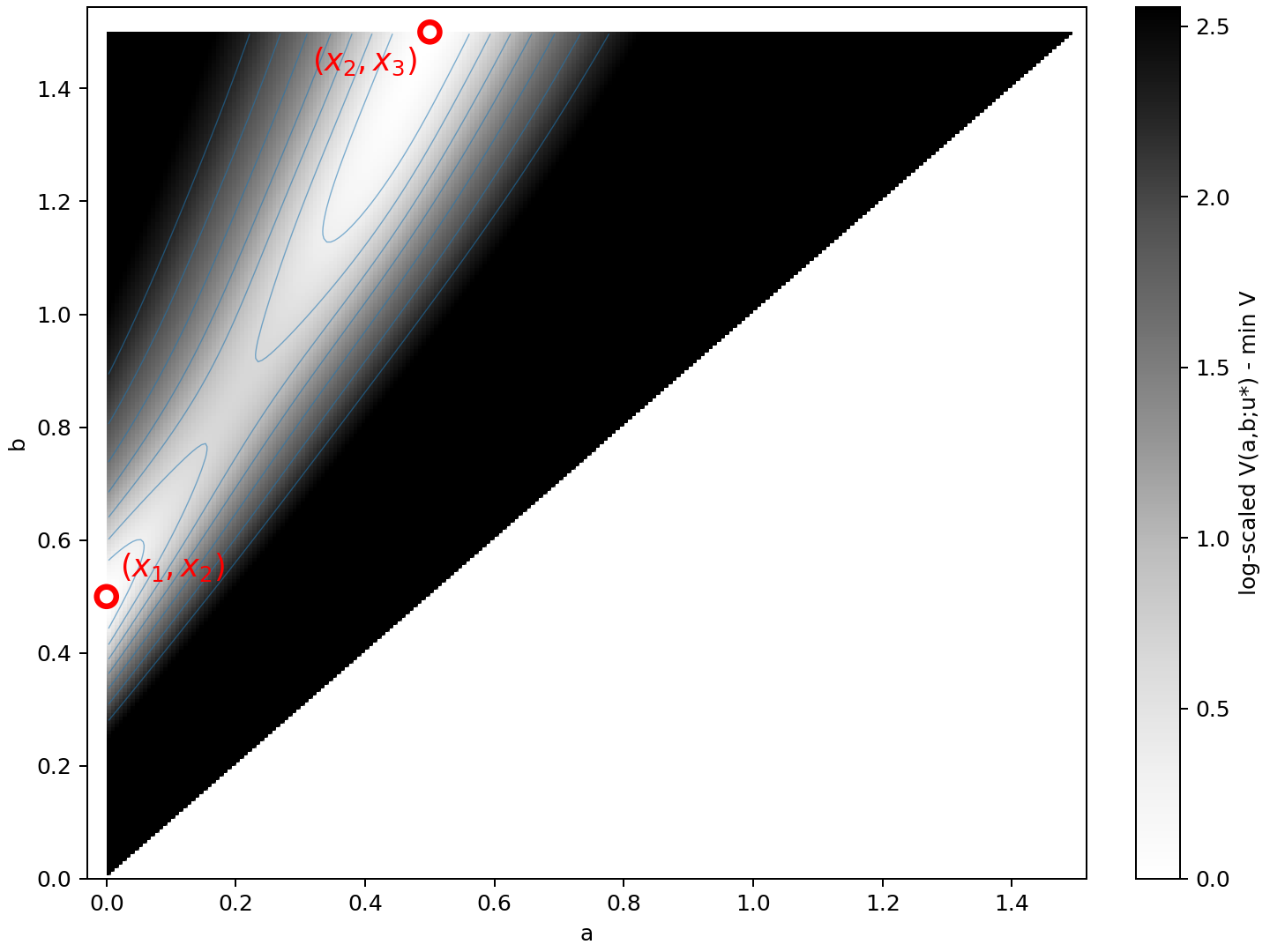}
\caption{
Interval expenditure under the locally supporting price \(p^\ast\).
The horizontal and vertical axes are the interval endpoints \(a\) and \(b\),
respectively. The red circles mark the assigned endpoint pairs
\((0,x_2)\) and \((x_2,\bar r)\). The grayscale reports
\(\log(1+(V-\min V)/c)\), where \(c\) is a visualization scale. Both assigned
parcels attain \(V=10\), and the grid search finds no lower-expenditure
interval.
}
\label{fig:theta-one-theorem2-heatmap}
\end{figure}

Beyond these local conditions, the grid search in Figure
\ref{fig:theta-one-theorem2-heatmap} finds no interval with
expenditure below \(10\), and neither does a search over intervals that
extend into agricultural land beyond \(\bar r\).  Numeraire consumption
\(u^\ast-\log s\) is positive at every interval covered by these searches,
since it is positive whenever \(s<e^{u^\ast}\approx3{,}256\), so the
nonnegativity constraint on numeraire consumption does not bind.  Since \(p^\ast\) is strictly convex, Lemma
\ref{lem:convex-interval-reduction} implies that no measurable assignment
does better than the best interval, so the numerical evidence indicates that,
in this example, \(p^\ast\) is also a supporting price.
\end{example}

\begin{rem}[Parcel payments and hedonic prices]
\label{rem:parcel-payments-hedonic}\normalfont
Fix $\Omega^\ast$ and a common utility level $u$.  Every budget-consistent
price $p$ satisfies
\[
    \Pi_i:=\int_{\Omega_i^\ast}p(r)\,dr
    =m\bigl\{w-Z(s_i^\ast,u)-t\,r_{c,i}^\ast\bigr\},
    \qquad i=1,\ldots,N.
\]
Given the equilibrium parcel, the total payment is a residual of income that depends on the price only through $u$.
Supporting prices at a given $u$ therefore differ only in how the payment for
each parcel is distributed over its locations.  This profile affects
households' incentives to deviate but no payment in equilibrium.  The range
of admissible utility levels, the level layer of rent indeterminacy, then
determines how the surplus is divided between households and landlords.

The right-hand side is the value at $(S,r_c)=(ms_i^\ast,r_{c,i}^\ast)$ of
\[
    R(S,r_c;u):=m\Bigl\{w-Z\Bigl(\frac{S}{m},u\Bigr)-t\,r_c\Bigr\},
\]
the largest payment for a parcel of measure $S$ and centroid $r_c$ that leaves
a household with utility $u$. As in hedonic models, bid rents are the corresponding marginal prices.  To
see this, write the bid function in terms of the measure and moment of a
parcel, $(S,M)=\int_A(1,r)\,dr$, which are the terminal values of the state
variables \eqref{eq:state_S} and \eqref{eq:state_M}:
\[
    \tilde R(S,M;u):=R\Bigl(S,\frac{M}{S};u\Bigr)
    =m\Bigl\{w-Z\Bigl(\frac{S}{m},u\Bigr)-t\,\frac{M}{S}\Bigr\}.
\]
A unit of land at location $r$ adds $(1,r)$ to $(S,M)$, so, in the spirit of
\citet{lancaster1966JournalofPoliticalEconomy}, it is a bundle of the two
characteristics that enter the household's expenditure.  The derivative of
$\tilde R$ in the direction $(1,r)$ is
\[
    \tilde R_S+\tilde R_M\,r
    =MRS(s,u)+\frac{t}{s}\,r_c-\frac{t}{s}\,r
    =MRS(s,u)-\frac{t}{s}(r-r_c),
\]
which is the bid rent \eqref{eq:bid_rent} at $r$ of a household with lot
size $s=S/m$ and centroid $r_c=M/S$.  At the equilibrium parcel of household
$i$, $\tilde R_S=m\lambda_{S_i}$ and $\tilde R_M=m\lambda_{M_i}$ by
\eqref{eq:costate_M} and \eqref{eq:costate_S}, so the bid rent values the
characteristics of each location at the household's shadow prices of land
and moment.  Only the $N$ bundles $(S,M)$ of the equilibrium parcels are
traded, however, so parcel payments reveal $\tilde R$ only at these points.

On the price side, the payment for an interval parcel is a function of its
size and centroid,
\[
    P_I(S,r_c):=\int_{r_c-S/2}^{r_c+S/2}p(r)\,dr,
    \qquad S/2\le r_c\le\hat r-S/2.
\]
Intervals close to $\Omega_i^\ast$ are among the small deviations that a
locally supporting price deters, so attaining utility $u$ with such an
interval requires an expenditure of at least $w$.  Hence $P_I\ge R$ near
$(ms_i^\ast,r_{c,i}^\ast)$, with equality at that point by budget
consistency.  As
in \citet{rosen1974JournalofPoliticalEconomy}, the price function lies above
the bid function and touches it at each traded parcel.  If $p$ is continuous
at the endpoints of an interior parcel $\Omega_i^\ast=[a,b]$, tangency means
that the partial derivatives of the two functions coincide there:
\[
    \frac{\partial P_I}{\partial S}=\frac{p(a)+p(b)}{2}
    =MRS(s_i^\ast,u)=\frac{\partial R}{\partial S},
    \qquad
    \frac{\partial P_I}{\partial r_c}=p(b)-p(a)
    =-mt=\frac{\partial R}{\partial r_c}.
\]
Solving gives $p(a)=MRS(s_i^\ast,u)+mt/2=\Psi_i^\ast(a,u)$ and
$p(b)=MRS(s_i^\ast,u)-mt/2=\Psi_i^\ast(b,u)$, which is the adjacent-boundary
condition of Lemma \ref{lem:fixed-utility-lot-sizes-dominance-regions}.  For
the innermost
parcel, the constraint $r_c\ge S/2$ binds, and only the outer boundary is
pinned down.  Rents inside the parcel affect $P_I$ at the traded point only
through budget consistency.

What differs from the hedonic model of
\citet{rosen1974JournalofPoliticalEconomy} is that the price function cannot
be chosen freely away from the traded parcels.  With identical consumers,
Rosen's hedonic price function can coincide with the common bid function
everywhere, so that consumers are indifferent among all products, traded or
not.
Here, in terms of the endpoints of an interval,
$\partial P_I/\partial a=-p(a)$ and $\partial P_I/\partial b=p(b)$, so
$\partial^2P_I/\partial a\,\partial b=0$, whereas
\[
    \frac{\partial^2R}{\partial a\,\partial b}=\frac{Z_{ss}(s,u)}{m}>0
\]
when $R$ is written in terms of the endpoints.  Hence no rent function makes
$P_I$ equal to $R$ on any neighborhood of a traded parcel, and prices must
exceed bids arbitrarily close to each traded parcel.\footnote{Since
$P_I-R=m\{V(a,b;u)-w\}$, with $V$ as in the proof of Proposition
\ref{prop:local-interval-stability}, the second-order conditions in that
proposition are sufficient conditions for the price function
to lie strictly above the bid function near the traded parcel, and the
off-diagonal terms of the relevant Hessian come only from the cross-partial
of $R$.}  The prices
of untraded intervals, moreover, are not separate objects: they are integrals
of the same rents that price the traded parcels, and are therefore determined
by the profile of rents within those parcels.  The difference stems from how
parcels are bought.  In the hedonic model, each consumer buys a single unit
of an indivisible product, whose price can be an arbitrary function of its
characteristics.  In the present model, by contrast, a household assembles its
parcel by buying land at many locations, much as consumers combine goods in
\citet{lancaster1966JournalofPoliticalEconomy}, so the price of a parcel is
the sum of the prices of the land it contains.

In particular, when $E_i(u)>w$, as under a strictly convex marginal rate of
substitution, the rent within a parcel cannot be the occupant's bid rent.
Since $\Psi_i^\ast$ is affine,
$\int_{\Omega_i^\ast}\Psi_i^\ast(r,u)\,dr=ms_i^\ast MRS(s_i^\ast,u)$, which
values the land in the parcel at its marginal rather than its total
valuation and exceeds $R(ms_i^\ast,r_{c,i}^\ast;u)$ by $m\{E_i(u)-w\}$ by
\eqref{eq:upper-envelope-expenditure}.  The profile layer is the freedom to
distribute this discount within the parcel.  Appendix
\ref{app:hedonic-estimation} discusses the implications for hedonic
estimation.\qed
\end{rem}

\section{Heterogeneous Households}
\label{sec:heterogeneous}

This section extends the benchmark to households that differ in preferences,
income, or commuting costs. In this setting, lot size alone no longer
determines spatial order; the relevant index is bid-rent steepness. We first
establish the ordering of individual households and then give conditions under
which these comparisons aggregate into contiguous blocks of household types.
We conclude with the income-heterogeneity specialization and a strategy for
constructing locally supporting prices.

\subsection{Finite-Type Environment}

We now extend the baseline model to a finite economy with heterogeneous
households. There are $K$ types, indexed by $k=1,\ldots,K$, and type $k$
contains $N_k\geq1$ households, indexed by $i=1,\ldots,N_k$. The total
population is $N=\sum_{k=1}^K N_k\geq2$. Each household continues to have mass
$m>0$; in the purely discrete normalization one may set $m=1$.

Households of type $k$ share a utility function $U_k(z,s)$, income $w_k$, and
marginal commuting cost $t_k>0$. Preferences satisfy the counterparts of
Assumptions \ref{ass:primitive} and \ref{ass:land-essentiality}. In a
Walrasian equilibrium, all households of the same type attain a common utility
level $u_k$, because any member of that type can imitate another member's
consumption plan. Utilities need not be equalized across different types.

Let $s_{k,i}>0$ and $r_{c,k,i}$ denote the per-capita lot size and centroid of
household $(k,i)$. Its bid-rent function at the equilibrium utility level is
\begin{equation}\label{eq:heterogeneous-bid-rent}
    \Psi_{k,i}(r,u_k)
    =
    MRS_k(s_{k,i},u_k)
    -\frac{t_k}{s_{k,i}}(r-r_{c,k,i}).
\end{equation}
Extending the bid-rent steepness in \eqref{eq:bid-rent-steepness} to
type-specific commuting costs, define
\begin{equation}\label{eq:individual-steepness}
    b_{k,i}:=\frac{t_k}{s_{k,i}}.
\end{equation}
Thus $\Psi_{k,i}$ is affine with slope $-b_{k,i}$. Heterogeneity in
preferences, income, or commuting technology affects spatial order through the
equilibrium collection of steepnesses $\{b_{k,i}\}$.

\subsection{Individual Sorting}
\label{subsec:finite-individual-sorting}

The first sorting result orders individual households and does not by itself
imply that each type occupies a contiguous territory. Without a restriction
across types such as Assumption \ref{ass:global-parcel-single-crossing} below,
members of two types may be interlaced even though every individual parcel is
contiguous.

\begin{proposition}[Individual steepness ordering]
\label{prop:individual-steepness-ordering}
Suppose that the equilibrium steepnesses
$b_{k,i}$ are pairwise distinct. Then the households' dominance regions are
nonempty intervals ordered from the CBD to the urban boundary in decreasing
order of $b_{k,i}$. In particular, if
$b_{k,i}>b_{\ell,j}$, household $(k,i)$ is located closer to the CBD than
household $(\ell,j)$.
\end{proposition}

\begin{proof}
Each bid-rent function in \eqref{eq:heterogeneous-bid-rent} is affine. The
upper envelope of finitely many affine functions is convex, so its slope is
non-decreasing in $r$. Because the slopes $-b_{k,i}$ are pairwise distinct,
the active bid-rent functions therefore appear in increasing order of their
slopes, or equivalently in decreasing order of their steepnesses $b_{k,i}$.
The single-crossing argument in Lemma
\ref{lem:nondegenerate-ordering} then implies that every active dominance
region is an interval. The counterparts of Assumption
\ref{ass:land-essentiality}, together with household optimality and market
clearing, imply that every household's dominance region has positive measure
and hence is nonempty.
\end{proof}

Within a type, $t_k$ is common, and hence
\begin{equation}\label{eq:within-type-size-steepness}
    b_{k,i}>b_{k,j}
    \quad\Longleftrightarrow\quad
    s_{k,i}<s_{k,j}.
\end{equation}
Consequently, Proposition \ref{prop:individual-steepness-ordering} orders the
members of each type by increasing lot size as distance from the CBD rises.
This is a relative ordering within the type; it does not yet exclude a member
of another type from locating between two type-$k$ households.

The ordering also determines each household's marginal rate of substitution.
Index households by their spatial order as $a=1,\ldots,N$, and let
$\kappa(a)$ denote the type of household $a$.  As in Lemma
\ref{lem:fixed-utility-lot-sizes-dominance-regions}, the bid rents of adjacent
households coincide at their common boundary, and the bid rent of the
outermost household equals the agricultural rent at the urban boundary.
Since the bid rent of a household of type $k$ differs from its marginal rate
of substitution by $mt_k/2$ at either end of its parcel, these conditions
are
\begin{equation}\label{eq:heterogeneous-adjacent-bid-matching}
\begin{aligned}
    &MRS_{\kappa(a)}(s_a,u_{\kappa(a)})
    -MRS_{\kappa(a+1)}(s_{a+1},u_{\kappa(a+1)})
    =\frac{m}{2}\{t_{\kappa(a)}+t_{\kappa(a+1)}\},
    &&a=1,\ldots,N-1,\\
    &MRS_{\kappa(N)}(s_N,u_{\kappa(N)})
    =p_A+\frac{m}{2}t_{\kappa(N)}.
\end{aligned}
\end{equation}
Summing them from the fringe inward gives
\begin{equation}\label{eq:heterogeneous-mrs-by-position}
    MRS_{\kappa(a)}(s_a,u_{\kappa(a)})
    =p_A+\frac{m}{2}t_{\kappa(a)}+m\sum_{a'>a}t_{\kappa(a')},
    \qquad a=1,\ldots,N.
\end{equation}
The sum $\sum_{a'>a}t_{\kappa(a')}$ runs over the households located outside
household $a$ and lies between $0$ and
$\sum_{\ell=1}^KN_\ell t_\ell-t_{\kappa(a)}$.
Whatever the order of households, the lot size of every household of type
$k$ therefore belongs to
\begin{equation}\label{eq:admissible-lot-sizes}
    \Sigma_k
    :=\Bigl\{s>0:\ p_A+\frac{m}{2}t_k\le MRS_k(s,u_k)\le p_A+m\Bigl(\sum_{\ell=1}^KN_\ell t_\ell-\frac{t_k}{2}\Bigr)
    \Bigr\}.
\end{equation}
We call $\Sigma_k$ the set of admissible lot sizes of type $k$.  It depends
on the parameters and on the utility level $u_k$, but not on the order of
households.

\subsection{Type-Level Sorting}
\label{subsec:global-parcel-single-crossing}

The individual steepness ordering does not by itself explain why members of a
type must occur consecutively. We now impose a condition directly on the
types' relative rankings of finite parcels. For a parcel, that is, a measurable set $P\subseteq[0,\hat r]$ with $|P|>0$,
write $s_P:=|P|/m$ for its per-capita lot size, $r_P:=r_c(P)$ for its
centroid, and $R_P:=\frac1m\int_P p(r)\,dr$ for its per-capita rent payment.

For two types $k$ and $\ell$ and two parcels $P,Q$ with $r_P<r_Q$, define
\begin{equation}\label{eq:general-parcel-single-crossing-index}
    \Gamma_{k\ell}(P,Q)
    :=
    \{Z_k(s_Q,u_k)-Z_k(s_P,u_k)\}
    -\{Z_\ell(s_Q,u_\ell)-Z_\ell(s_P,u_\ell)\}
    +(t_k-t_\ell)(r_Q-r_P).
\end{equation}
This index is the difference between the two types' relative costs of moving
from the inner parcel $P$ to the outer parcel $Q$. The common rent difference
$R_Q-R_P$ cancels and therefore does not enter
\eqref{eq:general-parcel-single-crossing-index}.

Let $\Sigma_k$ be the set of admissible lot sizes of type $k$ defined in
\eqref{eq:admissible-lot-sizes}.  We say that an interval $P$ lies to the
left of an interval $Q$ if $\sup P\le\inf Q$.

\begin{assumption}[Pairwise parcel single crossing]
\label{ass:global-parcel-single-crossing}
The types can be relabeled as $1,\ldots,K$ so that, for every $k<\ell$,
\begin{equation}\label{eq:pairwise-global-single-crossing}
    \Gamma_{k\ell}(P,Q)>0
\end{equation}
for every pair of intervals $P,Q\subseteq[0,\hat r]$ such that $P$ lies to
the left of $Q$, $s_P\in\Sigma_\ell$, $s_Q\in\Sigma_k$, and
\begin{equation}\label{eq:single-crossing-comparison-set}
    \frac{t_\ell}{s_P}>\frac{t_k}{s_Q}.
\end{equation}
\end{assumption}

Assumption \ref{ass:global-parcel-single-crossing} is a single-crossing
condition on the types' willingness to pay for inner locations. Because
$\Sigma_k$ and $\Gamma_{k\ell}$ are evaluated at the equilibrium utility
levels, it is not a condition on primitives alone. At its
equilibrium utility level, type $k$ is willing to pay at most
$\{Z_k(s_Q,u_k)+t_kr_Q\}-\{Z_k(s_P,u_k)+t_kr_P\}$ more in rent for the inner
parcel $P$ than for the outer parcel $Q$. The assumption states that this
maximal premium is larger for type $k$ than for any type $\ell>k$. Because
both types face the same rent difference $R_P-R_Q$, the type with the lower
label gains more, or loses less, from taking the inner of the two parcels.
Types with lower labels therefore locate closer to the CBD.

\begin{proposition}[Two-level spatial sorting]
\label{prop:two-level-sorting}
Suppose that the hypotheses of Proposition
\ref{prop:individual-steepness-ordering} and Assumption
\ref{ass:global-parcel-single-crossing} hold. Then the aggregate territory
occupied by each type is a contiguous interval, and the types are ordered
from the CBD outward as $1,\ldots,K$. Within each type's territory, its
households are ordered by increasing lot size.

Moreover, the equilibrium steepness ranges are endogenously separated. For
every $k<\ell$,
\begin{equation}\label{eq:endogenous-steepness-range-separation}
    \min_{1\leq i\leq N_k}b_{k,i}
    >
    \max_{1\leq j\leq N_\ell}b_{\ell,j}.
\end{equation}
In particular, \eqref{eq:endogenous-steepness-range-separation} holds for
every pair of adjacent types $k$ and $k+1$.
\end{proposition}

\begin{proof}
Fix $k<\ell$. Suppose, toward a contradiction, that a type-$\ell$ household
occupies an inner parcel $P$ and a type-$k$ household occupies an outer parcel
$Q$. By Proposition \ref{prop:individual-steepness-ordering}, $P$ and $Q$ are
intervals, $P$ lies to the left of $Q$, and the household at $P$ has the
steeper bid rent, so \eqref{eq:single-crossing-comparison-set} holds. Since
$s_P$ and $s_Q$ are equilibrium lot sizes of types $\ell$ and $k$,
$s_P\in\Sigma_\ell$ and $s_Q\in\Sigma_k$ by
\eqref{eq:admissible-lot-sizes}. Hence Assumption
\ref{ass:global-parcel-single-crossing} applies to the pair $(P,Q)$. Expenditure minimization by the type-$k$ household gives
\begin{equation}\label{eq:general-sc-inner-type-choice}
    Z_k(s_Q,u_k)+R_Q+t_kr_Q
    \leq
    Z_k(s_P,u_k)+R_P+t_kr_P,
\end{equation}
whereas expenditure minimization by the type-$\ell$ household gives
\begin{equation}\label{eq:general-sc-outer-type-choice}
    Z_\ell(s_P,u_\ell)+R_P+t_\ell r_P
    \leq
    Z_\ell(s_Q,u_\ell)+R_Q+t_\ell r_Q.
\end{equation}
Adding \eqref{eq:general-sc-inner-type-choice} and
\eqref{eq:general-sc-outer-type-choice} yields
$\Gamma_{k\ell}(P,Q)\leq0$, contradicting
\eqref{eq:pairwise-global-single-crossing}. Thus every household of type $k$
is located closer to the CBD than every household of type $\ell$.

Applying this comparison to every $k<\ell$ shows that the individual
intervals occur in type blocks ordered as $1,\ldots,K$. Proposition
\ref{prop:individual-steepness-ordering} implies that these nonempty individual
intervals are mutually disjoint, so the union of the intervals in each block
is itself an interval. Within a type, \eqref{eq:within-type-size-steepness}
then gives the increasing-lot-size ordering.

Finally, the same individual steepness ordering implies
\[
    b_{k,i}
    >
    b_{\ell,j}
    \qquad
    \text{for every }k<\ell,
    \quad i=1,\ldots,N_k,
    \quad j=1,\ldots,N_\ell.
\]
Taking the minimum over type $k$ and the maximum over type $\ell$ proves
\eqref{eq:endogenous-steepness-range-separation}.
\end{proof}

The term ``two-level sorting'' refers to the structure of the equilibrium
allocation, not to a two-stage process. The land market clears once, and the
resulting allocation is ordered at two levels. Across types, Assumption
\ref{ass:global-parcel-single-crossing} determines the order of the type
territories. Within each territory, Proposition
\ref{prop:individual-steepness-ordering} orders households by lot size. The
separation of steepness ranges in
\eqref{eq:endogenous-steepness-range-separation} is part of the conclusion of
Proposition \ref{prop:two-level-sorting}, not an additional assumption.

The assumption involves every pair of intervals, but when $t_k\ge t_\ell$ it
reduces to a condition on pairs of lot sizes alone, which is easier to check.
The index depends on
the two parcels only through their lot sizes $s_P$ and $s_Q$ and the
distance between their centroids: since $\partial Z_k/\partial s=-MRS_k$,
\begin{equation}\label{eq:parcel-single-crossing-decomposition}
    \Gamma_{k\ell}(P,Q)
    =(t_k-t_\ell)(r_Q-r_P)
    -\int_{s_P}^{s_Q}\bigl\{MRS_k(s,u_k)-MRS_\ell(s,u_\ell)\bigr\}\,ds.
\end{equation}
If $P$ lies to the left of $Q$, then $r_Q-r_P\ge m(s_P+s_Q)/2$, with
equality if and only if the two intervals are adjacent. Hence, if $t_k\ge t_\ell$,
the index is smallest when the two intervals are adjacent, and the
assumption holds for the pair of types $k<\ell$ if and only if
\begin{equation}\label{eq:parcel-single-crossing-sizes}
    \int_{s_P}^{s_Q}\bigl\{MRS_k(s,u_k)-MRS_\ell(s,u_\ell)\bigr\}\,ds
    <(t_k-t_\ell)\,\frac{m(s_P+s_Q)}{2}
\end{equation}
for all lot sizes $s_P\in\Sigma_\ell$ and $s_Q\in\Sigma_k$ with
$s_Q/s_P>t_k/t_\ell$ and $m(s_P+s_Q)\le\hat r$, the last condition ensuring
that two such intervals fit in $[0,\hat r]$. The right-hand side is the additional commuting cost that type $k$ bears,
relative to type $\ell$, when it takes the outer of two adjacent parcels. The
left-hand side is the difference between the two types' valuations of the
change in lot size from the inner to the outer parcel. Since
$s_Q/s_P>t_k/t_\ell\ge1$, the outer parcel is the larger one in every such
comparison.

Two special cases illustrate the condition. If the two types have the same
commuting cost, $t_k=t_\ell$, then \eqref{eq:parcel-single-crossing-sizes}
requires $\int_{s_P}^{s_Q}\{MRS_\ell(s,u_\ell)-MRS_k(s,u_k)\}\,ds>0$
whenever $s_P<s_Q$.
It holds if $MRS_k(s,u_k)<MRS_\ell(s,u_\ell)$ at every lot size $s$ between
the smallest and the largest of the lot sizes concerned: the type with the
lower marginal valuation of land holds smaller lots closer to the CBD, as in
the homogeneous benchmark. If instead the two types have the same marginal
valuation of land, $MRS_k(\cdot,u_k)=MRS_\ell(\cdot,u_\ell)$, the condition
reduces to $t_k>t_\ell$: the type with the higher commuting cost locates
closer to the CBD. In general, the assumption requires the difference in
commuting costs to dominate the difference in the valuation of land. If
$t_k<t_\ell$, the index falls as the two parcels move apart, and the
assumption has to be checked in the form
\eqref{eq:pairwise-global-single-crossing}.

\subsubsection{Specialization to Income Heterogeneity}
\label{subsubsec:income-heterogeneity}

We next specialize the sorting condition to income heterogeneity. There are two
income types, denoted by $H$ and $L$, with incomes $w_H>w_L$, and commuting
cost is proportional to income, $t_k=\tau w_k$ with $\tau>0$, as when
commuting takes time valued at the wage.  Thus $t_H>t_L$: a high-wage
household incurs a larger opportunity cost of commuting. To isolate income heterogeneity, suppose that both types have the
same general utility function $U(z,s)$ and expenditure function $Z(s,u)$,
although their equilibrium utility levels $u_H$ and $u_L$ may differ. The
decomposition \eqref{eq:parcel-single-crossing-decomposition} then becomes
\begin{equation}\label{eq:parcel-single-crossing-mrs}
    \Gamma_{HL}(P,Q)
    =(t_H-t_L)(r_Q-r_P)
    -\int_{s_P}^{s_Q}\{MRS(s,u_H)-MRS(s,u_L)\}\,ds.
\end{equation}
A higher wage thus has two possible effects on location. The first
term is an opportunity-cost effect: a higher wage raises $t_k=\tau w_k$ and
thus the cost of an outward commute. The second term is an income effect: a
higher wage may raise the demand for land and make a larger outer parcel
more attractive.

Suppose first that type $H$ is labeled first. Since $t_H/t_L=w_H/w_L$,
Assumption \ref{ass:global-parcel-single-crossing} holds if and only if
\eqref{eq:parcel-single-crossing-sizes} holds with $k=H$ and $\ell=L$, that
is, for all lot sizes $s_P\in\Sigma_L$ and $s_Q\in\Sigma_H$ with
$s_Q/s_P>w_H/w_L$ and $m(s_P+s_Q)\le\hat r$. The comparison
concerns a low-wage household at an inner parcel and a high-wage household
at an outer parcel that is larger by more than the income ratio. If the
opportunity-cost effect dominates the income effect in every such
comparison, Proposition \ref{prop:two-level-sorting} places the high-wage
block inside.

Suppose instead that type $L$ is labeled first. The comparisons then concern
a high-wage household at an inner parcel of size $s_P\in\Sigma_H$ and a
low-wage household at an outer parcel of size $s_Q\in\Sigma_L$ with
$s_P/s_Q<w_H/w_L$. In these comparisons the opportunity-cost effect works
against the assumption, and it does so more strongly the farther apart the
two parcels are. The assumption holds, however, when there is no such
comparison, for example, if even the smallest admissible lot size of type
$H$ exceeds $w_H/w_L$ times the largest admissible lot size of type $L$. Lot
size then rises more than proportionally with income, every high-wage
household has a flatter bid rent than every low-wage household, and the
high-wage block lies outside. In this
case type-level sorting already follows from Proposition
\ref{prop:individual-steepness-ordering}. If neither labeling satisfies the
assumption, households are still ordered by bid-rent steepness, but the
types may be interlaced.

Assumption \ref{ass:global-parcel-single-crossing} is the finite-parcel
counterpart of the familiar income-elasticity condition in continuum
bid-rent models \citep[Proposition 2.2]{fujita1989}. In a continuum model, let $s(w,r)$ denote the lot demanded by
a household with income $w$ at location $r$, define the steepness of its bid rent by
$b(w,r):=w\tau/s(w,r)$, and define
$\varepsilon_{s,w}(w,r):=\partial\log s(w,r)/\partial\log w$. Then
\(\partial\log b(w,r)/\partial\log w=1-\varepsilon_{s,w}(w,r)\).
If the elasticity is uniformly below one, high-wage households therefore have
steeper bid rents at every location, and the standard continuum argument, in
which steeper bid rents win locations closer to the CBD, places them in the inner area. If the elasticity is uniformly above one, the order is reversed.\footnote{Under Cobb--Douglas preferences,
$\varepsilon_{s,w}=1$, and with commuting costs proportional to income, any
arrangement of the income groups, including a mixed one, is an equilibrium of
the continuum model \citep[Proposition 2.2(iii)]{fujita1989}.  In the present
model, $\Gamma_{HL}$ may then change sign across pairs of parcels, and we
have not examined which arrangements arise in equilibrium.}

In the present finite economy, however, a household chooses a parcel of
positive length rather than land at a single location. A local elasticity
condition therefore does not by itself determine the ranking of arbitrary
finite parcels. Assumption \ref{ass:global-parcel-single-crossing} imposes the
corresponding single-crossing property directly on every relevant pair of
parcels. It thus plays the same role as the continuum elasticity condition,
without being implied by that condition.

Finally, the no-income-effect case provides a simple primitive example in
which the positive sign is automatic. If \(U(z,s)=z+v(s)\),
then $MRS(s,u)=v'(s)$ does not depend on $u$, and the integral in
\eqref{eq:parcel-single-crossing-mrs} vanishes. Hence
\(\Gamma_{HL}(P,Q)=(t_H-t_L)(r_Q-r_P)>0\).
High-wage households therefore form the inner block, low-wage households form
the outer block, and the two-level pattern follows without imposing
slope-range separation directly.

\subsection{Sufficient Conditions for Locally Supporting Prices}
\label{subsec:heterogeneous-supporting-price-route}

We conclude this section by describing how the construction of Theorem
\ref{thm:supporting-price-existence} may be extended to a fixed candidate
ordering of heterogeneous households.  We do not establish an existence
theorem for the heterogeneous case.  The construction takes the
spatial ordering as given and assumes heterogeneous counterparts of five
elements of the homogeneous case: the lot-size profile of Lemma
\ref{lem:fixed-utility-lot-sizes-dominance-regions}, the anchor utility of
Lemma \ref{lem:anchor-utility}, the monotonicity of upper-envelope
expenditures of Lemma \ref{lem:upper-envelope-expenditure-monotonicity}, now
within each type, the slack condition \eqref{eq:midpoint-expenditure-slack},
and the curvature condition of Assumption \ref{ass:midpoint-curvature}.  In
the homogeneous case, the first three follow from primitive assumptions,
whereas the last two are hypotheses of Theorem
\ref{thm:supporting-price-existence}.  Here, the lot sizes of
inner types shift the centroids of all outer types, so the anchor utilities
of the types solve a genuinely multidimensional system.  A numerical example shows that the
key sorting, bracketing, and local-curvature conditions can hold
simultaneously, and constructs an explicit locally supporting price.  As in
the homogeneous case, the construction itself rules out only small deviations,
although in the example a grid search finds no profitable large deviation
either.

Index households by their proposed spatial order as $a=1,\ldots,N$, and let
$\kappa(a)$ denote the type of household $a$, as in Section
\ref{subsec:finite-individual-sorting}.  Conditional on this order, the lot
sizes satisfy the adjacent-boundary and fringe equations
\eqref{eq:heterogeneous-adjacent-bid-matching}.
Under the backward recursion from the fringe, the Jacobian of this system
with respect to $(s_1,\ldots,s_N)$ is triangular, and it is nonsingular if
the derivatives $MRS_{k,s}(s_{k,i},u_k)$ are nonzero.  The implicit function
theorem then gives a locally unique and continuously differentiable lot-size
vector $\boldsymbol s=\boldsymbol s(\boldsymbol u)$, where
$\boldsymbol u=(u_1,\ldots,u_K)$, and the parcel boundaries and centroids are
continuously differentiable functions of $\boldsymbol u$ as well.

For the allocation induced by $\boldsymbol u$, define the upper-envelope
expenditure of household $(k,i)$ by
\(E_{k,i}(\boldsymbol u):=Z_k(s_{k,i}(\boldsymbol u),u_k)
+s_{k,i}(\boldsymbol u)MRS_k(s_{k,i}(\boldsymbol u),u_k)
+t_k r_{c,k,i}(\boldsymbol u)\).
Although only $u_k$ enters $Z_k$ and $MRS_k$ directly,
$E_{k,i}$ generally depends on the full vector $\boldsymbol u$.  Changing an
inner type's lot sizes shifts the centroids of all types located outside it.

Let $(k,1)$ denote the innermost household within type $k$, and introduce the
anchor-expenditure map
\(G(\boldsymbol u):=\bigl(E_{1,1}(\boldsymbol u),\ldots,
          E_{K,1}(\boldsymbol u)\bigr)^{\mathsf T}\).
The heterogeneous counterpart of the starting point in Theorem
\ref{thm:supporting-price-existence} is a vector
$\boldsymbol u^0$ satisfying the $K$ anchor equations
\begin{equation}\label{eq:heterogeneous-anchor-equations}
    G(\boldsymbol u^0)=\boldsymbol w,
    \qquad
    \boldsymbol w=(w_1,\ldots,w_K)^{\mathsf T}.
\end{equation}
Suppose that $D G(\boldsymbol u^0)$ is nonsingular.  The implicit function
theorem then gives, for each sufficiently small $\varepsilon>0$, a utility
vector $\boldsymbol u^\varepsilon$ near $\boldsymbol u^0$ with
$G(\boldsymbol u^\varepsilon)=\boldsymbol w+\varepsilon\boldsymbol 1$, at
which all type-specific anchor expenditures lie strictly above their
respective incomes.  Strict steepness ordering, if it holds at $\boldsymbol u^0$, continues to
hold at $\boldsymbol u^\varepsilon$ for small $\varepsilon$, because it
consists of finitely many strict inequalities between continuous functions
of $\boldsymbol u$.  The construction takes the spatial order as given and
does not use Assumption \ref{ass:global-parcel-single-crossing}.

For the lower side of the required expenditure bounds, define the midpoint
expenditure
\(L_{k,i}(\boldsymbol u):=E_{k,i}(\boldsymbol u)-\frac12T_{k,i}(\boldsymbol u)\),
where
\(T_{k,i}(\boldsymbol u):=\frac{1}{m}\int_{\Omega_{k,i}(\boldsymbol u)}
      \{\Psi_{k,i}(r,u_k)-\Psi_{-(k,i)}(r,\boldsymbol u)\}\,dr\).
Suppose that at the anchor vector
\begin{equation}\label{eq:heterogeneous-midpoint-slack}
    L_{k,i}(\boldsymbol u^0)<w_k
    \qquad
    \text{for every }k\text{ and }i.
\end{equation}
If the within-type counterpart of upper-envelope expenditure monotonicity
holds, so that $E_{k,1}(\boldsymbol u)\le E_{k,i}(\boldsymbol u)$ for every
member $i$ of type $k$, the upper inequalities extend from the anchors to all
households at $\boldsymbol u^\varepsilon$, while
\eqref{eq:heterogeneous-midpoint-slack} is preserved by continuity.  Hence,
for sufficiently small $\varepsilon>0$, $\boldsymbol u^\ast:=\boldsymbol
u^\varepsilon$ satisfies
\begin{equation}\label{eq:heterogeneous-strict-bracketing}
    L_{k,i}(\boldsymbol u^\ast)
    <w_k
    <E_{k,i}(\boldsymbol u^\ast)
    \qquad
    \text{for every }k\text{ and }i.
\end{equation}

Condition \eqref{eq:heterogeneous-strict-bracketing} brackets the required rent
integral on every parcel between the midpoint and upper schedules.  Write
\(q_{k,i}(r):=\frac12\{\Psi_{k,i}(r,u_k)+\Psi_{-(k,i)}(r,\boldsymbol u)\}\)
for the midpoint schedule.  Since the bid rent $\Psi_{k,i}$ integrates to
$ms_{k,i}MRS_k(s_{k,i},u_k)$ over the parcel, the definitions of $E_{k,i}$ and
$L_{k,i}$ show that \eqref{eq:heterogeneous-strict-bracketing} is equivalent
to
\(\int_{\Omega_{k,i}}q_{k,i}(r)\,dr
    <m\{w_k-Z_k(s_{k,i},u_k)-t_k r_{c,k,i}\}
    <\int_{\Omega_{k,i}}\Psi_{k,i}(r,u_k)\,dr\).
Thus a parcel-specific convex combination of the midpoint and upper schedules
can match each type-specific budget.

Continuity and budget consistency alone do not yet give a locally supporting price.
To obtain a price that is $C^1$ and strictly convex on $[0,\bar r]$, as in
Theorem \ref{thm:supporting-price-existence}, one must
construct smooth midpoint and upper-envelope approximations that share both
their boundary values and their boundary derivatives.  With
\(b_a:=t_{\kappa(a)}/s_a\) and \(c_{a+1}:=-(b_a+b_{a+1})/2\),
the heterogeneous parcelwise SOC takes the form
\begin{equation}\label{eq:heterogeneous-midpoint-soc}
    Z_{\kappa(a),ss}(s_a,u_{\kappa(a)})
       (c_{a+1}-c_a)
    -m c_a c_{a+1}>0.
\end{equation}
If these inequalities and the appropriate endpoint conditions hold, smooth
midpoint and upper-envelope approximations can be chosen with the same boundary
values and slopes. Their parcel-specific convex combinations then paste into a
price that is $C^1$ on $[0,\bar r]$ despite the different budget-matching
weights. The resulting price is first-order supporting, budget consistent, and
satisfies the strict local interval SOCs.  Since it is strictly convex, Lemma
\ref{lem:local-measurable-optimality} then makes it locally supporting, which
gives a conditional heterogeneous analogue of Theorem
\ref{thm:supporting-price-existence}.

\begin{example}[Two-level sorting and price bracketing]
\label{ex:heterogeneous-quasilinear-numerical}
Consider the income-heterogeneity model of Section
\ref{subsubsec:income-heterogeneity}, with two households of each type,
\(N_H=N_L=2\), \(w_H=20\), and \(w_L=10\).
Set \(m=1\), \(\tau=1/10\), and \(p_A=1/2\), so that \(t_H=2\) and
\(t_L=1\).
Both types have log-quasilinear preferences,
\(U(z,s)=z+\log s\), \(Z(s,u)=u-\log s\), and \(MRS(s,u)=1/s\).
Because there is no income effect on the marginal valuation of land, this
quasilinear specification gives
\(\Gamma_{HL}(P,Q)>0\) for every pair \(r_P<r_Q\).  The type order is
therefore \(H,H,L,L\).

Starting at the fringe and applying
\eqref{eq:heterogeneous-adjacent-bid-matching}, the marginal rates of
substitution in this spatial order are \(\left(\frac{11}{2},\frac72,2,1\right)\).
Since \(MRS=1/s\), the corresponding lot sizes and bid-rent steepnesses are
\[
\begin{array}{c|c|c|c|c}
 a&\text{type}&s_a&r_{c,a}&b_a=t_{\kappa(a)}/s_a\\
\hline
1&H&2/11&1/11&11\\
2&H&2/7&25/77&7\\
3&L&1/2&221/308&2\\
4&L&1&113/77&1
\end{array}
\]
Thus the steepnesses are strictly decreasing from the CBD outward.  The
example exhibits both levels of sorting: the high-income block is inside the
low-income block, and the two households within each type have distinct lot
sizes and steepnesses.

For these fixed parcels, upper-envelope expenditure is
\(E_a(\boldsymbol u)=u_{\kappa(a)}-\log s_a+1+t_{\kappa(a)}r_{c,a}\).
The two anchor equations, using the innermost member of each type, give
\[
    u_H^0=20+\log(2/11)-1-\frac{2}{11}=17.1134,
    \qquad
    u_L^0=10+\log(1/2)-1-\frac{221}{308}=7.5893.
\]
At this anchor vector, the midpoint expenditure is strictly below income for
every household.  Increase both type utilities by \(0.01\), so that
\(u_H^\ast=u_H^0+0.01\) and \(u_L^\ast=u_L^0+0.01\).
The resulting midpoint expenditures \(L_a\), incomes, and upper-envelope
expenditures \(E_a\) are
\[
\begin{array}{c|c|c|c|c}
 a&\text{type}&L_a&w_{\kappa(a)}&E_a\\
\hline
1&H&19.9769&20&20.01\\
2&H&19.9802&20&20.0255\\
3&L& 9.9579&10&10.01\\
4&L& 9.9419&10&10.0669
\end{array}
\]
and hence \(L_a<w_{\kappa(a)}<E_a\) for \(a=1,\ldots,4\).

Finally, using the steepnesses \(b_a\) in the first table, choose the common
midpoint boundary-slope vector
\((c_1,c_2,c_3,c_4,c_5)=(-13,-9,-4.5,-1.5,-0.5)\).
It satisfies the required interlacing inequalities.  Since
\(Z_{ss}(s,u)=1/s^2\), the four margins in
\eqref{eq:heterogeneous-midpoint-soc} are
\[
\begin{array}{c|c|c}
 a&Z_{ss}(s_a,u_{\kappa(a)}^\ast)
 &Z_{ss}(c_{a+1}-c_a)-c_ac_{a+1}\\
\hline
1&30.25&4.000\\
2&12.25&14.625\\
3& 4.00&5.250\\
4& 1.00&0.250
\end{array}
\]
and are all strictly positive.  Thus this finite economy satisfies the
sorting, strict bracketing, and midpoint-SOC conditions required by the
construction described above.

These conditions can be turned into an explicit price.  The construction above
needs smooth approximations, as in Example \ref{ex:numerical-supporting-price},
because the upper envelope and the midpoint schedule are themselves piecewise
linear.  The former has kinks at the parcel boundaries, the latter has kinks
inside the parcels, and neither is strictly convex, so their parcelwise
combinations cannot be used directly.  Here we instead build a single
function that is \(C^1\) and strictly convex on \([0,\bar r]\) by
construction, and use the two
schedules only as bounds.  This also allows every condition to be verified
exactly.  On each parcel,
set \(p^\ast(r)=\Psi_a(r,u_{\kappa(a)}^\ast)-\varphi_a(r-x_a)\), where
\(\varphi_a\) is concave, vanishes at both ends of the parcel, and has end
slopes such that the slopes of \(p^\ast\) at the two ends of \(\Omega_a\)
equal \(c_a\) and \(c_{a+1}\).  Such a \(\varphi_a\) lies between zero and its
two end tangents.  With the slopes \(c\), these tangents never exceed
\(\frac12(\Psi_a-\Psi_{-a})\), the discount of the midpoint schedule, and
coincide with it except on the innermost parcel.  The upper envelope and the
midpoint schedule thus correspond to the two limiting cases of
\(\varphi_a\), and every admissible \(\varphi_a\) yields a price between them.  In particular,
\(\Psi_{-a}\le p^\ast\le\Psi_a\), so \(p^\ast\) is first-order supporting.
Instead of mixing two approximations, we take \(\varphi_a\) to be piecewise
quadratic, with a negative second derivative that is constant on each of three
subintervals, and choose the three curvature levels to solve three linear
equations.  These equations impose the end slopes and budget consistency,
\(\int_{\Omega_a}\varphi_a(r-x_a)\,dr=m\{E_a-w_{\kappa(a)}\}\), and each
solution is strictly positive.  Three subintervals suffice here, but in
general more may be needed to keep every curvature level positive.  Write
\(\varphi_a''=-\omega_{a,j}\) on the \(j\)th subinterval and \(S_a=ms_a\) for
the length of the parcel.  The break points between the subintervals, chosen
on a grid so that all three curvature levels are positive, and the curvature
levels are
\[
\begin{array}{c|c|c|c|c|c}
 a&\text{type}&\text{break points}/S_a
 &\omega_{a,1}&\omega_{a,2}&\omega_{a,3}\\
\hline
1&H&0.19,\ 0.81&28.61&17.95&28.61\\
2&H&0.07,\ 0.79&27.92&10.95&28.16\\
3&L&0.03,\ 0.98&161.10&0.494&34.87\\
4&L&0.16,\ 0.84&1.736&0.654&1.736
\end{array}
\]
Together with \(\varphi_a(0)=0\) and \(\varphi_a'(0)=-b_a-c_a\), these
values determine \(p^\ast\).  The
resulting price, shown in Figure
\ref{fig:heterogeneous-price}, is \(C^1\) and strictly convex on
\([0,\bar r]\), with
\(p^\ast(0)=\Psi_1(0)\) and \(p^\ast(\bar r)=p_A\).  Since its boundary slopes
are \(c\), the margins in the table above are exactly the determinant
conditions of Proposition \ref{prop:local-interval-stability} and Lemmas
\ref{lem:appendix-innermost-local-stability} and
\ref{lem:appendix-outermost-local-stability}, whose diagonal conditions also
hold.  Lemma \ref{lem:local-measurable-optimality} then implies that
\(p^\ast\) is a locally supporting price.

\begin{figure}[H]
\centering
\includegraphics[width=.9\textwidth]{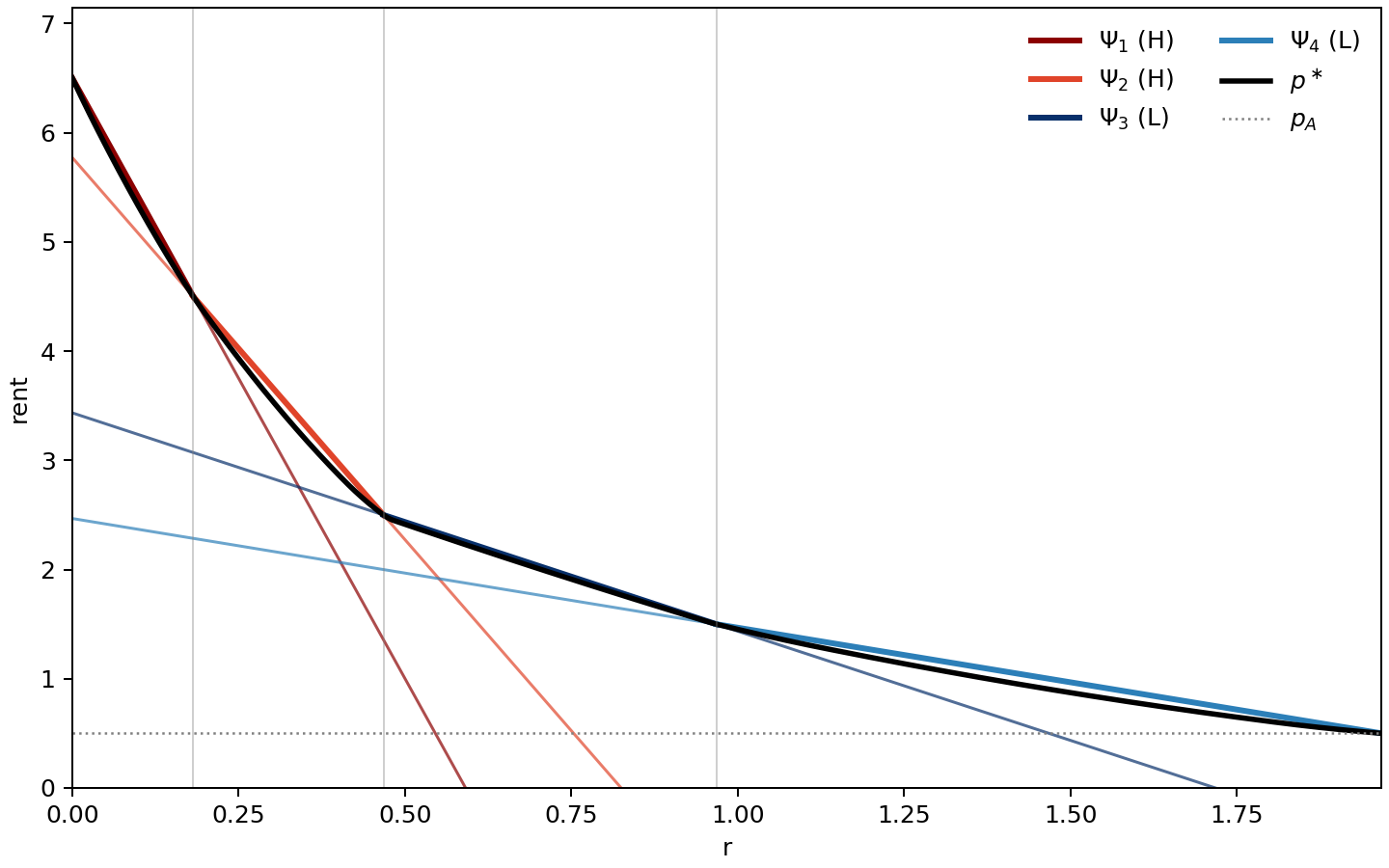}
\caption{
A locally supporting price for the heterogeneous example.
The red and blue lines are the bid rents of the high-wage and low-wage
households, drawn thicker on each household's own parcel, and the solid black
curve is \(p^\ast\).  The vertical lines mark the parcel boundaries.
}
\label{fig:heterogeneous-price}
\end{figure}

Figure \ref{fig:heterogeneous-heatmap} reports the interval expenditure
\(V_k(a,b)\) of each type on \([0,\bar r]\).  On a \(1200\times1200\) grid over
\([0,2\bar r]\), which also covers intervals extending into agricultural land,
no interval costs less than the type's income.  As in Example
\ref{ex:numerical-supporting-price}, numeraire consumption is positive at
every interval on the grid.  The minimum is attained at the
parcels of that type, between which households of the same type are
indifferent, while the parcels of the other type cost at least \(0.15\) more.
Since \(p^\ast\) is strictly convex, Lemma \ref{lem:convex-interval-reduction}
again implies that no measurable assignment does better than the best
interval.  The numerical evidence therefore indicates that, in this example,
\(p^\ast\) is also a supporting price.

\begin{figure}[H]
\centering
\includegraphics[width=\textwidth]{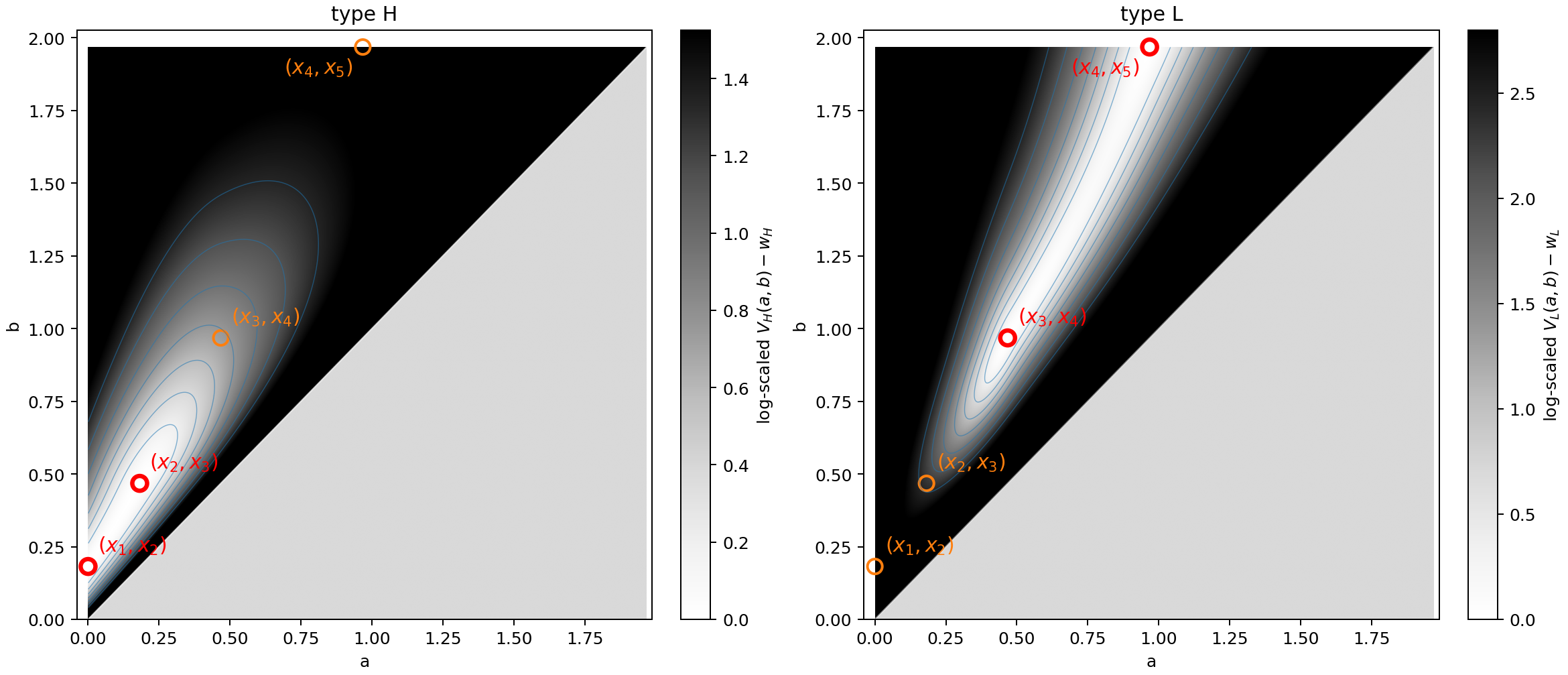}
\caption{
Interval expenditure under \(p^\ast\) for the high-wage type (left) and the
low-wage type (right).  The horizontal and vertical axes are the interval
endpoints \(a\) and \(b\).  Red circles mark the parcels of the type shown,
and orange circles mark those of the other type.  The grayscale reports a
log-scaled transformation of \(V_k(a,b)-w_k\), and the light gray triangle is
the infeasible region \(b\le a\).
}
\label{fig:heterogeneous-heatmap}
\end{figure}
\end{example}

\section{Welfare}
\label{sec:welfare}

We now state a welfare result that applies to both the homogeneous benchmark
and the heterogeneous extension.  Index households by \(i=1,\ldots,N\), and
allow household \(i\) to have utility function \(U_i\), income \(w_i\), and
marginal commuting cost \(t_i\).

Following the Herbert--Stevens formulation in
\citet[Section 3.4]{fujita1989}, welfare comparisons account for both resident
utilities and the surplus accruing to the rest of the economy, including
absentee landlords.\footnote{An alternative is a Benthamite planner who
maximizes the sum of utilities while leaving the rest of the economy at least
a given surplus. Because two
households cannot occupy the same location, this planner may give identical
households different utilities, for example higher utility to those farther
from the CBD \citepalias[Section 3]{berliant1992InternationalEconomicReview}.
Equilibrium, which treats identical households equally, need not maximize this
objective, although it is efficient in the sense of Theorem
\ref{thm:first-welfare}.}  For a feasible allocation \(a=\{z_i,f_i\}_{i=1}^N\),
define its real resource cost by
\begin{equation}\label{eq:welfare-real-resource-cost}
 C(a)
 :=\sum_{i=1}^N m\{z_i+t_i r_{c,i}\}
 +p_A\int_0^{\hat r}\sum_{i=1}^N f_i(r)\,dr,
\end{equation}
the sum of numeraire consumption, commuting costs, and the opportunity cost of
land withdrawn from agriculture.  The surplus accruing to the rest of the
economy is then \(\mathcal S(a):=m\sum_i w_i-C(a)\).

\begin{theorem}[First welfare theorem]\label{thm:first-welfare}
Let \(a^\ast\) be a Walrasian equilibrium allocation, with homogeneous or
heterogeneous households.  Then \(a^\ast\) minimizes \(C\) among feasible
allocations that give every household at least its equilibrium utility, and
any such allocation that makes at least one household strictly better off
costs strictly more than \(a^\ast\).
\end{theorem}
\begin{proof}
Let \(a^*=\{z_i^*,f_i^*\}_{i=1}^N\) be an equilibrium allocation with rent
\(p\), and set \(q(r):=p(r)-p_A\).  Then \(q\geq0\) on \([0,\hat r]\) and
\(q=0\) beyond \(\bar r\).  Suppose that a feasible allocation
\(\widetilde a\) gives every household at least its equilibrium utility.
Household optimality and strict monotonicity in the numeraire imply
\[
 m\widetilde z_i
 +\int_0^{\hat r}p(r)\widetilde f_i(r)\,dr
 +m t_i\widetilde r_{c,i}
 \geq mw_i
\]
for every \(i\), with a strict inequality whenever household \(i\) is strictly
better off.  Summing these inequalities, using \(p=p_A+q\), and using the
equilibrium budgets and land-market clearing gives
\begin{equation}\label{eq:first-welfare-summed-budgets}
 C(\widetilde a)
 +\int_0^{\hat r}q(r)\sum_i\widetilde f_i(r)\,dr
 \geq
 C(a^*)+\int_0^{\hat r}q(r)\,dr.
\end{equation}
Feasibility gives \(\sum_i\widetilde f_i(r)\leq1\), while \(q\geq0\).
Consequently, \(C(\widetilde a)\geq C(a^*)\), with a strict inequality if at
least one household is strictly better off.
\end{proof}

Thus, making at least one resident household strictly better off without
making any resident worse off necessarily reduces \(\mathcal S\).  In this
sense, the equilibrium is Pareto efficient for residents and the rest of the
economy taken together.

The proof uses only household optimality, land-market feasibility, and
\(p\ge p_A\).  It requires neither identical households nor a convex choice
set; the choice set here is not convex, because land assignments are
indicator functions.

\section{Conclusion}
\label{sec:conclusion}

This paper has studied a finite-household monocentric city in which households
choose arbitrary measurable land assignments.  Under centroid-based commuting
costs, bid rents are affine in location, and equilibrium eliminates
fragmentation: every equilibrium parcel is an interval, and parcels are
ordered by lot size from the CBD.  The enlarged choice set leaves this
structure, familiar from the interval model of
\citetalias{berliant1992InternationalEconomicReview}, intact but changes the
supporting-price problem.  Rents are confined pointwise to a band between
competing and own bid rents, the upper envelope of bid rents generally
violates budget consistency, and first-order support and budget consistency
do not guarantee household optimality.  We gave sufficient conditions for a
strictly convex rent that satisfies both and rules out all small deviations,
and we showed that equilibria are efficient whenever they exist.

Several questions remain open.  First, we have established only local
optimality; Section \ref{subsubsec:interval-local-stability} describes the
step that remains for global optimality.  Existence in the heterogeneous
case, for which we have only outlined a strategy for constructing prices, raises the same issue
together with the multidimensional anchor problem.

Second, the model does not say which supporting rent the market selects, either in its level or in its profile within parcels (Remark \ref{rem:parcel-payments-hedonic}). A theory of rent determination, for instance through bargaining between landlords and households, landlord pricing, or auctions for parcels, could clarify how both are selected. The literature on markets for indivisible goods suggests possible approaches to level selection. There, natural mechanisms implement extreme points of the set of equilibrium prices: an ascending auction approximately reaches the minimum equilibrium prices \citep{demange1986JournalofPoliticalEconomy}, and in a discrete land market with fixed lots, a Dutch auction yields the maximal equilibrium rent vector \citep{miyake2003Reg.Sci.UrbanEcon.}. Selection of the within-parcel profile raises an additional question that has no counterpart when fixed lots are traded as indivisible units.

Third, the implications of the model for empirical work remain to be worked
out.  Appendix \ref{app:hedonic-estimation} gives a preliminary discussion
of what parcel data reveal, but an econometric framework requires introducing
sources of error, which the deterministic model lacks, and characterizing
the set of structural parameters consistent with observed parcel boundaries
and payments, in the spirit of the compatibility analysis of
\citet{ekeland2010EconTheory}.  A sharp characterization would also require
global household optimality, and heterogeneous households raise further
identification problems.  The profile itself matters for rules that price
parts of parcels directly.  Partial takings, in
which the government acquires part of a parcel, are common in eminent domain
\citep{bell2017ColumbiaLawReview}.  Compensation based on location-specific
rents then depends on which profile prevails, so the selection of rents
within parcels, raised above as an open question, has direct consequences for
policy.

\appendix
\numberwithin{lemma}{section}
\numberwithin{equation}{section}
\renewcommand{\theHlemma}{\thesection.\arabic{lemma}}
\renewcommand{\theHequation}{\thesection.\arabic{equation}}
\makeatletter
\renewcommand{\@seccntformat}[1]{\@nameuse{appprefix@#1}\@nameuse{the#1}\quad}
\newcommand{\appprefix@section}{Appendix~}
\renewcommand{\@hangfrom}[1]{\noindent#1}
\makeatother

\section{Endpoint Stability}
\label{app:endpoint-stability}

\begin{lemma}[Local stability of the innermost parcel]
\label{lem:appendix-innermost-local-stability}
Let \(p\) be a strictly convex first-order supporting price for
\(\Omega^\ast\) at \(u^\ast\) whose right derivative \(p'_+(0)\) exists and is
finite, and consider the innermost parcel \(\Omega_1^\ast=[0,x_2^\ast]\). Write \(Z_1:=Z_{ss}(s_1^\ast,u^\ast)\)
and, for \(\tau\in\{-,+\}\), define
\[
    A_+:=Z_1-m p'_+(0),
    \qquad
    B_\tau:=Z_1+m p'_\tau(x_2^\ast).
\]
If \(B_->0\) and \(B_+>0\), and either
\[
    p(0)<\Psi_1^\ast(0),
    \qquad\text{or}\qquad
    p(0)=\Psi_1^\ast(0),\ \ A_+>0,\ \ A_+B_+-Z_1^2>0,
\]
then \(\Omega_1^\ast\) is a strict local minimum of the interval expenditure
problem among feasible deviations.
\end{lemma}

\begin{proof}
At the assigned interval, the derivative with respect to the left endpoint is
\[
    V_a(0,x_2^\ast;u^\ast)
    =
    \frac{\Psi_1^\ast(0)-p(0)}{m}
    =:\gamma\ge0,
\]
whereas boundary matching at \(x_2^\ast\) gives
\(V_b(0,x_2^\ast;u^\ast)=0\). For a feasible deviation
\(\Delta a\ge0\), the directional expansion is therefore
\[
\begin{aligned}
    \Delta V
    ={}&
    \gamma\Delta a+\frac{1}{2m^2}
    \left[
    A_+(\Delta a)^2
    +B_\tau(\Delta b)^2
    -2Z_1\Delta a\Delta b
    \right]
    +o(\|(\Delta a,\Delta b)\|^2),
\end{aligned}
\]
where \(\tau=\operatorname{sgn}(\Delta b)\). If \(p(0)<\Psi_1^\ast(0)\), then
\(\gamma>0\), so the linear term rules out all sufficiently small deviations with
\(\Delta a>0\); along \(\Delta a=0\), \(B_\tau>0\) makes the quadratic term
strictly positive whenever \(\Delta b\ne0\).

If instead \(p(0)=\Psi_1^\ast(0)\), then \(\gamma=0\). If \(\Delta b\le0\), the cross term is
nonnegative, so \(A_+>0\) and \(B_->0\) suffice. If \(\Delta b\ge0\), the
quadratic form is positive definite by \(A_+>0\), \(B_+>0\), and
\(A_+B_+-Z_1^2>0\). Thus the assigned interval is a strict local minimum in
either case.
\end{proof}

\begin{lemma}[Local stability of the outermost parcel]
\label{lem:appendix-outermost-local-stability}
Let \(p\) be a strictly convex first-order supporting price for
\(\Omega^\ast\) at \(u^\ast\) whose left derivative \(p'_-(\bar r)\) exists and
is finite, and consider the outermost parcel \(\Omega_N^\ast=[x_N^\ast,\bar r]\). Write \(Z_N:=Z_{ss}(s_N^\ast,u^\ast)\)
and, for \(\sigma\in\{-,+\}\), define
\[
    A_\sigma:=Z_N-m p'_\sigma(x_N^\ast),
    \qquad
    B_-:=Z_N+m p'_-(\bar r).
\]
If \(A_+>0\), \(A_->0\), \(B_->0\), and \(A_-B_--Z_N^2>0\), then
\(\Omega_N^\ast\) is a strict local minimum of the interval expenditure
problem among feasible deviations; in particular, these inequalities hold
if \(p'_-(\bar r)=0\).
\end{lemma}

\begin{proof}
At the assigned interval, first-order support gives
\[
    V_a(x_N^\ast,\bar r;u^\ast)
    =
    \frac{\Psi_N^\ast(x_N^\ast)-p(x_N^\ast)}{m}
    =0,
    \qquad
    V_b(x_N^\ast,\bar r;u^\ast)
    =
    \frac{p(\bar r)-\Psi_N^\ast(\bar r)}{m}
    =0.
\]
For a deviation \(a=x_N^\ast+\Delta a\), \(b=\bar r+\Delta b\), the
directional expansion is
\[
\begin{aligned}
    \Delta V
    ={}&
    \frac{1}{2m^2}
    \left[
    A_\sigma(\Delta a)^2
    +B_\tau(\Delta b)^2
    -2Z_N\Delta a\Delta b
    \right]
    +o(\|(\Delta a,\Delta b)\|^2),
\end{aligned}
\]
where \(\sigma=\operatorname{sgn}(\Delta a)\), \(\tau=\operatorname{sgn}(\Delta b)\),
and \(B_+:=Z_N+m p'_+(\bar r)=Z_N\) because \(p=p_A\) beyond the urban
boundary.  Consider first \(\Delta b\le0\). If
\(\Delta a\ge0\), the cross term is nonnegative, and \(A_+>0\) and
\(B_->0\) make the quadratic term strictly positive for every nonzero
feasible direction. If
\(\Delta a\le0\), set \(x=-\Delta a\ge0\) and \(y=-\Delta b\ge0\); the
quadratic form becomes \(A_-x^2+B_-y^2-2Z_Nxy\), which is positive for
every \((x,y)\ne(0,0)\) by \(A_->0\), \(B_->0\), and
\(A_-B_--Z_N^2>0\).  If \(\Delta b>0\), the household extends its parcel
into agricultural land, and the quadratic form
\(A_\sigma(\Delta a)^2+Z_N(\Delta b)^2-2Z_N\Delta a\Delta b\) is positive
definite because \(A_\sigma Z_N-Z_N^2=-mZ_Np'_\sigma(x_N^\ast)>0\), where
\(p'_\sigma(x_N^\ast)<0\) is shown below. Thus the quadratic term is positive
on every nonzero direction, proving strict local optimality.

For the final claim, the secant slope over the outermost parcel is
\(\frac{p(\bar r)-p(x_N^\ast)}{\bar r-x_N^\ast}
=-\frac{mt}{m s_N^\ast}=-\frac{t}{s_N^\ast}<0\), so convexity gives
\(p'_-(x_N^\ast)\le p'_+(x_N^\ast)\le -\frac{t}{s_N^\ast}<0\), hence
\(A_->0\) and \(A_+>0\). Moreover, \(p'_-(\bar r)=0\) gives
\(B_-=Z_N>0\), and
\(A_-B_--Z_N^2=-mZ_Np'_-(x_N^\ast)>0\).
Thus all the sufficient conditions above are satisfied.
\end{proof}

\section{Auxiliary Lemmas for Supporting Prices}
\label{app:auxiliary-lemmas}

The following two lemmas supply the price schedules used in Theorem
\ref{thm:supporting-price-existence} to reconcile household budgets with local
stability. Lemma \ref{lem:appendix-upper-envelope-approximations} provides
prices arbitrarily close to the upper envelope, allowing rent payments to
remain near their upper bound. Lemma
\ref{lem:appendix-midpoint-approximations} provides prices arbitrarily close
to the midpoint schedule, allowing larger rent discounts. The upper-envelope approximations may leave rents too high for some budgets, while the midpoint approximations may leave them too low. For the payments between these two benchmarks considered in the theorem, the two lemmas together allow each rent payment to be matched to its budget while preserving first-order support and local stability.

\begin{lemma}[Admissible slopes and upper-envelope approximations]
\label{lem:appendix-upper-envelope-approximations}
Under Assumption \ref{ass:mrs-convexity}, write \(b_i^\ast:=t/s_i^\ast\) and
\(Z_i:=Z_{ss}(s_i^\ast,u^\ast)>0\). Then there exist numbers
\(c_1,\ldots,c_{N+1}\) satisfying
\[
    c_1<-b_1^\ast<c_2<-b_2^\ast<\cdots<c_N<-b_N^\ast<c_{N+1}<0
\]
and
\begin{equation}\label{eq:joint-slope-soc}
    Z_i(c_{i+1}-c_i)-m c_i c_{i+1}>0,
    \qquad i=1,\ldots,N.
\end{equation}
Moreover, for any such slope vector, there is a sequence of functions
\(\overline p_n:[0,\bar r]\to\mathbb R\) with the following properties:
\begin{enumerate}[label=(\roman*)]
    \item \(\overline p_n\) is \(C^1\) on \([0,\bar r]\), and its restriction
    \(\overline h_{i,n}:=\overline p_n|_{\Omega_i^\ast}\) is \(C^2\) and
    strictly convex;
    \item \(\overline h_{i,n}(x_i^\ast)=\Psi_i^\ast(x_i^\ast)\),
    \(\overline h_{i,n}(x_{i+1}^\ast)=\Psi_i^\ast(x_{i+1}^\ast)\), and
    \[
        \overline h_{i,n}'(x_i^\ast)=c_i,
        \qquad
        \overline h_{i,n}'(x_{i+1}^\ast)=c_{i+1};
    \]
    \item the full supporting-price bounds hold:
    \[
        \Psi_{-i}^\ast(r)
        \le \overline h_{i,n}(r)
        \le \Psi_i^\ast(r),
        \qquad r\in\Omega_i^\ast;
    \]
    \item all directional second-order conditions for household \(i\) hold
    strictly, including those involving derivatives from adjacent parcels;
    and
    \[
        \max_{r\in\Omega_i^\ast}
        \{\Psi_i^\ast(r)-\overline h_{i,n}(r)\}
        \longrightarrow0
        \quad\text{for every }i.
    \]
\end{enumerate}
Consequently, for every parcel,
\[
    \inf_{\overline p}
    \int_{\Omega_i^\ast}
    \{\Psi_i^\ast(r)-\overline p(r)\}\,dr
    =0,
    \qquad i=1,\ldots,N,
\]
where the infimum is taken over the upper-envelope approximation class above
and is not attained.
\end{lemma}

\begin{proof}
We first construct the common boundary slopes. Convexity of
\(g(s):=MRS(s,u^\ast)\), together with \(Z_i=-g'(s_i^\ast)\) and the adjacent
boundary equation, gives
\(mt=g(s_i^\ast)-g(s_{i+1}^\ast)=\int_{s_i^\ast}^{s_{i+1}^\ast}\{-g'(s)\}\,ds
\le Z_i(s_{i+1}^\ast-s_i^\ast)\).  Since \(b_i^\ast=t/s_i^\ast\), this is
equivalent to \(Z_i(b_i^\ast-b_{i+1}^\ast)-m b_i^\ast b_{i+1}^\ast\ge0\) for
\(i=1,\ldots,N-1\).  For \(v<0\), define \(\phi_i(v):=v/\{1-(m/Z_i)v\}\).  The
function \(\phi_i\) is strictly increasing and satisfies \(\phi_i(v)>v\), and
the preceding inequality says \(\phi_i(-b_i^\ast)\le-b_{i+1}^\ast\).  Choose
\(c_1<-b_1^\ast\).  Having chosen \(c_i<-b_i^\ast\), choose \(c_{i+1}\) with
\(\max\{-b_i^\ast,\phi_i(c_i)\}<c_{i+1}<-b_{i+1}^\ast\) for
\(i=1,\ldots,N-1\), which is possible because
\(\phi_i(c_i)<\phi_i(-b_i^\ast)\le-b_{i+1}^\ast\).  Finally choose
\(\max\{-b_N^\ast,\phi_N(c_N)\}<c_{N+1}<0\).  These choices give the stated
ordering, and \(c_{i+1}>\phi_i(c_i)\) is equivalent to
\eqref{eq:joint-slope-soc}.

Now fix any slope vector satisfying these conditions. For each \(i\), set
\(a=x_i^\ast\), \(d=x_{i+1}^\ast\), \(L_i=d-a\),
\(\alpha_i:=-b_i^\ast-c_i>0\), and \(\delta_i:=c_{i+1}+b_i^\ast>0\).  For
sufficiently large integers \(n\), let
\(k_{i,n}:=(\delta_i/\alpha_i)(n+1)-1\) and \(y:=(r-a)/L_i\), and prescribe
the derivative of \(\overline h_{i,n}\) by
\[
    \overline h_{i,n}'(r)
    :=
    -b_i^\ast
    -\alpha_i(1-y)^n
    +\delta_i y^{k_{i,n}}.
\]
For large \(n\), \(k_{i,n}>1\), and
\(\overline h_{i,n}''(r)=L_i^{-1}\{\alpha_i n(1-y)^{n-1}
+\delta_i k_{i,n}y^{k_{i,n}-1}\}>0\), so \(\overline h_{i,n}'\) is continuous
and strictly increasing, with \(\overline h_{i,n}'(a)=-b_i^\ast-\alpha_i=c_i\)
and \(\overline h_{i,n}'(d)=-b_i^\ast+\delta_i=c_{i+1}\).  Define
\(\overline h_{i,n}(r):=\Psi_i^\ast(a)+\int_a^r\overline h_{i,n}'(q)\,dq\).
The definition of \(k_{i,n}\) gives
\(\int_a^d\overline h_{i,n}'(r)\,dr
=L_i\{-b_i^\ast-\alpha_i/(n+1)+\delta_i/(k_{i,n}+1)\}=-L_ib_i^\ast\), so
\(\overline h_{i,n}\) matches \(\Psi_i^\ast\) at both endpoints.  Because
adjacent bid rents agree at each \(x_i^\ast\) and the adjacent restrictions
have the same derivative \(c_i\), the restrictions \(\overline h_{i,n}\) paste
together to form a globally \(C^1\) function \(\overline p_n\).

Moreover, \(\overline h_{i,n}\) is strictly convex and agrees with the affine
function \(\Psi_i^\ast\) at both endpoints, so it lies below the chord joining
those endpoint values, and \(\Psi_i^\ast(r)-\overline h_{i,n}(r)\ge0\).  Direct
integration gives
\(\Psi_i^\ast(r)-\overline h_{i,n}(r)=\{L_i\alpha_i/(n+1)\}
\{1-(1-y)^{n+1}-y^{k_{i,n}+1}\}\).  Since \(k_{i,n}\ge0\) and \(0\le y\le1\),
this gap is at most \(L_i\alpha_i/(n+1)\).  Together, these bounds imply that
\(\overline h_{i,n}\to\Psi_i^\ast\) uniformly on \(\Omega_i^\ast\).

It remains to verify the lower envelope. For \(1<i<N\),
\(\Psi_{-i}^\ast\) is the maximum of the affine functions
\(\Psi_{i-1}^\ast\) and \(\Psi_{i+1}^\ast\).  At the left endpoint,
\(\overline h_{i,n}=\Psi_{i-1}^\ast\) and
\(\overline h_{i,n}'\ge c_i>-b_{i-1}^\ast=(\Psi_{i-1}^\ast)'\).  At the right
endpoint, \(\overline h_{i,n}=\Psi_{i+1}^\ast\) and
\(\overline h_{i,n}'\le c_{i+1}<-b_{i+1}^\ast=(\Psi_{i+1}^\ast)'\).  These two
derivative inequalities imply \(\overline h_{i,n}\ge\Psi_{i-1}^\ast\) and
\(\overline h_{i,n}\ge\Psi_{i+1}^\ast\) throughout the parcel.  The same
argument applies to \(i=1\) using the right-hand competitor \(\Psi_2^\ast\),
and to \(i=N\) using \(\Psi_{N-1}^\ast\) at the left endpoint and the constant
agricultural rent at the fringe; the latter comparison uses \(c_{N+1}<0\).
Therefore \(\Psi_{-i}^\ast\le \overline h_{i,n}\) for every \(i\).

Finally, global \(C^1\) matching means that both one-sided derivatives of
\(\overline p_n\) at \(x_i^\ast\) equal \(c_i\).  For household \(i\), all
directional second-order matrices therefore coincide and have diagonal entries
\(A_i=Z_i-mc_i>0\) and \(B_i=Z_i+mc_{i+1}\).  Their determinant is
\(A_iB_i-Z_i^2=m\{Z_i(c_{i+1}-c_i)-mc_ic_{i+1}\}>0\) by
\eqref{eq:joint-slope-soc}.  In particular \(B_i>0\), and the matrix is
positive definite.  This verifies all coupled directional SOCs, including the
appropriate one-sided conditions for the boundary parcels.

The uniform convergence proved above implies convergence of the integral
gap to zero on every parcel. If the value zero were attained for any parcel
\(\Omega_i^\ast\), the nonnegative continuous gap would imply
\(\overline p=\Psi_i^\ast\) throughout that parcel, contradicting the required
strict convexity of the restriction. Hence the infimum is not attained for
any parcel.
\end{proof}

\begin{lemma}[Midpoint approximations with common boundary slopes]
\label{lem:appendix-midpoint-approximations}
Maintain the notation of Lemma
\ref{lem:appendix-upper-envelope-approximations}, and suppose that Assumption
\ref{ass:midpoint-curvature} holds at \(u^\ast\). Set
\[
    c_i:=-\frac{b_{i-1}^\ast+b_i^\ast}{2},
    \quad i=2,\ldots,N,
    \qquad
    c_{N+1}:=-\frac{b_N^\ast}{2}.
\]
Then there exists \(c_1<-b_1^\ast\) such that \((c_1,\ldots,c_{N+1})\)
satisfies the slope ordering of Lemma
\ref{lem:appendix-upper-envelope-approximations} and
\eqref{eq:joint-slope-soc}.
Moreover, fix any such \(c_1\) and define
\[
    q_i(r)
    :=
    \frac12\{\Psi_i^\ast(r)+\Psi_{-i}^\ast(r)\},
    \qquad r\in\Omega_i^\ast.
\]
Then there is a globally \(C^1\) sequence
\(\underline p_n:[0,\bar r]\to\mathbb R\), with restrictions
\(\underline h_{i,n}:=\underline p_n|_{\Omega_i^\ast}\), such that:
\begin{enumerate}[label=(\roman*)]
    \item each \(\underline h_{i,n}\) is \(C^2\) and strictly convex and has
    endpoint derivatives \(c_i,c_{i+1}\); it agrees with \(q_i\) at both
    endpoints when \(i\ge2\), while
    \(\underline h_{1,n}(x_2^\ast)=q_1(x_2^\ast)\);
    \item
    \[
        \Psi_{-i}^\ast(r)
        \le\underline h_{i,n}(r)
        \le\Psi_i^\ast(r),
        \qquad r\in\Omega_i^\ast,
    \]
    all coupled directional SOCs hold strictly, and
    \(\underline h_{1,n}(0)<\Psi_1^\ast(0)\);
    \item
    \[
        \max_{r\in\Omega_i^\ast}
        |\underline h_{i,n}(r)-q_i(r)|
        \longrightarrow0
        \quad\text{for every }i.
    \]
\end{enumerate}
In particular, for every parcel,
\[
    \inf_{\underline p}
    \int_{\Omega_i^\ast}\underline p(r)\,dr
    =
    \int_{\Omega_i^\ast}q_i(r)\,dr,
\]
where the infimum is taken over the midpoint-approximation class above and is
not attained.
\end{lemma}

\begin{proof}
We first verify the conditions on the slope vector. Since
\(b_1^\ast>b_2^\ast>\cdots>b_N^\ast>0\), each midpoint slope lies strictly
between the adjacent bid-rent slopes:
\(-b_{i-1}^\ast<c_i<-b_i^\ast\) for \(i=2,\ldots,N\), and
\(-b_N^\ast<c_{N+1}<0\). For \(i=2,\ldots,N-1\), direct substitution of
\(b_j^\ast=t/s_j^\ast\) shows that
\(Z_i(c_{i+1}-c_i)-mc_ic_{i+1}>0\)
is exactly the corresponding interior inequality in Assumption
\ref{ass:midpoint-curvature}; for \(i=N\), it is exactly the last inequality
in that assumption. For \(i=1\), write
\(Z_1(c_2-c_1)-mc_1c_2=Z_1c_2-c_1(Z_1+mc_2)\).  The first inequality in the
assumption is equivalent to \(Z_1+mc_2>0\), so the right-hand side is
positive for every sufficiently negative \(c_1\).  Choosing such a
\(c_1<-b_1^\ast\) gives the slope ordering and \eqref{eq:joint-slope-soc}.

Now fix such a \(c_1\). For every \(i\ge2\), \(q_i\) is the
V-shaped maximum of the two affine supporting lines at the endpoints of
\(\Omega_i^\ast\), with slopes \(c_i\) and \(c_{i+1}\). We smooth the
corner by replacing the derivative jump with a normalized Gaussian transition,
which preserves the endpoint slopes and, being strictly increasing, gives
strict convexity. Specifically, fix \(a=x_i^\ast\), \(d=x_{i+1}^\ast\) and, for
\(\xi\in\mathbb R\), define
\[
    F_{n,\xi}(r)
    :=
    \frac{\displaystyle\int_a^r e^{-n(z-\xi)^2}\,dz}
         {\displaystyle\int_a^d e^{-n(z-\xi)^2}\,dz}.
\]
This function is smooth and strictly increasing on \([a,d]\), with endpoint
values \(0\) and \(1\); the parameter \(n\) controls the concentration of the
transition, and \(\xi\) its location. Because \(-b_i^\ast\in(c_i,c_{i+1})\),
continuity in \(\xi\) allows us to choose \(\xi_{i,n}\) so that
\(\int_a^d\{c_i+(c_{i+1}-c_i)F_{n,\xi_{i,n}}(r)\}\,dr=-(d-a)b_i^\ast\), which
will ensure that the endpoint values match.  Define
\(\underline h_{i,n}(r):=q_i(a)+\int_a^r\{c_i+(c_{i+1}-c_i)
F_{n,\xi_{i,n}}(z)\}\,dz\).  Then \(\underline h_{i,n}\) is \(C^2\) and
strictly convex, has endpoint derivatives \(c_i,c_{i+1}\), and agrees with
\(q_i\), hence also with \(\Psi_i^\ast\), at both endpoints.  As
\(n\to\infty\), the transition concentrates, and
\(\underline h_{i,n}\to q_i\) uniformly on \(\Omega_i^\ast\).  Moreover,
convexity places \(\underline h_{i,n}\) above both of its endpoint supporting
lines, and hence above \(q_i\).  Since it is convex and has the same endpoint
values as the affine function \(\Psi_i^\ast\), it also lies below
\(\Psi_i^\ast\).  Thus
\(\Psi_{-i}^\ast\le q_i\le\underline h_{i,n}\le\Psi_i^\ast\) for \(i\ge2\).

The first parcel requires a different treatment because there is no
competing household on its inner side: \(\Psi_{-1}^\ast=\Psi_2^\ast\), so
\(q_1\) is affine with constant slope
\(c_2=-(b_1^\ast+b_2^\ast)/2\), rather than having an interior corner to
smooth. The approximation must nevertheless have the prescribed left-end
slope \(c_1<c_2\). A strictly convex function matching \(q_1\) in value and
slope at \(x_2^\ast\) lies strictly above this tangent line to its left, so
it cannot also match \(q_1(0)\). Since \(0\) is not a shared parcel boundary,
we leave that endpoint value free and match only the value at
\(x_2^\ast\).

To construct the approximation, take \(a=0\), \(d=x_2^\ast\), and \(\xi=0\)
in the smoothing function above, and define \(\underline h_{1,n}\) by
\(\underline h_{1,n}'(r):=c_1+(c_2-c_1)F_{n,0}(r)\) and
\(\underline h_{1,n}(d)=q_1(d)\).  Since \(F_{n,0}(0)=0\) and
\(F_{n,0}(d)=1\), the derivative increases strictly from \(c_1\) to \(c_2\).
Moreover, \(F_{n,0}(r)\to1\) for every fixed \(r>0\), so the change in slope
becomes concentrated near \(0\).  The derivative therefore converges in
\(L^1\) to the constant \(c_2\), and the common value at \(d\) implies that
\(\underline h_{1,n}\) converges uniformly to \(q_1\).

It remains to verify the bid-rent bounds.  For the lower bound, use the common value at \(d\) and \(\underline h_{1,n}'\le c_2=q_1'\)
to obtain
\(\underline h_{1,n}(r)-q_1(r)=\int_r^d\{c_2-\underline h_{1,n}'(z)\}\,dz
\ge0\).  For the upper bound, the two affine schedules satisfy
\(\Psi_1^\ast(r)-q_1(r)=(c_2+b_1^\ast)(d-r)\), where \(c_2+b_1^\ast>0\).  On
\([0,d/2]\), this gap is bounded below by \((c_2+b_1^\ast)d/2>0\), so uniform
convergence to \(q_1\) gives \(\underline h_{1,n}<\Psi_1^\ast\) there for all
sufficiently large \(n\).  Near \(d\), the gap vanishes, so uniform
convergence alone does not establish the upper bound.  Instead, since
\(\underline h_{1,n}'\) is increasing and
\(\underline h_{1,n}'(d/2)\to c_2>-b_1^\ast\), for all sufficiently large
\(n\) we have \(\underline h_{1,n}'(r)>-b_1^\ast\) throughout \([d/2,d]\).
The common value at \(d\) then gives
\(\Psi_1^\ast(r)-\underline h_{1,n}(r)=\int_r^d\{\underline h_{1,n}'(z)
+b_1^\ast\}\,dz\ge0\) for \(r\in[d/2,d]\).  Combining the two parts of the
interval, and recalling \(q_1\ge\Psi_{-1}^\ast\), yields, for all
sufficiently large \(n\),
\(\Psi_{-1}^\ast\le\underline h_{1,n}\le\Psi_1^\ast\) and
\(\underline h_{1,n}(0)<\Psi_1^\ast(0)\).

Relabeling a sufficiently far tail of each sequence, the restrictions paste
together because their values and derivatives agree at every shared boundary.
They therefore define a globally \(C^1\) sequence \(\underline p_n\).

Finally, the endpoint slopes of \(\underline p_n\) are the same \(c_i\)'s as
those of the upper-envelope sequence in Lemma
\ref{lem:appendix-upper-envelope-approximations}. The
positive-definiteness calculation in that lemma therefore applies without
change and verifies all coupled directional SOCs for \(\underline p_n\).

Uniform convergence and \(\underline h_{i,n}\ge q_i\) imply the stated
parcelwise infimum. If it were attained, the nonnegative continuous gap would
vanish throughout the parcel, so the restriction would equal the piecewise
affine function \(q_i\), contradicting its required strict convexity. Hence the
infimum is not attained.
\end{proof}

\section{Implications for Hedonic Estimation}
\label{app:hedonic-estimation}

This appendix discusses what parcel transaction data generated by the
homogeneous-household model reveal.  Suppose that the econometrician observes
the equilibrium parcel boundaries
$0=x_1^\ast<x_2^\ast<\cdots<x_{N+1}^\ast=\bar r$, and hence the lot sizes
$s_i^\ast$ and centroids $r_{c,i}^\ast$, together with the parcel payments
$\Pi_i$ of Remark \ref{rem:parcel-payments-hedonic} and the values of $w$,
$m$, and $p_A$.  Throughout, $u$ denotes the realized
common utility level, and we abstract from the mapping between capitalized
sale prices and the flow rents in the model.

Two empirical objectives can be distinguished.  The first is to recover
$Z(\cdot,u)$ and the commuting-cost parameter $t$, which determine the bid
function.  The second is to recover or bound the market prices of parcels,
including those that are not traded.  The bid and market price functions
coincide at traded parcels but need not coincide elsewhere, so recovering
the former does not by itself determine the latter.
Section \ref{app:estimating-bid} derives restrictions on $Z(\cdot,u)$ and
$t$ without selecting a within-parcel rent profile.  The existence of a
profile that supports the observed allocation nevertheless restricts these
objects.  Section \ref{app:estimating-price} then distinguishes the prices
and spatial gradients determined by observed payments and equilibrium
restrictions from those that depend on the remaining freedom in the
profile.

\subsection{Estimating the Bid Function}
\label{app:estimating-bid}

\paragraph{Equality restrictions.}
Budget consistency gives
\begin{equation}\label{eq:app-value}
    Z(s_i^\ast,u)=w-\frac{\Pi_i}{m}-t\,r_{c,i}^\ast,
    \qquad i=1,\ldots,N,
\end{equation}
and Lemma \ref{lem:fixed-utility-lot-sizes-dominance-regions} gives
\begin{equation}\label{eq:app-slope}
    MRS(s_i^\ast,u)=p_A+\Bigl(N-i+\frac12\Bigr)mt,
    \qquad i=1,\ldots,N.
\end{equation}
Neither restriction involves the profile of rents within parcels.  Since
$Z_s=-MRS$, they determine, for a given $t$, the values and the slopes of
$Z(\cdot,u)$ at the $N$ observed lot sizes.  The bid function can therefore
be estimated without recovering location-specific rents.\footnote{Condition
\eqref{eq:app-slope} plays the role of the boundary-indifference
conditions in models with ordered alternatives, such as the rent equation of
\citet{kaneko2006JournalofUrbanEconomics}, which
\citet{kaneko2017SpatialInteractionModels} bring to rental-housing data, and
the sorting model of \citet{epple1999JournalofPoliticalEconomy}.  Because
\eqref{eq:app-value} and \eqref{eq:app-slope} give the values and the
derivatives of $Z(\cdot,u)$ at the same points, the estimation problem also
resembles that of a cost function observed together with its derivatives
\citep{hall2007Ann.Statist.}.}

\paragraph{The commuting cost.}
Payments alone do not identify $t$.  Because each lot size is traded at a
single centroid, payments do not separate the effect of size from that of
location: for any $t'$, replacing $Z(s_i^\ast,u)$ by
$Z(s_i^\ast,u)+(t-t')\,r_{c,i}^\ast$ preserves
\eqref{eq:app-value}.\footnote{\citet{colwell1997JournalofUrbanEconomics}
make the empirical counterpart of this point and separate the two effects
using variation in lot size at a given distance, which equilibrium parcels
here lack.}  The slope restriction \eqref{eq:app-slope} and the
shape of $Z$ restrict $t$.  Because $MRS(\cdot,u)$ is decreasing, $Z(\cdot,u)$
is convex, so the secant slope of $Z(\cdot,u)$ between adjacent lot sizes lies
between its slopes at the two endpoints.  Since
$r_{c,i+1}^\ast-r_{c,i}^\ast=m(s_i^\ast+s_{i+1}^\ast)/2$, conditions
\eqref{eq:app-value} and \eqref{eq:app-slope} then require
\begin{equation}\label{eq:app-secant}
    p_A+\Bigl(N-i-\frac12\Bigr)mt
    \le
    \frac{\Pi_{i+1}-\Pi_i}{m(s_{i+1}^\ast-s_i^\ast)}
    +\frac{m(s_i^\ast+s_{i+1}^\ast)}{2(s_{i+1}^\ast-s_i^\ast)}\,t
    \le
    p_A+\Bigl(N-i+\frac12\Bigr)mt
\end{equation}
for $i=1,\ldots,N-1$.  Each inequality is linear in $t$, so the values of $t$
that satisfy \eqref{eq:app-secant} form an interval.  Under a parametric
specification of $Z$ and a rank condition, the $2N$ restrictions
\eqref{eq:app-value} and \eqref{eq:app-slope} can point-identify $t$.

\paragraph{Existence of a supporting price.}
Although the rent profile does not enter \eqref{eq:app-value} and
\eqref{eq:app-slope}, its existence restricts the parameters.  By
\eqref{eq:first-order-budget-bounds} and the discussion that follows it, a
first-order supporting and budget-consistent price exists at $u$ if and only
if $E_i(u)-T_i(u)\le w\le E_i(u)$ for all $i$.  By \eqref{eq:bid_rent} and
\eqref{eq:app-slope}, the bid-rent functions are determined by $t$ and the
observed parcels,
\[
    \Psi_j^\ast(r,u)
    =p_A+\Bigl(N-j+\frac12\Bigr)mt-\frac{t}{s_j^\ast}\,(r-r_{c,j}^\ast),
    \qquad j=1,\ldots,N,
\]
and by \eqref{eq:app-value}, $E_i(u)-w=s_i^\ast MRS(s_i^\ast,u)-\Pi_i/m$.
The bounds therefore become
\begin{equation}\label{eq:app-band}
    \int_{\Omega_i^\ast}\Psi_{-i}^\ast(r,u)\,dr
    \;\le\;\Pi_i\;\le\;
    \int_{\Omega_i^\ast}\Psi_i^\ast(r,u)\,dr
    =ms_i^\ast\Bigl\{p_A+\Bigl(N-i+\frac12\Bigr)mt\Bigr\},
    \qquad i=1,\ldots,N,
\end{equation}
which depend only on $t$ and the data, not on $Z$.  The right inequality
states that the average rent on each parcel does not exceed the occupant's
marginal rate of substitution, and it is equivalent to
\[
    t\ \ge\ \frac{\Pi_i/(ms_i^\ast)-p_A}{(N-i+1/2)\,m}.
\]
Since $\Psi_j^\ast(r,u)$ is linear in $t$ for each $r$, the integral of
$\Psi_{-i}^\ast=\max_{j\ne i}\Psi_j^\ast$ is convex in $t$, so the left
inequality also defines an interval of $t$.  Local support further
restricts the slopes of the rent at the parcel boundaries through
second-order conditions, for which Proposition
\ref{prop:local-interval-stability} gives sufficient conditions, and global
optimality adds further restrictions.  The values of $t$ that satisfy
\eqref{eq:app-secant} and \eqref{eq:app-band} therefore form an outer set for
the commuting cost.

\paragraph{Estimators based on the continuum condition.}
Structural applications of the monocentric model equate the rent with the bid
rent at every location.  Applied to parcel data, this treats the average rent
$\Pi_i/(ms_i^\ast)$ as $MRS(s_i^\ast,u)$, so that the right inequality in
\eqref{eq:app-band} binds on every parcel, or equivalently $E_i(u)=w$ for all
$i$.  This is the condition of Corollary
\ref{cor:upper-envelope-trapezoid-exact}, and by Lemma
\ref{lem:upper-envelope-expenditure-monotonicity} it fails whenever
$MRS(\cdot,u)$ is strictly convex.  An estimator that imposes the continuum
condition therefore either violates \eqref{eq:app-slope}, so that the
observed parcels are not the dominance regions of the estimated bid
functions, or, under Assumption \ref{ass:mrs-convexity}, requires
$MRS(\cdot,u)$ to be affine between the observed lot sizes.  In practice,
structural applications take $t$ from external data on the monetary and time
costs of commuting and add to the continuum condition an error term,
interpreted as unobserved amenities.  \citet{caruso2015DatainBrief} do so
with land rents and lot sizes, and
\citet{liotta2022RegionalScienceandUrbanEconomics} use the same approach for
housing rents per unit of floor space, which would require a model of housing
production to relate to the present one.  If the present model generates the
data, the average rent on parcel $i$ falls short of $MRS(s_i^\ast,u)$ by
$(E_i(u)-w)/s_i^\ast$, and this gap enters the residual.  Since the gap
depends on the size and position of the parcel, the residual need not
satisfy the usual orthogonality conditions with respect to regressors such
as distance from the CBD.

\paragraph{Dependence on the model.}
The data are affected by the two differences from the interval model of
\citetalias{berliant1992InternationalEconomicReview} in different ways.  The first
is where commuting cost is measured, and it affects the equality restrictions
\eqref{eq:app-value} and \eqref{eq:app-slope}.  By Lemma
\ref{lem:bf-correspondence}, on intervals the present model coincides with the
front-based model under the rent $p+mt/2$, except at the urban fringe.  Under
front-based commuting, \eqref{eq:app-slope} becomes
$MRS(s_i^\ast,u)=p_A+(N-i)mt$, and the commuting cost in \eqref{eq:app-value}
becomes $t\,x_i^\ast$.  The second is the choice set, and it affects the
inequality restrictions that come from the existence of a supporting price,
such as \eqref{eq:app-band}.  In the interval model, a supporting price needs
to deter only interval deviations, and the pointwise bid-rent bounds of Lemma
\ref{lem:foc-support-bid-rent-bounds} need not hold.  Holding the commuting
specification fixed, a supporting price here must also deter deviations to
arbitrary measurable sets, so the set of supporting prices is a subset of
that in the interval model, and the set of parameters consistent with the
data is weakly smaller: the inequality restrictions are more informative, but
are also more likely to be rejected.

\paragraph{Limitations.}
Because all households attain the same utility level, the data reveal
$Z(\cdot,u)$ only at the realized $u$.  Counterfactual equilibria at other
utility levels require a functional-form assumption or data from several
markets.  With heterogeneous households, parcel prices trace the upper
envelope of the bid functions of different types rather than a common bid
function, so recovering each type's preferences requires additional
identifying structure.  One approach is the second step of the procedure of
\citet{rosen1974JournalofPoliticalEconomy}, in which the implicit prices
estimated from the price function are regressed on the characteristics and on
buyer attributes to recover marginal bid functions; it faces the
identification problems discussed by \citet{brown1982Econometrica},
\citet{epple1987JournalofPoliticalEconomy}, and
\citet{ekeland2004JournalofPoliticalEconomy}.  Equilibrium sorting models
such as that of \citet{epple1999JournalofPoliticalEconomy} provide another one.

\subsection{Estimating the Market Price Function}
\label{app:estimating-price}

The market price function is of interest for valuing specified parcels and
studying how land prices vary with location and size.  Some of these
objectives concern only existing parcels and their unions, such as the
acquisition prices examined by \citet{brooks2016AmericanEconomicJournal:EconomicPolicy}
in their study of land assembly.  Such prices need not distinguish among
within-parcel rent profiles that generate the same parcel payments: under
additive pricing, the price of a union of existing parcels is the sum of
their payments.

Restrictions on the within-parcel profile become relevant when the parcels of
interest include parts of existing parcels.  Land-value surfaces, such as
those estimated by \citet{mcmillen1996JournalofUrbanEconomics} and used by
\citet{ahlfeldt2018TheReviewofEconomicsandStatistics} to study building
heights and construction technology, assign values to lots of given
dimensions at arbitrary locations, and such lots generally straddle parcel
boundaries.  The literature on land price and lot size, which asks how the
price of a lot varies with its size at a given location, concerns such
parcels as well, and explains the curvature of this relation by the costs of
subdivision and assembly
\citep{brownstone1991TheReviewofEconomicsandStatistics,colwell1993TheReviewofEconomicsandStatistics}.
Partial takings compensated at location-specific rents are
another example (Section \ref{sec:conclusion}).  Subdivision policies affect
which parts of parcels households can acquire:
\citet{white1988JournalofUrbanEconomics} studies minimum lot sizes and
subdivision costs, while \citet{wrenn2015RegionalScienceandUrbanEconomics}
examine regulatory delays in subdivision development.  These studies do not
themselves impose within-parcel supporting-price conditions.  In an extension
of the present model, such policies would change the deviations that
supporting rents must deter, and hence the set of admissible profiles, even
when existing parcel payments are unchanged.  Such an analysis
need not recover a unique profile; it may instead use the restrictions common
to all supporting profiles.

This subsection distinguishes the price information determined by observed
parcel payments and equilibrium restrictions from that which depends on the
within-parcel profile.  Parcel payments identify cumulative rent at parcel
boundaries, and the bid-rent bounds determine rents at those boundaries and
restrict rents between them.  Recovering a complete rent surface requires a
choice of interpolation, but some prices and gradients can be determined or
bounded without selecting a particular profile.  In particular, land-value
gradients measured over windows that span many parcels are nearly
independent of the profile.

Let $F(x):=\int_0^x p(r)\,dr$.  The price of an interval $[a,b]$ is
$P_I=F(b)-F(a)$, the function introduced in Remark
\ref{rem:parcel-payments-hedonic}.  When households may choose measurable
parcels, the price relevant to their choice of size and centroid is the
minimum of $\int_A p(r)\,dr$ over sets $A$ with the given measure and
centroid, which is at most $P_I$.  We focus on the price of a given interval,
which is the object that an appraisal of a lot of given dimensions reports.

\paragraph{What the data determine.}
Since the equilibrium parcels partition $[0,\bar r]$,
$F(x_i^\ast)=\sum_{j<i}\Pi_j$ for every $i$.  The cumulative rent is
therefore identified at the parcel boundaries, and so is the price of any
interval whose endpoints are parcel boundaries.

\paragraph{A restriction on the price function.}
In terms of size and centroid,
$P_I(S,r_c)=F(r_c+S/2)-F(r_c-S/2)$, so wherever $p$ is differentiable at the
endpoints,
\begin{equation}\label{eq:app-wave}
    \frac{\partial^2P_I}{\partial r_c^2}
    =p'(b)-p'(a)
    =4\,\frac{\partial^2P_I}{\partial S^2}.
\end{equation}
An estimated price function that is flexible in size and location must
satisfy \eqref{eq:app-wave} if parcel prices are additive over locations; for
a quadratic in $(S,r_c)$, the coefficient on $r_c^2$ must be four times that
on $S^2$.  Moreover, at a fixed centroid, the curvature of the price in lot
size reflects only the curvature of the rent profile,
$\partial^2P_I/\partial S^2=\{p'(b)-p'(a)\}/4$.  At an interior traded parcel
at which $p$ is differentiable at both endpoints, local support makes
$P_I-R$ minimal, and since $R$ is linear in $r_c$,
$\partial^2P_I/\partial r_c^2\ge0$.  By \eqref{eq:app-wave},
$\partial^2P_I/\partial S^2\ge0$ there as well.  If the rent is convex, as
are the rents constructed in Theorem \ref{thm:supporting-price-existence},
the same holds at every interval: at a fixed centroid,
$\partial P_I/\partial S=\{p(b)+p(a)\}/2$ is nondecreasing in $S$, so the
price of an interval is convex in its size.  The bid function, by contrast,
satisfies $\partial^2R/\partial S^2=-Z_{ss}(s,u)/m<0$.  Additive pricing thus
provides a benchmark for the literature on land price and lot size, which
has widely found land prices to be concave in parcel size
\citep[see][and the references therein]{colwell1999TheJournalofRealEstateFinanceandEconomics}.
Concavity is weaker, with some evidence of convexity, within the central
business district \citep{colwell1999TheJournalofRealEstateFinanceandEconomics},
and unit prices rise with plot size among small and otherwise similar plots
\citep{lin2000LandEconomics}.  Comparing these findings with the model
requires specifying which function a regression recovers: payments for
traded parcels lie on the bid function, which is concave in size, whereas
the prices of untraded intervals with a given centroid are convex in size
when the rent is convex, and transactions in a single city do not distinguish the extensions of
the two functions beyond the traded parcels.

\paragraph{Rents at and near parcel boundaries.}
Although the data do not determine the rent profile, the bid-rent bounds
restrict it.  As shown in Remark \ref{rem:bf-prices-excluded}, the bounds close at
the interior parcel boundaries.  For a rent that satisfies them at every
location, as in the convention of footnote \ref{fn:pointwise-foc},
\eqref{eq:app-slope} therefore gives
\begin{equation}\label{eq:app-boundary-rents}
    p(x_i^\ast)=p_A+(N-i+1)mt,\qquad i=2,\ldots,N+1.
\end{equation}
The slope of $F$ at the boundaries is thus identified once $t$ is.  The bounds
also restrict the rent near the boundaries.  Consider a parcel $\Omega_k^\ast$
with $1<k<N$.  On $\Omega_k^\ast$, the rent lies below $\Psi_k^\ast$, which
decreases from $p(x_k^\ast)$, and above $\Psi_{k+1}^\ast$, which exceeds
$p(x_{k+1}^\ast)$ to the left of $x_{k+1}^\ast$.  Hence
\begin{equation}\label{eq:app-within-parcel}
    p(x_{k+1}^\ast)\le p(r)\le p(x_k^\ast),
    \qquad r\in\Omega_k^\ast,
\end{equation}
so the rent within such a parcel lies between its two boundary rents, which
differ by $mt$.  Near an interior boundary $x_{k+1}^\ast$, the same bounds
imply that the one-sided difference quotients of $p$ at $x_{k+1}^\ast$ lie in
$[-t/s_k^\ast,-t/s_{k+1}^\ast]$: every supporting rent declines through an
interior boundary at a rate between the bid-rent slopes of the two adjacent
households.

\paragraph{The profile is selected by the interpolation.}
Between the boundaries, the shape of $F$, that is, the rent profile, is
determined neither by the data nor by the model beyond the bid-rent bounds,
budget consistency, the second-order conditions, and global optimality.  Any method that
recovers location-specific values from parcel totals therefore selects a
profile.  Pycnophylactic interpolation \citep{tobler1979JournaloftheAmericanStatisticalAssociation}
and area-to-point kriging \citep{kyriakidis2004GeographicalAnalysis} preserve
the total of each area, and \citet{kyriakidis2004GeographicalAnalysis} shows
that the choropleth map, in which the value is constant within each area, is
a special case of his framework.  By the previous paragraph, a rent that is
constant within a parcel is never a supporting price, so the choropleth
completion is excluded.  A smooth mass-preserving interpolation is continuous
at the boundaries, as a supporting rent must be, but need not satisfy the
bid-rent bounds.

\paragraph{Gradients at different scales.}
Since $\partial P_I/\partial r_c=p(b)-p(a)$, the gradient of the unit price of
an interval $[a,b]$ of size $S$ with respect to its location is
$\{p(b)-p(a)\}/S$.  Suppose that $a\in[x_j^\ast,x_{j+1}^\ast)$ and
$b\in[x_k^\ast,x_{k+1}^\ast)$ with $1<j\le k<N$.  By
\eqref{eq:app-boundary-rents} and \eqref{eq:app-within-parcel},
$p(a)\in[p(x_j^\ast)-mt,\,p(x_j^\ast)]$,
$p(b)\in[p(x_k^\ast)-mt,\,p(x_k^\ast)]$, and
$p(x_k^\ast)-p(x_j^\ast)=-(k-j)mt$.  Hence, whatever the profile,
\[
    \frac{p(b)-p(a)}{S}=-\frac{(k-j)\,mt}{S}+\varepsilon,
    \qquad
    |\varepsilon|\le\frac{mt}{S}.
\]
The error vanishes when $a$ and $b$ are parcel boundaries.  Then
$S=m\sum_{l=j}^{k-1}s_l^\ast$, and the gradient equals $-t/\bar s_{jk}$,
where $\bar s_{jk}$ is the mean lot size of parcels $j,\ldots,k-1$: the
bid-rent slope at the mean lot size, which for a single parcel is the
bid-rent slope of its occupant.  In general, if $j<k$, the window width cancels
in the ratio of the error to the first term,
\[
    \frac{|\varepsilon|}{(k-j)\,mt/S}
    \le\frac{mt/S}{(k-j)\,mt/S}
    =\frac{1}{k-j},
\]
so the error is at most a fraction $1/(k-j)$ of the first term in absolute
value.  Land-value gradients over windows that span
many parcels are therefore essentially profile-free, whereas gradients
measured within a parcel ($j=k$) are determined by the selected profile.

\paragraph{Prices of nearby untraded intervals.}
Suppose that the rent is convex, so that its one-sided derivatives exist.
For an interior parcel $[x_i^\ast,x_{i+1}^\ast]$, a second-order expansion
gives
\[
    P_I(x_i^\ast+\Delta a,\,x_{i+1}^\ast+\Delta b)
    =\Pi_i+p(x_{i+1}^\ast)\Delta b-p(x_i^\ast)\Delta a
    +\frac12\bigl\{p'(x_{i+1}^\ast)\Delta b^2-p'(x_i^\ast)\Delta a^2\bigr\}
    +o(\Delta a^2+\Delta b^2),
\]
with one-sided derivatives according to the signs of $\Delta a$ and
$\Delta b$.  The first-order terms are pinned down by
\eqref{eq:app-boundary-rents}, and the derivatives in the second-order terms
are bounded by the bid-rent slopes of the adjacent households.  The price of
an untraded interval near a traded parcel is thus profile-free to first order
and bounded at second order.

\bibliographystyle{ecca}
\bibliography{references}

@article{ahlfeldt2018TheReviewofEconomicsandStatistics,
  title = {Tall {{Buildings}} and {{Land Values}}: {{Height}} and {{Construction Cost Elasticities}} in {{Chicago}}, 1870--2010},
  shorttitle = {Tall {{Buildings}} and {{Land Values}}},
  author = {Ahlfeldt, Gabriel M. and McMillen, Daniel P.},
  year = 2018,
  month = dec,
  journal = {The Review of Economics and Statistics},
  volume = {100},
  number = {5},
  pages = {861--875},
  issn = {0034-6535, 1530-9142},
}

@article{ali2019WorldBankEcon.Rev.,
  title = {Costs and {{Benefits}} of {{Land Fragmentation}}: {{Evidence}} from {{Rwanda}}},
  shorttitle = {Costs and {{Benefits}} of {{Land Fragmentation}}},
  author = {Ali, Daniel Ayalew and Deininger, Klaus and Ronchi, Loraine},
  year = 2019,
  month = oct,
  journal = {The World Bank Economic Review},
  volume = {33},
  number = {3},
  pages = {750--771},
  issn = {0258-6770, 1564-698X},
}

@article{baldwin2019ECTA,
  title = {Understanding {{Preferences}}: ``{{Demand Types}}'', and the {{Existence}} of {{Equilibrium With Indivisibilities}}},
  shorttitle = {Understanding {{Preferences}}},
  author = {Baldwin, Elizabeth and Klemperer, Paul},
  year = 2019,
  journal = {Econometrica},
  volume = {87},
  number = {3},
  pages = {867--932},
  issn = {0012-9682},
}

@article{bell2017ColumbiaLawReview,
  title = {Partial Takings},
  author = {Bell, Abraham and Parchomovsky, Gideon},
  year = 2017,
  journal = {Columbia Law Review},
  volume = {117},
  number = {8},
  pages = {2043--2094}
}

@article{berliant1985RegionalScienceandUrbanEconomics,
  title = {Equilibrium Models with Land},
  author = {Berliant, Marcus},
  year = 1985,
  month = jun,
  journal = {Regional Science and Urban Economics},
  volume = {15},
  number = {2},
  pages = {325--340},
  issn = {01660462},
}

@article{berliant2026,
  title = {Evil {{Deeds}} in {{Urban Economics}}},
  author = {Berliant, Marcus},
  year = 2026,
}

@article{berliant2004JournalofMathematicalEconomics,
  title = {A Foundation of Location Theory: Existence of Equilibrium, the Welfare Theorems, and Core},
  shorttitle = {A Foundation of Location Theory},
  author = {Berliant, Marcus and Dunz, Karl},
  year = 2004,
  month = aug,
  journal = {Journal of Mathematical Economics},
  volume = {40},
  number = {5},
  pages = {593--618},
  issn = {03044068},
}

@article{berliant1992InternationalEconomicReview,
  title = {Alonso's {{Discrete Population Model}} of {{Land Use}}: {{Efficient Allocations}} and {{Competitive Equilibria}}},
  shorttitle = {Alonso's {{Discrete Population Model}} of {{Land Use}}},
  author = {Berliant, Marcus and Fujita, Masahisa},
  year = 1992,
  month = aug,
  journal = {International Economic Review},
  volume = {33},
  number = {3},
  eprint = {2527125},
  eprinttype = {jstor},
  pages = {535},
  issn = {00206598},
}

@article{berliant2007IntEconomicReview,
  title = {Equilibrium and the {{Core}} in {{Alonso}}'s {{Discrete Population Model}} of {{Land Use}}},
  author = {Berliant, Marcus and Raa, Thijs Ten},
  year = 2007,
  month = feb,
  journal = {International Economic Review},
  volume = {48},
  number = {1},
  pages = {235--246},
  issn = {0020-6598, 1468-2354},
}

@article{berliant2008RegionalScienceandUrbanEconomics,
  title = {When Worlds Collide: {{Different}} Comparative Static Predictions of Continuous and Discrete Agent Models with Land},
  shorttitle = {When Worlds Collide},
  author = {Berliant, Marcus and Sabarwal, Tarun},
  year = 2008,
  month = sep,
  journal = {Regional Science and Urban Economics},
  volume = {38},
  number = {5},
  pages = {438--444},
  issn = {01660462},
}

@article{berliant2013RegionalScienceandUrbanEconomics,
  title = {Rational Expectations in Urban Economics},
  author = {Berliant, Marcus and Yu, Chia-Ming},
  year = 2013,
  month = mar,
  journal = {Regional Science and Urban Economics},
  volume = {43},
  number = {2},
  pages = {197--208},
  issn = {01660462},
}

@article{blarel1992WorldBankEconRev,
  title = {The {{Economics}} of {{Farm Fragmentation}}: {{Evidence}} from {{Ghana}} and {{Rwanda}}},
  shorttitle = {The {{Economics}} of {{Farm Fragmentation}}},
  author = {Blarel, Benoit and Hazell, Peter and Place, Frank and Quiggin, John},
  year = 1992,
  journal = {The World Bank Economic Review},
  volume = {6},
  number = {2},
  pages = {233--254},
  issn = {0258-6770, 1564-698X},
}

@article{brooks2016AmericanEconomicJournal:EconomicPolicy,
  title = {From {{Today}}'s {{City}} to {{Tomorrow}}'s {{City}}: {{An Empirical Investigation}} of {{Urban Land Assembly}}},
  shorttitle = {From {{Today}}'s {{City}} to {{Tomorrow}}'s {{City}}},
  author = {Brooks, Leah and Lutz, Byron},
  year = 2016,
  month = aug,
  journal = {American Economic Journal: Economic Policy},
  volume = {8},
  number = {3},
  pages = {69--105},
  issn = {1945-7731, 1945-774X},
}

@article{brown1982Econometrica,
  title = {On the {{Estimation}} of {{Structural Hedonic Price Models}}},
  author = {Brown, James N. and Rosen, Harvey S.},
  year = 1982,
  month = may,
  journal = {Econometrica},
  volume = {50},
  number = {3},
  eprint = {1912614},
  eprinttype = {jstor},
  pages = {765},
  issn = {00129682},
}

@article{brownstone1991TheReviewofEconomicsandStatistics,
  title = {Zoning, {{Returns}} to {{Scale}}, and the {{Value}} of {{Undeveloped Land}}},
  author = {Brownstone, David and Vany, Arthur De},
  year = 1991,
  month = nov,
  journal = {The Review of Economics and Statistics},
  volume = {73},
  number = {4},
  eprint = {2109409},
  eprinttype = {jstor},
  pages = {699},
  issn = {00346535},
}

@article{caruso2015DatainBrief,
  title = {Housing Land Transaction Data and Structural Econometric Estimation of Preference Parameters for Urban Economic Simulation Models},
  author = {Caruso, Geoffrey and Cavailh{\`e}s, Jean and Peeters, Dominique and Thomas, Isabelle and Frankhauser, Pierre and Vuidel, Gilles},
  year = 2015,
  month = dec,
  journal = {Data in Brief},
  volume = {5},
  pages = {447--452},
  issn = {23523409},
}

@article{chiappori2010EconTheory,
  title = {Hedonic Price Equilibria, Stable Matching, and Optimal Transport: Equivalence, Topology, and Uniqueness},
  shorttitle = {Hedonic Price Equilibria, Stable Matching, and Optimal Transport},
  author = {Chiappori, Pierre-Andr{\'e} and McCann, Robert J. and Nesheim, Lars P.},
  year = 2010,
  month = feb,
  journal = {Economic Theory},
  volume = {42},
  number = {2},
  pages = {317--354},
  issn = {0938-2259, 1432-0479},
}

@article{colwell1997JournalofUrbanEconomics,
  title = {The {{Structure}} of {{Urban Land Prices}}},
  author = {Colwell, Peter F. and Munneke, Henry J.},
  year = 1997,
  month = may,
  journal = {Journal of Urban Economics},
  volume = {41},
  number = {3},
  pages = {321--336},
  issn = {00941190},
}

@article{colwell1999TheJournalofRealEstateFinanceandEconomics,
  title = {Land {{Prices}} and {{Land Assembly}} in the {{CBD}}},
  author = {Colwell, Peter F. and Munneke, Henry J.},
  year = 1999,
  month = mar,
  journal = {The Journal of Real Estate Finance and Economics},
  volume = {18},
  number = {2},
  pages = {163--180},
  issn = {0895-5638, 1573-045X},
}

@article{colwell1993TheReviewofEconomicsandStatistics,
  title = {A {{Comment}} on {{Zoning}}, {{Returns}} to {{Scale}}, and the {{Value}} of {{Undeveloped Land}}},
  author = {Colwell, Peter F. and Sirmans, C. F.},
  year = 1993,
  month = nov,
  journal = {The Review of Economics and Statistics},
  volume = {75},
  number = {4},
  eprint = {2110041},
  eprinttype = {jstor},
  pages = {783},
  issn = {00346535},
}

@article{demange1985Econometrica,
  title = {The {{Strategy Structure}} of {{Two-Sided Matching Markets}}},
  author = {Demange, Gabrielle and Gale, David},
  year = 1985,
  month = jul,
  journal = {Econometrica},
  volume = {53},
  number = {4},
  eprint = {1912658},
  eprinttype = {jstor},
  pages = {873},
  issn = {00129682},
}

@article{demange1986JournalofPoliticalEconomy,
  title = {Multi-{{Item Auctions}}},
  author = {Demange, Gabrielle and Gale, David and Sotomayor, Marilda},
  year = 1986,
  month = aug,
  journal = {Journal of Political Economy},
  volume = {94},
  number = {4},
  pages = {863--872},
  issn = {0022-3808, 1537-534X},
}

@article{ekeland2010EconTheory,
  title = {Optimal Transportation and the Falsifiability of Incompletely Specified Economic Models},
  author = {Ekeland, Ivar and Galichon, Alfred and Henry, Marc},
  year = 2010,
  month = feb,
  journal = {Economic Theory},
  volume = {42},
  number = {2},
  pages = {355--374},
  issn = {0938-2259, 1432-0479},
}

@article{ekeland2004JournalofPoliticalEconomy,
  title = {Identification and {{Estimation}} of {{Hedonic Models}}},
  author = {Ekeland, Ivar and Heckman, James~J. and Nesheim, Lars},
  year = 2004,
  month = feb,
  journal = {Journal of Political Economy},
  volume = {112},
  number = {S1},
  pages = {S60-S109},
  issn = {0022-3808, 1537-534X},
}

@article{epple1987JournalofPoliticalEconomy,
  title = {Hedonic {{Prices}} and {{Implicit Markets}}: {{Estimating Demand}} and {{Supply Functions}} for {{Differentiated Products}}},
  shorttitle = {Hedonic {{Prices}} and {{Implicit Markets}}},
  author = {Epple, Dennis},
  year = 1987,
  month = feb,
  journal = {Journal of Political Economy},
  volume = {95},
  number = {1},
  pages = {59--80},
  issn = {0022-3808, 1537-534X},
}

@article{epple1999JournalofPoliticalEconomy,
  title = {Estimating {{Equilibrium Models}} of {{Local Jurisdictions}}},
  author = {Epple, Dennis and Sieg, Holger},
  year = 1999,
  month = aug,
  journal = {Journal of Political Economy},
  volume = {107},
  number = {4},
  pages = {645--681},
  issn = {0022-3808, 1537-534X},
}

@book{fujita1989,
  title = {Urban {{Economic Theory}}: {{Land Use}} and {{City Size}}},
  author = {Fujita, Masahisa},
  year = 1989,
  publisher = {Cambridge University Press},
  address = {Cambridge}
}

@article{hall2007Ann.Statist.,
  title = {Nonparametric Estimation When Data on Derivatives Are Available},
  author = {Hall, Peter and Yatchew, Adonis},
  year = 2007,
  month = feb,
  journal = {The Annals of Statistics},
  volume = {35},
  number = {1},
  pages = {300--323},
  issn = {0090-5364, 2168-8966},
}

@book{hayashi2021,
  title = {Microeconomic {{Theory}} for the {{Social Sciences}}},
  author = {Hayashi, Takashi},
  year = 2021,
  publisher = {Springer Singapore},
  address = {Singapore},
  isbn = {978-981-16-3540-3 978-981-16-3541-0},
}

@article{kaneko1983JournalofUrbanEconomics,
  title = {Housing Markets with Indivisibilities},
  author = {Kaneko, Mamoru},
  year = 1983,
  month = jan,
  journal = {Journal of Urban Economics},
  volume = {13},
  number = {1},
  pages = {22--50},
  issn = {00941190},
}

@incollection{kaneko2017SpatialInteractionModels,
  title = {An {{Equilibrium-Econometric Analysis}} of {{Rental Housing Markets}} with {{Indivisibilities}}},
  booktitle = {Spatial {{Interaction Models}}},
  author = {Kaneko, Mamoru and Ito, Tamon},
  editor = {Mallozzi, Lina and D'Amato, Egidio and Pardalos, Panos M.},
  year = 2017,
  volume = {118},
  pages = {193--223},
  publisher = {Springer International Publishing},
  address = {Cham},
  isbn = {978-3-319-52653-9 978-3-319-52654-6}
}

@article{kaneko2006JournalofUrbanEconomics,
  title = {Duality in Comparative Statics in Rental Housing Markets with Indivisibilities},
  author = {Kaneko, Mamoru and Ito, Tamon and Osawa, Yu-ichi},
  year = 2006,
  month = jan,
  journal = {Journal of Urban Economics},
  volume = {59},
  number = {1},
  pages = {142--170},
  issn = {00941190},
}

@article{kyriakidis2004GeographicalAnalysis,
  title = {A {{Geostatistical Framework}} for {{Area}}-to-{{Point Spatial Interpolation}}},
  author = {Kyriakidis, Phaedon C.},
  year = 2004,
  journal = {Geographical Analysis},
  volume = {36},
  number = {3},
  pages = {259--289},
  issn = {1538-4632},
}

@article{lafountain2008IntJofEconomicTheory,
  title = {Core Equivalence for Residential Land Use Models},
  author = {LaFountain, Courtney},
  year = 2008,
  month = dec,
  journal = {International Journal of Economic Theory},
  volume = {4},
  number = {4},
  pages = {459--481},
  issn = {1742-7355, 1742-7363},
}

@article{lancaster1966JournalofPoliticalEconomy,
  title = {A {{New Approach}} to {{Consumer Theory}}},
  author = {Lancaster, Kelvin J.},
  year = 1966,
  month = apr,
  journal = {Journal of Political Economy},
  volume = {74},
  number = {2},
  pages = {132--157},
  issn = {0022-3808, 1537-534X},
}

@article{lin2000LandEconomics,
  title = {The {{Relationship}} between the {{Price}} of {{Land}} and {{Size}} of {{Plot When Plots Are Small}}},
  author = {Lin, Tzu-Chin and Evans, Alan W.},
  year = 2000,
  month = aug,
  journal = {Land Economics},
  volume = {76},
  number = {3},
  eprint = {3147036},
  eprinttype = {jstor},
  pages = {386},
  issn = {00237639},
}

@article{liotta2022RegionalScienceandUrbanEconomics,
  title = {Testing the Monocentric Standard Urban Model in a Global Sample of Cities},
  author = {Liotta, Charlotte and Vigui{\'e}, Vincent and Lepetit, Quentin},
  year = 2022,
  month = nov,
  journal = {Regional Science and Urban Economics},
  volume = {97},
  pages = {103832},
  issn = {01660462},
}

@book{mas-colell1995,
  title = {Microeconomic Theory},
  author = {{Mas-Colell}, Andreu and Whinston, Michael Dennis and Green, Jerry R.},
  year = 1995,
  publisher = {Oxford University Press},
  address = {New York},
  isbn = {978-0-19-507340-9 978-0-19-510268-0},
  lccn = {HB172 .M6247 1995},
}

@article{mcmillen1996JournalofUrbanEconomics,
  title = {One {{Hundred Fifty Years}} of {{Land Values}} in {{Chicago}}: {{A Nonparametric Approach}}},
  shorttitle = {One {{Hundred Fifty Years}} of {{Land Values}} in {{Chicago}}},
  author = {McMillen, Daniel P.},
  year = 1996,
  month = jul,
  journal = {Journal of Urban Economics},
  volume = {40},
  number = {1},
  pages = {100--124},
  issn = {00941190},
}

@article{milgrom2026Rev.Econ.Stud.,
  title = {A {{Walrasian Mechanism}} with {{Markups}} for {{Nonconvex Markets}}},
  author = {Milgrom, Paul and Watt, Mitchell},
  year = 2026,
  month = may,
  journal = {Review of Economic Studies},
  volume = {93},
  number = {3},
  pages = {1995--2020},
  issn = {0034-6527, 1467-937X},
}

@article{miyake2003Reg.Sci.UrbanEcon.,
  title = {Precise Computation of a Competitive Equilibrium of the Discrete Land Market Model},
  author = {Miyake, Mitsunobu},
  year = 2003,
  journal = {Regional Science and Urban Economics},
  volume = {33},
  number = {6},
  pages = {721--743},
  issn = {01660462}
}

@article{rosen1974JournalofPoliticalEconomy,
  title = {Hedonic {{Prices}} and {{Implicit Markets}}: {{Product Differentiation}} in {{Pure Competition}}},
  shorttitle = {Hedonic {{Prices}} and {{Implicit Markets}}},
  author = {Rosen, Sherwin},
  year = 1974,
  month = jan,
  journal = {Journal of Political Economy},
  volume = {82},
  number = {1},
  pages = {34--55},
  issn = {0022-3808, 1537-534X},
}

@article{shapley1971IntJGameTheory,
  title = {The Assignment Game {{I}}: {{The}} Core},
  shorttitle = {The Assignment Game {{I}}},
  author = {Shapley, L. S. and Shubik, M.},
  year = 1971,
  month = dec,
  journal = {International Journal of Game Theory},
  volume = {1},
  number = {1},
  pages = {111--130},
  issn = {0020-7276, 1432-1270},
}

@article{tobler1979JournaloftheAmericanStatisticalAssociation,
  title = {Smooth {{Pycnophylactic Interpolation}} for {{Geographical Regions}}},
  author = {Tobler, Waldo R.},
  year = 1979,
  month = sep,
  journal = {Journal of the American Statistical Association},
  volume = {74},
  number = {367},
  pages = {519--530},
  issn = {0162-1459, 1537-274X},
}

@article{white1988JournalofUrbanEconomics,
  title = {Large Lot Zoning and Subdivision Costs: {{A}} Test},
  shorttitle = {Large Lot Zoning and Subdivision Costs},
  author = {White, James R.},
  year = 1988,
  month = may,
  journal = {Journal of Urban Economics},
  volume = {23},
  number = {3},
  pages = {370--384},
  issn = {00941190},
}

@article{wrenn2015RegionalScienceandUrbanEconomics,
  title = {Time Is Money: {{An}} Empirical Examination of the Effects of Regulatory Delay on Residential Subdivision Development},
  shorttitle = {Time Is Money},
  author = {Wrenn, Douglas H. and Irwin, Elena G.},
  year = 2015,
  month = mar,
  journal = {Regional Science and Urban Economics},
  volume = {51},
  pages = {25--36},
  issn = {01660462},
}

\end{document}